\documentclass[10pt,pre,showpacs,amsmath,amssymb,floatfix,onecolumn,nofootinbib,superscriptaddress]{revtex4-1}
\pdfoutput=1 
\usepackage[english]{babel}
\usepackage{graphicx}% Include figure files
\usepackage{epsfig}% Include figure files
\usepackage{dcolumn}% Align table columns on decimal point
\usepackage{bm}% bold math
\usepackage{float}
\makeatletter
\let\newfloat\newfloat@ltx
\makeatother
\usepackage{algorithm}
\usepackage{algpseudocode}
\usepackage{color}
\usepackage{amsthm}
\usepackage{amsfonts}
\usepackage{hyperref}
\usepackage{bbm}
\usepackage[toc,page]{appendix}
\usepackage{caption}
\usepackage{subcaption}
\usepackage{paralist}

\begin{document}

\title{Optimization of the dynamic transition in the continuous coloring problem}

\author{Angelo Giorgio Cavaliere}
\affiliation{Dipartimento di Fisica, Universit\`{a} ``La Sapienza'', P.le A. Moro 5, 00185, Rome, Italy}

\author{Thibault Lesieur}
\affiliation{Dipartimento di Fisica, Universit\`{a} ``La Sapienza'', P.le A. Moro 5, 00185, Rome, Italy}

%\affiliation{
%$^1$ Sapienza University of Rome \\
%$^*$ To whom correspondence shall be sent: %angelog.cavaliere@gmail.com, lesieur.thibault@gmail.com\\
%}

\author{Federico Ricci-Tersenghi}
\affiliation{Dipartimento di Fisica, Universit\`{a} ``La Sapienza'', P.le A. Moro 5, 00185, Rome, Italy}
\affiliation{INFN, Sezione di Roma1, and CNR--Nanotec, Rome unit, P.le A. Moro 5, 00185, Rome, Italy}

\begin{abstract}
\textbf{
Random constraint satisfaction problems can exhibit a phase where the number of constraints per variable $\alpha$ makes the system solvable in theory on the one hand, but also makes the search for a solution hard, meaning that common algorithms such as Monte-Carlo method fail to find a solution. The onset of this hardness is deeply linked to the appearance of a dynamical phase transition where the phase space of the problem breaks into an exponential number of clusters. The exact position of this dynamical phase transition is not universal with respect to the details of the Hamiltonian one chooses to represent a given problem. In this paper, we develop some theoretical tools in order to find a systematic way to build a Hamiltonian that maximizes the dynamic $\alpha_{\rm d}$ threshold.
To illustrate our techniques, we will concentrate on the problem of continuous coloring, where one tries to set an angle $x_i \in [0;2\pi]$ on each node of a network in such a way that no adjacent nodes are closer than some threshold angle $\theta$, that is $\cos(x_i - x_j) \leq \cos\theta$. This problem can be both seen as a continuous version of the discrete graph coloring problem or as a one-dimensional version of the the Mari-Krzakala-Kurchan (MKK) model. The relevance of this model stems from the fact that continuous constraint satisfaction problems on sparse random graphs remain largely unexplored in statistical physics. We show that for sufficiently small angle $\theta$ this model presents a random first order transition and compute the dynamical, condensation and Kesten-Stigum transitions; we also compare the analytical predictions with Monte Carlo simulations for values of $\theta = 2\pi/q$, ~$q \in \mathbb{N}$. Choosing such values of $q$ allows us to easily compare our results with the renowned problem of discrete coloring. 
}
\end{abstract}

\date{\today}
\maketitle

\tableofcontents

\newpage

\section{Introduction}

In this work we study in detail a constraint satisfaction problem with continuous variables and defined on sparse random graphs, that goes under the name of \emph{continuous coloring}.
The problem is essentially the following: given a sparse random graph $G=(V,E)$, which can be \emph{e.g.} Erd{\H o}s-R\'enyi or random regular, the aim is to assign a `color' $x_i\in[0,2\pi)$ to each vertex in $V$, such that the colors assigned to any pair of connected vertices $(ij)\in E$ are different enough, that is they satisfy $\cos(x_i-x_j) \le \cos\theta$ for some threshold value $\theta$.

This model can be regarded as a continuous version of the well-known problem of discrete coloring of random graphs \cite{zdeborovaPhaseTransitionsColoring2007}. Moreover, it also corresponds to the 1-dimensional version of the Mari-Kurchan-Krzakala (MKK) model~\cite{mariJammingGlassTransitions2009}, a mean-field approximation to models of hard spheres showing jamming. In the MKK model, $d$-dimensional hard spheres living in a $d$-dimensional box with periodic boundary conditions interact with only a finite number of other particles, according to an underlying sparse graph network. From this point of view, continuous coloring can thus be interpreted also as an `angle-packing' problem, with the obvious identification of $\theta$ with the diameter of the particles~\cite{krzakalaLandscapeAnalysisConstraint2007}.

The MKK model for $d=2$ was numerically studied in~\cite{mariJammingGlassTransitions2009}, exhibiting the presence of a random first order transition (RFOT) when increasing the diameter of the spheres (the packing fraction), for sufficiently high connectivities. A generalized version of the MKK model, accounting also for $p$-body interactions, was studied in~\cite{mezardSolutionSolvableModel2011} for different values of $p$ and $d$ through the cavity (or belief propagation) formalism. The case $d=1$ and $p=2$, which we are interested in this work, was shown to undergo a RFOT for large connectivities. However, the dynamic or clustering threshold was not computed. 

Let us start explaining why we believe it is very useful to study this model.
The continuous coloring problem possesses \emph{all} the following features:
\begin{compactitem}
\item it is a constraint satisfaction problem (CSP),
\item having continuous variables,
\item defined on a sparse random graph,
\item showing a random first order transition (RFOT).
\end{compactitem}

The most famous and well-studied model showing a RFOT is the spherical $p$-spin model: in this case one has $N$ real (unbounded) variables $x_i$, subject only to a global constraint $\sum_i x_i^2 = N$. Unfortunately, this model is well defined only on very dense graphs, because as soon as one makes the interactions slightly sparser the model ground state condensates on a small subset of the variables, becoming meaningless from the physical point of view. In order to have a very sparse model, one needs to avoid the above condensation phenomenon, and this can be achieved either adding a Lagrange multiplier to each real variable, or more simply by using variables defined on a bounded domain. Among the latter models we have all CSPs with discrete variables (\emph{e.g.} $q$-col, $k$-SAT, $k$-XORSAT). Willing to use continuous variables with bounded domain, the simplest way is to choose models with vector spins, \emph{e.g.}\ XY or Heisenberg spins. The continuous coloring problem we study can indeed be seen as an XY model where variables are unit norm vectors of 2 components, $\vec{s}_i=(\cos(x_i),\sin(x_i))$, the constraints that we enforce being written as $\vec{s}_i\cdot\vec{s}_j \le \cos\theta$. A recent work addresses the glass transition of this kind of rotational degrees of freedom, but focusing on the opposite limit of a large number of vector components~\cite{yoshinoDisorderfreeSpinGlass2018}. 

The results of~\cite{mezardSolutionSolvableModel2011} are particularly encouraging, since it is not obvious at all that taking a discrete CSP, like $q$-coloring, and transforming it to a version with continuous variables, the physics, in this case the RFOT, is preserved. The change in the symmetry of variables, from $\mathbb{Z}_q$ to $O(2)$, is drastic and may change the nature of the phase transition. In the successful case we do find a model with all the above features (and this is what we are going to confirm in Section~\ref{sec:model}), we have in our hand a model which is very useful in many aspects. Let us list those aspects we find more interesting:
\begin{compactitem}
\item being defined on a locally tree-like graph, the model can be solved analytically via the cavity method, and the location of phase transitions can be computed with high accuracy;
\item having bounded variables interacting in a sparse way, the numerical simulation of the model can be made very efficiently, and thus a meaningful comparison between numerical results and analytical solution can be done;
\item being a CSP with continuous variables, and where the density of constraints can be varied continuously, the model would show a jamming transition, for which it can be considered as a very simple sparse mean-field model for jamming (please notice that dense mean-field models for jamming, as the perceptron, have problems when generalized to the sparse case\footnote{The delicate point is once more given by the fact that one generally considers continuous models with unbounded variables. In the case of the perceptron, if a variable is subjected to $K$ constraints, the probability for all the random obstacles' components relative to that variable to have the same sign, \emph{e.g.} to be all positive, is $2^{-K}$. In this case that variable can satisfy all of its constraints by taking arbitrarily large values (irrespective of other variables) and the model is ill defined. While in the perceptron $K=O(N)$ and this event does never occur, in the diluted case $K$ is finite and the model is ill defined with a finite probability.});
\item having a RFOT, we expect the model to exhibit a dynamical phase transition and, below the dynamic transition temperature, the energy relaxation to get stuck at some threshold energy connected in some way to the topological properties of the energy landscape (\emph{e.g.} to the spectrum of the energy Hessian at the stationary points) that can be computed, for sufficiently smooth interaction potentials, thanks to the continuous nature of the variables;
\item being defined on a sparse random graph, one can change the mean degree, thus testing very different regimes from the dense one to the very sparse one, thus better understanding how much of the classical (dense) mean-field physical behaviour is preserved in the sparse regime (this is particularly relevant to understand why it is so difficult to find good glass models in low dimensions where the number of nearest neighbours becomes small).
\end{compactitem}
\vspace{0.3mm}
To the best of our knowledge, the above aspects have not been investigated in detail in a \emph{single} model before.

\vspace{1.1mm}
The second part of this work will focus on a very interesting aspect, namely the possibility of optimizing the dynamical threshold by reweighting properly the solutions of the problem.
It is well known that in complex CSP, like the one we are studying, solutions may have very different features. For example, in discrete random CSP, solutions organize in clusters of very different sizes and of different nature (\emph{e.g.} with or without frozen variables) \cite{krzakalaGibbsStatesSet2007,zdeborovaPhaseTransitionsColoring2007,montanariClustersSolutionsReplica2008}. Solutions that dominate the thermodynamics may be very different from the ones which are found via the best solving algorithms \cite{krzakalaLandscapeAnalysisConstraint2007}, and this makes difficult the connection between thermodynamics and the behaviour of dynamical processes searching for solutions. Reweighting the solutions, that is giving them a non-uniform weight, is a simple way to count the atypical solutions that would not weigh enough in the uniform measure (which is usually adopted in order to derive the equilibrium phase diagram of random CSPs). 

This program has been pursued in the literature following slightly different approaches: ~\cite{baldassiSubdominantDenseClusters2015,baldassiLocalEntropyMeasure2016,baldassiUnreasonableEffectivenessLearning2016} favour solutions which are surrounded by a higher number of other solutions in configuration space; the relevance of subdominant clusters of solutions with high internal entropy has been recently pointed out also in~\cite{zhaoMaximallyFlexibleSolutions2020}; on the contrary~\cite{braunsteinLargeDeviationsWhitening2016} uses the number of frozen variables inside each cluster in order to weight differently the solutions. The simplest approach one can resort to is, however, to directly bias the model Hamiltonian. We will perform an optimization of the interaction potential with the aim of postponing as more as possible the dynamical phase transition. 
This has been done recently in~\cite{budzynskiBiasedLandscapesRandom2019} for the discrete CSP of bicoloring random hypergraphs (a generalization to larger interaction range for the bias is considered in~\cite{budzynskiBiasedMeasuresRandom2020}) and was done in~\cite{sellittoThermodynamicDescriptionColloidal2013,maimbourgGeneratingDensePackings2018} for models of hard spheres in infinite dimensions.
The idea behind this optimization is that, by working with the reweighted interaction potential, the long range correlations leading to the ergodicity breaking at the dynamic phase transition will appear later, and algorithms should find solutions in an easier way in these `biased landscapes'. The innovative aspect in the present work with respect  to~\cite{budzynskiBiasedLandscapesRandom2019} is that the optimization of the potential is performed in a semi-automatic way (similarly in spirit to the method of~\cite{maimbourgGeneratingDensePackings2018}), by modifying the interaction potential which is a function in $[0,2\pi)$, so formally with an infinite number of parameters (in practice we discretize it with a very large number of points).

\subsection{Main results and structure of the paper}

In order to help the reader, we start with a summary of the main results, referring to the parts of the paper were they are discussed in detail.

Section~\ref{sec:model} is devoted to the definition of the Continuous Coloring problem (CCP) and the corresponding model Hamiltonian, that will allow us to perform a statistical physics study both in the temperature $T$ and in the constraints per variable ratio $\alpha$. Special attention is devoted to highlight the differences and similarities with the Discrete Coloring problem (DCP). 
In particular, given that solutions to the DCP are a subset of the solutions to the CCP, one would expect the latter to have a much larger entropy of solutions and thus being simpler to solve.
Our first, unexpected result is that the computation of the phase diagrams we perform in Section~\ref{sec:phasediag} tells us a different story: the dynamical phase transition in the CCP, when present, happens much before that in the DCP, so there is a large range of $\alpha$ values where finding a solution to DCP is easy, while in CCP ergodicity is already broken and some classes of algorithms may undergo a dynamical arrest in the search for solutions.

We explain the apparent paradox by interpreting the different coloring problems as Bayes optimal inference problems (see section~\ref{sec:planted_model}, \emph{Planted model}) only differing in the a priori distribution, which is flat on $[0,2\pi)$ in the CCP, while it is the sum of delta functions in the DCP. Although it is true that a flat prior produces a coloring problem with a larger entropy of solutions, it is also evident that it will tend to set variables so that constraints are satisfied in a very loose (\emph{i.e.} inefficient) way.
We conclude Section~\ref{sec:model} by anticipating the result of our optimization of the model Hamiltonian and showing the new phase diagram in the biased measure.

Section~\ref{sec:numerical_determination} discusses the numerical strategies adopted in order to obtain accurate estimates of the thermodynamic thresholds. In particular, we address the numerical solution via population dynamics of the saddle point equation derived under the one-step replica symmetry breaking (1RSB) ansatz with Parisi parameter $m=1$. This provides independent estimates of $\alpha_d$ by considering both the divergence of the overlap relaxation time for $\alpha\to\alpha_d^{-}$ and the birth of instability in the non-trivial solution for $\alpha\to\alpha_d^{+}$. The equilibrium relaxation time from Monte Carlo simulations on real instances is also considered, and found in reasonable agreement with the analytical predictions.

In Section~\ref{sec:optimizing} we present how we solved the problem of maximizing the dynamical phase transition by modifying optimally the interaction potential. With respect to previous works on the subject, where the optimization was performed on very few parameters, we managed to optimize the entire interaction potential, which is a function in $[0,2\pi)$ (we discretize the interval with a very large number of points for numerical convenience), in a semi-automatic way. We find that the best optimization procedure relies on trying to maximize the complexity slightly above the dynamical threshold. In this way we can increase the dynamical threshold by more than 8\%. Finally, we show how the optimization of the dynamical phase transition through the complexity maximization procedure can be performed 
without the need to solve the belief propagation equations, but only relying on Monte Carlo measurements.

We relegate to the appendices most of the technical details. In particular, in Appendix~\ref{appendix:inference_def} we give a detailed interpretation of the planted model of section~\ref{sec:planted_model} as a Bayesian inference problem and 
define a mixed model which continuously interpolates between the DCP and CCP, while in Appendix~\ref{section:BP_CCP} we write the Belief Propagation equations valid in the planted setting. 

\section{The model} \label{sec:model}

\subsection{Definition of the model}
In the continuous coloring, $N$ real angular variables $x_i\in[0,2\pi)$, $i=1,\dotso,N$ interact via a sparse graph $G=(V,E)$ composed by a set $V$ of variable nodes (vertices), with size $\lvert V\lvert = N$, and a set $E$ of $M$ pairwise interactions between variables, $\lvert E \lvert = M$. In this work, graphs are random instances drawn from the Erd{\H o}s-R\'enyi ensemble, with average connectivity $\alpha=M/N$ (the average degree of variable nodes is hence equal to $2\alpha$). Neighbours on the graph have to satisfy the excluded-volume constraint $\cos(x_i-x_j)\leq\cos\theta$, for some excluded angle (or `diameter') $\theta\in[0,\pi)$.

We associate to the constraint satisfaction problem a Hamiltonian which exactly counts the number of violated constraints, that in the case of the continuous coloring reads
\begin{equation}
    \mathcal{H}_{\rm flat} \left(\{x\}\right)= \sum_{(ij)\in E} \mathbb{I}\big(\cos(x_i-x_j-\omega_{ij}) > \cos\theta \big),
    \label{eq:H_flat}
\end{equation}
where $\mathbb{I}(\cdots)$ is equal to 1 if the condition in the argument is satisfied, and 0 otherwise. The subscript `flat' refers to the fact that $\mathcal{H}_{\rm flat}=0$ on all the solutions to the problem. In the limit $\beta\to\infty$, 
the associated Boltzmann-Gibbs measure $Z^{-1} e^{-\beta\mathcal{H}_{\rm flat}}$ is hence uniform over the whole space of solutions, while configurations not simultaneously satisfying all the constraints have zero statistical weight. 

The parameters $\omega_{ij}\in[0,2\pi)$ are quenched random shifts living on each directed edge $i\to j$, with $\omega_{ji}=2\pi-\omega_{ij}$. The original problem, which is recovered for $\omega_{ij}=0$, is found to undergo a `crystallization' phenomenon when increasing connectivity (or packing fraction, by increasing $\theta$), or equivalently when lowering temperature, as already noted in~\cite{mezardSolutionSolvableModel2011}. In the crystal phase, variables condense around a discrete set of values on the interval $[0,2\pi)$, mimicking a $q'$-coloring packing, with $2\pi/q'>\theta$. On a real tree, with open boundary conditions, the homogeneous $\omega_{ij}=0$ and non-homogeneous models are equivalent, since the latter can be mapped onto the former by the following transformation: $\forall (ij)\in E$,  $\omega_{ij}\to 0$, $x_j \to x_j + \omega_{ij}$ and $x_k \to x_k + \omega_{ij}$ for all the $\{x_k\}$ belonging to the subtree starting from $j$ (that is excluding the whole branch containing $i$). However, in the case of random graphs, where loops are always present (even if very large), or on a finite tree with fixed boundary conditions on the leaves, this construction is not guaranteed to hold for every choice of the $\omega_{ij}$'s: loops, together with random shifts, induce frustration of the periodic order.

In order to suppress the crystal phase,~\cite{mariJammingGlassTransitions2009} adopted a small degree of polydispersity. Here we will follow two strategies. In the analytical derivation of the 1RSB belief propagation equations for the homogeneous case $\omega_{ij}=0$, we will impose translational (rotational) invariance: this allows to simplify the equations when the replica parameter is equal to $m=1$, obtaining a RS-like scheme at the price of introducing \emph{planted} random shifts in the messages (see Appendix~\ref{section:BP_CCP} for details on the derivation). By imposing translational invariance, one can disregard the ordered solution while maintaining the tree-like recursive structure of the BP equations. On the other hand, we will also support the BP predictions with direct numerical Monte Carlo simulations of the non-homogeneous model with planted random shifts in Section~\ref{sec:ZeroEnergyMC}.

The parameter $\theta\in[0,\pi)$ represents half the excluded angle around each `particle', playing the same role of the diameter in hard spheres systems. In order to make direct contact with the very well known $q$-coloring, it is useful to consider `discretized' values of $\theta$ defined as $\theta=\frac{2\pi}{q}$, with $q\in\mathbb{N}_{>1}$. The meaning of this relation is straightforward: in the $q$-coloring, colors $u\in\{1,\dotso,q\}$ can be associated to $q$-states Potts angular variables $x=\frac{2\pi}{q}(u-1)$, naturally satisfying $\cos(x_i-x_j)\leq\cos(2\pi/q)$ if $x_i\neq x_j$. In the following, we will express the diameter $\theta$ in terms of integer $q$'s. This mapping allows for the following observation: given a solution to the discrete coloring with $q$ colors, it also represents a solution to the continuous coloring with excluded angle $\theta=\frac{2\pi}{q}$ (actually, the solution is valid for any $\theta'\leq\frac{2\pi}{q}$). Conversely, the set of solutions to the continuous coloring for $\theta=\frac{2\pi}{q}$ contains all the solutions to the $q'$-coloring with $q'\leq q$. 

\subsubsection{Discretization}
\begin{figure}[htbp]
    \centering
    \includegraphics[width=.3\textwidth]{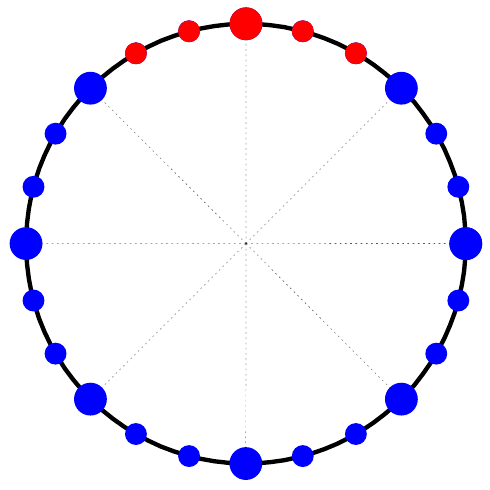}
    \caption{Discretization of precision $d=3$ for the continuous coloring. The excluded angle (on each side of the particle) is equal to $\theta=2\pi/q$ in the continuous limit, with $q=8$. As a consequence of discretization, $2d-1$ states are forbidden (red points), among the total $p=qd=24$ possible ones. Big dots, together with the dashed lines, are meant only to identify the $q$ multiples of $\theta$ along the circumference. The increase of precision $d$ with $q$ fixed corresponds to the proliferation of small dots inside each sector.}
    \label{fig:discretization}
\end{figure}

In order to numerically solve the BP equations, a discretization of the interval $[0,2\pi)$ is needed. To this end, we introduce a clock approximation with $p$ states by imposing
\begin{equation}
    x_i = \frac{2\pi}{p}t_i, \;\;\;\; t_i\in\{0,\dotso,p-1\}, \,\forall i \in V.
    \label{eq:discretization}
\end{equation}
Calling $d$ the number of clock states inside the excluded region $0\leq x<\theta$, \emph{i.e} the discretization precision, then it follows $p=qd$ (as we said, we restrict to integer $q$ for convenience). Each variable precludes $2d-1$ values to its neighbours, as depicted in figure~\ref{fig:discretization}.

\subsubsection{Biased measure}
A constraint satisfaction problem can be recast as the study of the $\beta\to\infty$ limit of a Boltzmann-Gibbs distribution which in some generality takes the form
\begin{align}
P(\{x_i\}) =  \frac{1}{Z(\beta)} \exp \big(- \beta H_0(\{x_i\}) + H_1(\{x_i\})\big).
\label{boltzmann-gibbs}
\end{align}
The function $H_0$ enforces the hard constraints and is defined by the condition to be equal to zero on satisfied configurations \emph{only}. The simplest choice, that we will adopt throughout the paper, is $H_0=\mathcal{H}_{\rm flat}$ given by eq.~\eqref{eq:H_flat}. Note that a different definition of $H_0$ (\emph{e.g.} a linear or quadratic potential) only affects the behaviour of the model as long as temperature is involved or, even at zero temperature, in the UNSAT phase $\alpha>\alpha_{\rm sat}$, but does not influence the equilibrium $\beta\to\infty$ limit in the SAT phase.

We call $H_1=O(1)$ for $\beta\to\infty$ the `soft' part of the interaction, as it alters the statistical weight of only the acceptable configurations when $\beta\to\infty$. In this way, it is possible to sample such configurations with a measure that is not necessarily flat (uniform over the solutions), as in the case of $H_1=0$. We limit ourselves to the introduction of a first-neighbours bias as in~\cite{budzynskiBiasedLandscapesRandom2019}. This implies that our bias factorizes, and one only needs to replace the local interaction term $f^{\rm flat}(x_i,x_j) \propto \exp[-\beta \,\mathbb{I}\left(\cos(x_i-x_j-\omega_{ij}) > \cos\theta \right)]$ in all the equations with a suitable biasing function $f(x_i,x_j)$, which for convenience we also take to satisfy translational invariance as the original flat Hamiltonian $H_0$. From the practical point of view, then, there is no substantial difference in solving the biased problem rather than the original one, and all the relevant equilibrium transition lines can be straightforwardly computed for any choice of the biased interaction $f$.

In this paper, we are particularly interested in the so called dynamic or clustering transition $\alpha_d$. By tuning $f$, \emph{i.e.} $H_1$, one can move the dynamical threshold to higher $\alpha$ values up to $\alpha_d^{\rm opt}=\alpha_d(f^{\rm opt})$. On the contrary, the location of the SAT/UNSAT transition, that is where the volume of the configurations that satisfy all the constraints becomes zero in the thermodynamic limit, does not depend on $H_1$ nor on the particular choice of $H_0$ by definition.

\subsubsection{Planted model}
\label{sec:planted_model}
Throughout this work, we will make extensive use of the planting technique~\cite{krzakalaHidingQuietSolutions2009}. According to this procedure, one first draws a planted configuration $\{x_i^0\}$ from some prior probability distribution, which is assumed to be factorized, $P_X(\{x_i\})=\prod_{i}P_X(x_i)$. Then, a random graph can be constructed in two ways, depending on whether random shifts are explicitly considered or not. In the latter case, one can run through all the pairs of vertices $(ij)$ and add an edge with probability $2\alpha f(x_i^0,x_j^0;\beta)/(N-1)$, where $f(x_i,x_j;\beta)=e^{-\beta \mathcal{H}(x_i,x_j)}/\int dx dy P_X(x)P_X(y)e^{-\beta \mathcal{H}(x,y)}$ and $\mathcal{H}(x_i,x_j)$ is the pairwise term in the Hamiltonian of the model (it can refer to $\mathcal{H}_{\rm flat}$ given by eq.~\eqref{eq:H_flat}, or to any biasing function as well). 
This ensures that the average degree $c$ of variable nodes is always equal to $2\alpha$
\begin{equation}
    c = (N-1) \int dx dy P_{X}(x) P_{X}(y) \frac{2\alpha f(x,y)}{(N-1)} = 2\alpha.
\end{equation}
However, we would also like different nodes to follow the same degree distribution. To this end, one can derive the average degree $c(x)$ conditional on the knowledge that $x_i^0=x$,
\begin{equation}
    c(x) = (N-1) \int dy P_{X}(y) \frac{2\alpha f(x,y)}{(N-1)} = 2\alpha \int dy P_{X}(y) f(x,y).
\end{equation}
Then, we must also ask for $\int dy P_{X}(y) f(x,y)=1$, so that $c(x)=2\alpha$ independently from $x$. One also recognizes that a bona fide Erd{\H o}s-R{\'e}nyi random graph ensemble is recovered for $f=1$.  

In the presence of random shifts, we can plant the system in a slightly different manner: given the configuration $\{x_i^0\}$ and a graph realization extracted from the proper distribution (our choice is the Erd{\H o}s-R{\'e}nyi ensemble), we associate a random variable $\omega_{ij}\in[0,2\pi)$ to each directed edge $i\to j$, and $\omega_{ji}=2\pi-\omega_{ij}$ on $j\to i$, with probability $p_{ij}(\omega_{ij})=e^{-\beta\mathcal{H}(x_i-x_j-\omega_{ij})}/\int d\omega e^{-\beta\mathcal{H}(x_i-x_j-\omega)}$. In the following, we will use this second method in order to plant the system when running Monte Carlo numerical simulations (we indeed require random shifts to suppress crystallization). Notice that one can choose, without loss of generality, $\{x_i^0\}=0$, this being equivalent to a set of local gauge transformations (valid for any $i$): $x_i\to x_i-x_i^0$, $\omega_{ij}\to\omega_{ij}-x_i^0$ ~$\forall j \in \partial i$.

Both the planted models are different from the original ones (with or without the random shifts), since planting a solution changes the graph ensemble, and, generally, its properties. This is particularly evident in the $\beta \to \infty$ limit: via planting, one can always construct for arbitrary $\alpha$, even beyond the SAT/UNSAT threshold, a pair graph-configuration of exactly zero energy. However, before the condensation transition $\alpha_c$, the planted ensembles are equivalent (provided that the original random model displays a uniform paramagnetic BP fixed point) to the original ones: this is known as \emph{quiet planting}~\cite{krzakalaHidingQuietSolutions2009}.
We can understand this fact by considering that the superimposition of a single planted cluster in the region $\alpha_d<\alpha<\alpha_c$, which is already dominated by exponentially many other clusters, should be thermodynamically undetectable as long as the planted cluster exhibits the properties of typical $m=1$ clusters from the random ensemble. This provides a very powerful tool to study the average properties of the whole clustered phase $\alpha_d<\alpha<\alpha_c$ by focusing on a single, easy to build planted cluster. 

Finally, both the DCP and CCP can be derived from the same planted model of Hamiltonian~\eqref{eq:H_flat} by allowing for a different definition of the prior distribution $P_X$ over $x\in[0,2\pi)$, namely $P_X(x)=\frac{1}{q}\sum_{k=0}^{q-1}\delta\left(x-\frac{2\pi k}{q}\right)$ for DCP and $P_X(x)=\frac{1}{2\pi}$ for CCP, where $q$ is the number of colors. This enables us to treat them in a unified Bayesian inference setting by defining a mixed model in Appendix~\ref{appendix:inference_def} which continuously interpolates between DCP and CCP. We also give an explicit definition of the interaction function $f(x_i,x_j;\beta)$ in all the different cases, also taking into account the discretization of the interval $[0,2\pi)$.

\subsection{Phase diagram}
\label{sec:phasediag}

\subsubsection{Transitions in random CSP}
\label{ExplainTransitions}
One of the achievement of the spin glass theory is the understanding of the nature and the structure of the phase space of constraint satisfaction problems as the number of constraints per variable $\alpha$ increases~\cite{krzakalaGibbsStatesSet2007}. This description allows us to understand both the random model and the planted model, which for $\alpha < \alpha_c$ have the same physical properties. For simplicity, we will describe this phase diagram in the case of a system that presents both a paramagnetic solution to the cavity equations along with a random first order transition.

\begin{itemize}
\item For $\alpha < \alpha_d$ the Gibbs measure is a Bethe measure~\cite{mezardInformationPhysicsComputation2012}. For both the planted and the random ensembles there is only one fixed point to the BP equations, the paramagnetic fixed point, and sampling from the Gibbs distribution can be achieved by Monte Carlo algorithms in polynomial time. The threshold $\alpha_d$ is usually referred to as dynamical or clustering transition, because at $\alpha_d$ the phase space breaks into an exponential number of pure states (or clusters), which Monte Carlo is unable to sample uniformly in a polynomial time.

\item For $\alpha_d < \alpha <\alpha_c$ the Gibbs measure is decomposed into an exponential number of pure states, each one corresponding to a different fixed point to the BP equations. BP initialised  in the planted solution will converge to a fixed point that is different from the paramagnetic fixed point. 
 The complexity $\Sigma$ is a measure of the number of pure states that dominate the Gibbs measure. In practice there is no known way to sample one of these pure states in polynomial time. %What is interesting with the concept of planting is, however, that the planted solution is a typical draw of the posterior distribution, and as a consequence the pure state associated with it is also typical. 
 Nevertheless, one can estimate the complexity as
\begin{align}
N\Sigma = F(\{{\nu}_{\rm para}\}) - F(\{{\nu}_{\rm planted}\}),
\label{eq:complexity_def}
\end{align}
where $F$ is the Bethe free entropy as in eq.~\eqref{FreeEnergy_Equation_Reconstruction_Sym}, $\{\nu_{\rm para}\}$ is the paramagnetic BP fixed point (we denote by $\{\nu\}$ a set of BP messages) and $\{\nu_{\rm planted}\}$ is the fixed point obtained by initializing BP in the planted solution.
As $\alpha$ increases $\Sigma$ diminishes up to the point $\alpha_c$, the condensation transition, where $\Sigma=0$. 

\item For $\alpha > \alpha_c$ it is possible to distinguish between the planted and the random model:
\begin{itemize}
\item in the random model the Gibbs distribution is dominated by the sum of a sub-exponential number of Bethe measures~\cite{mezardInformationPhysicsComputation2012}. The Gibbs distribution is not a Bethe measure anymore.
\item in the planted problem the Gibbs distribution is dominated by one Bethe measure (up to some symmetry) corresponding to the planted solution.
\end{itemize}
\end{itemize}

In this paper we make use of the planting technique in order to derive the position of the dynamical transition $\alpha_d$. This is equivalently done by inizializing Monte Carlo in a planted state or by solving the 1RSB $m=1$ cavity equations discussed in Appendix~\ref{section:BP_CCP}. Other important thresholds are the following:
\begin{itemize}
\item $\alpha_{KS}$: it is the value of $\alpha$ for which the paramagnetic solution ceases to be stable. It is known as the Kesten-Stigum transition in the tree reconstrunction literature~\cite{kestenAdditionalLimitTheorems1966,mosselInformationFlowTrees2003,jansonRobustReconstructionTrees2004}, or also as the de Almeida-Thouless local instability of the RS solution in the context of spin glasses~\cite{almeidaStabilitySherringtonKirkpatrickSolution1978}. We compute it analytically at the end of this section.
\item the \emph{rigidity} transition, that in constraint satisfaction problems with discrete variables, such as the $q$-coloring, signals the point beyond which all the dominant pure states contain a finite fraction of \emph{frozen} variables. A variable is frozen if it can take only one value in all the configurations within a given cluster. This transition can be located before or after $\alpha_c$.
\item the coloring threshold (COL/UNCOL or more generally SAT/UNSAT transition): it is the value of $\alpha$ beyond which no proper solution exists with probability one in the thermodynamic limit.
\end{itemize}

The stability analysis of the paramagnetic solution can be carried out and one finds that the Kesten-Stigum transition of the system is 
linked to the top eigenvalue of the following matrix~\cite{decelleAsymptoticAnalysisStochastic2011}
\begin{align}
T_{ab} &= n_a\left(f_{ab} - 1 \right),
\end{align}
where $\{n_a\}\in\mathbb{R}^p$ is our discretized notation for the prior $P_X(x)$ and the matrix $\{f_{ab}\}\in\mathbb{R}^{p\times p}$ is the discretized version of the interaction function $f(x,y;\beta)$, see Appendix ~\ref{sec:inference_discretized_version} for details.
The instability condition of the paramagnetic fixed point is found to be
\begin{align}
2\alpha \lambda^2 > 1,
\end{align}
where $\lambda$ is the top eigenvalue of $T$.
For the DCP and CCP the top eigenvalue of $T$ can be derived explicitly.
This yields us with the following transition lines
\begin{align}
\alpha_{\rm{KS}}^{\rm{DCP}} &= (q-1)^{2}/2, \label{KS_Potts}
\\
\alpha_{\rm KS}^{\rm CCP} &= \frac{1}{2}\left[\frac{\sin(\pi(1 - 2/q))}{\pi(1 - 2/q)} \right]^{-2} \underset{q\gg1}{=} O(q^2/8), \label{KS_ContinuousColoring}
\\
\alpha_{\rm KS}^{d-{\rm CCP}} &= \frac{1}{2}\left[\frac{\sin(((q-2)d+1)\pi/(dq))}{\sin(\pi/(dq))((q-2)d+1)}\right]^{-2}. \label{KS_DiscreteModel}
\end{align}

Eq.~\eqref{KS_Potts} represents the Kesten-Stigum transition line for the $q$-coloring, as was already derived in \cite{zdeborovaPhaseTransitionsColoring2007}.
Eq.~\eqref{KS_ContinuousColoring} is the Kesten-Stigum transition line for the continuous coloring model where $\theta=2\pi/q$, if one could deal with continuous messages.
Finally, eq.~\eqref{KS_DiscreteModel} represents the Kesten-Stigum transition line for the discretized version \eqref{eq:discretization} of the continuous coloring, for arbitrary values of $q$ and of the discretization precision $d$. By taking $d = 1$ or $d \rightarrow +\infty$, one respectively recovers \eqref{KS_Potts} and \eqref{KS_ContinuousColoring}. It is worth noticing the presence of a factor $1/4$ between the Kesten-Stigum bounds for the continuous and the discrete coloring, $\alpha_{\rm KS}^{\rm CCP} \approx \alpha_{\rm KS}^{\rm DCP}/4$.

\subsubsection{Uniform measure: Continuous vs q-coloring}
\label{phasediag-qcol}
\begin{figure}[t]
    \centering
    \begin{subfigure}[b]{.49\columnwidth}
        \includegraphics[width=\textwidth]{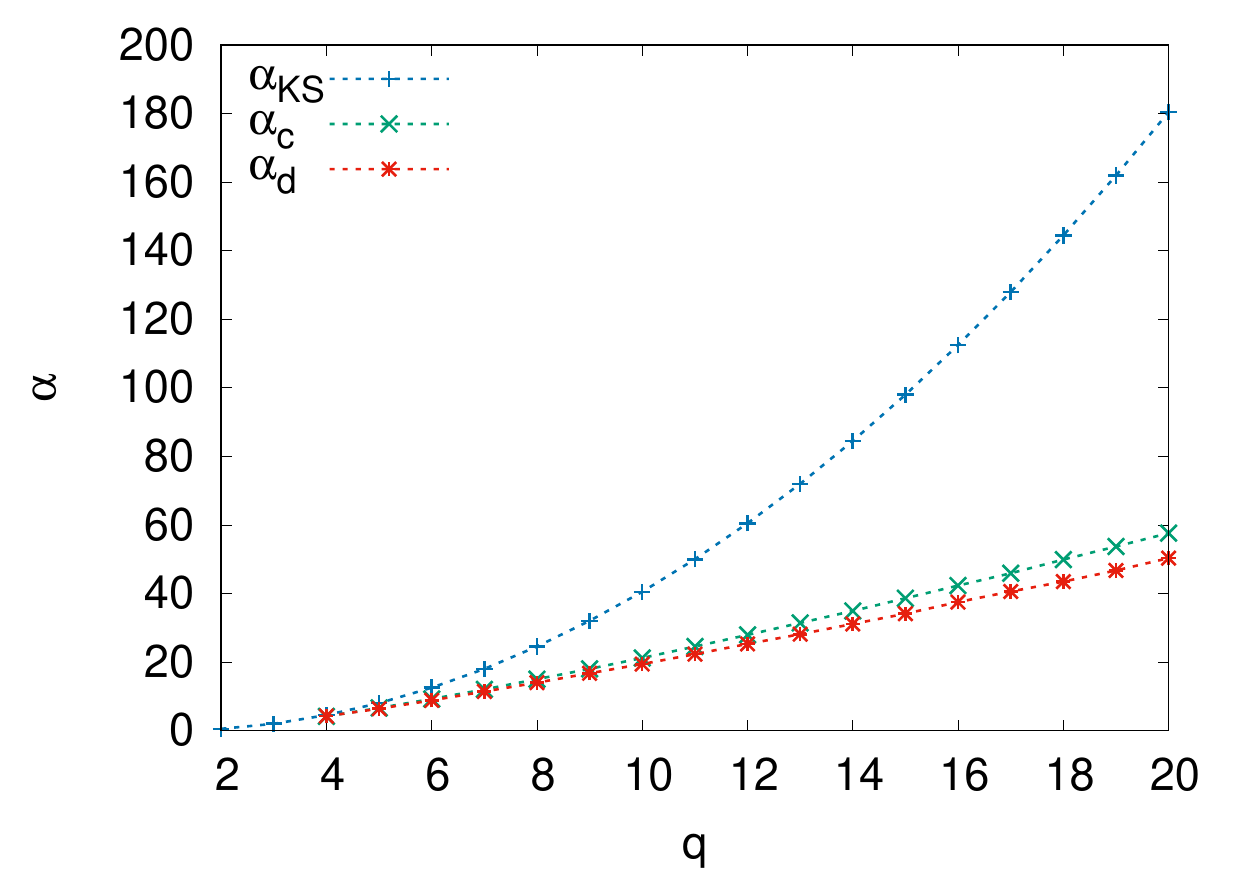}
        \caption{$q$-coloring}
    \end{subfigure}
    \hfill
    \begin{subfigure}[b]{.49\columnwidth}
        \includegraphics[width=\textwidth]{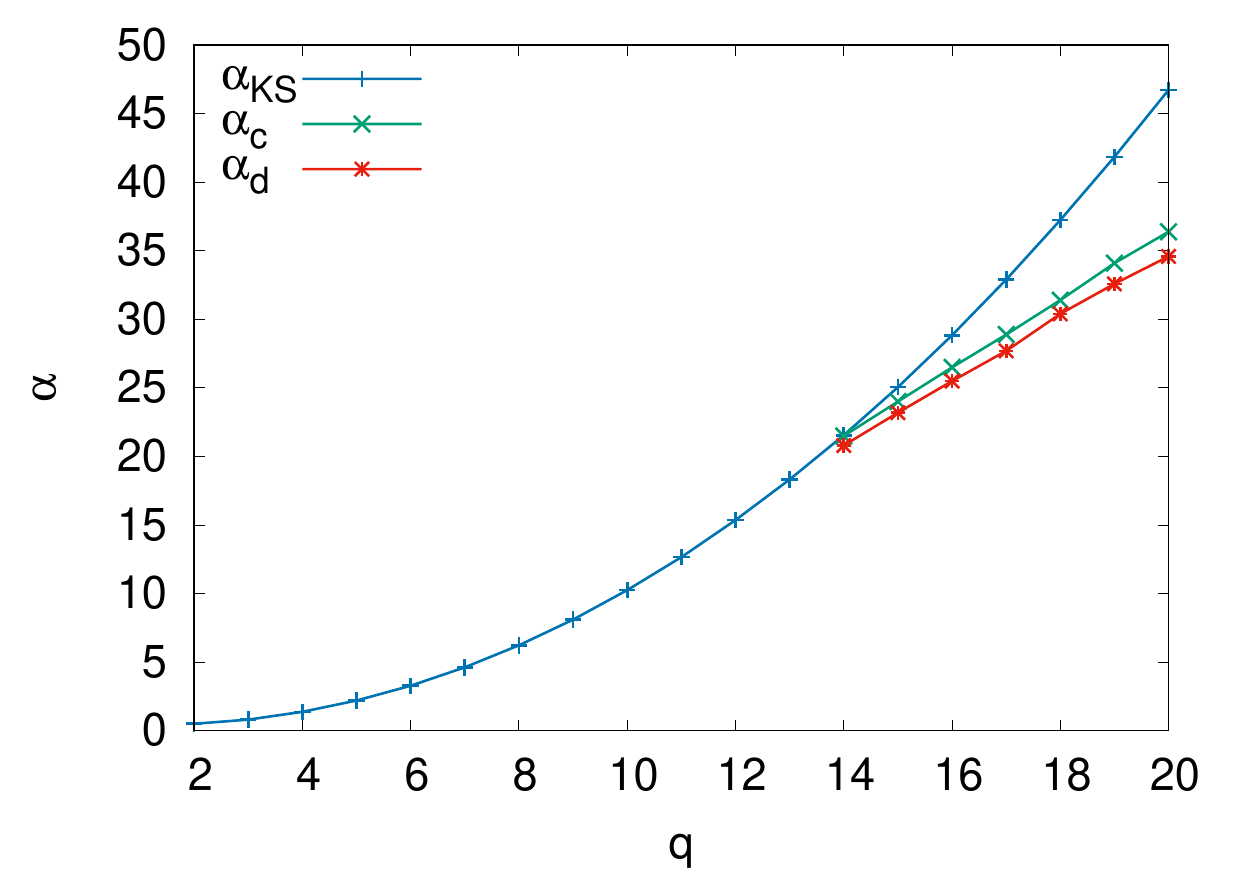}
        \caption{Continuous coloring, $d=10$}
    \end{subfigure}
    \hfill
    \begin{subfigure}[b]{.49\columnwidth}
        \includegraphics[width=\textwidth]{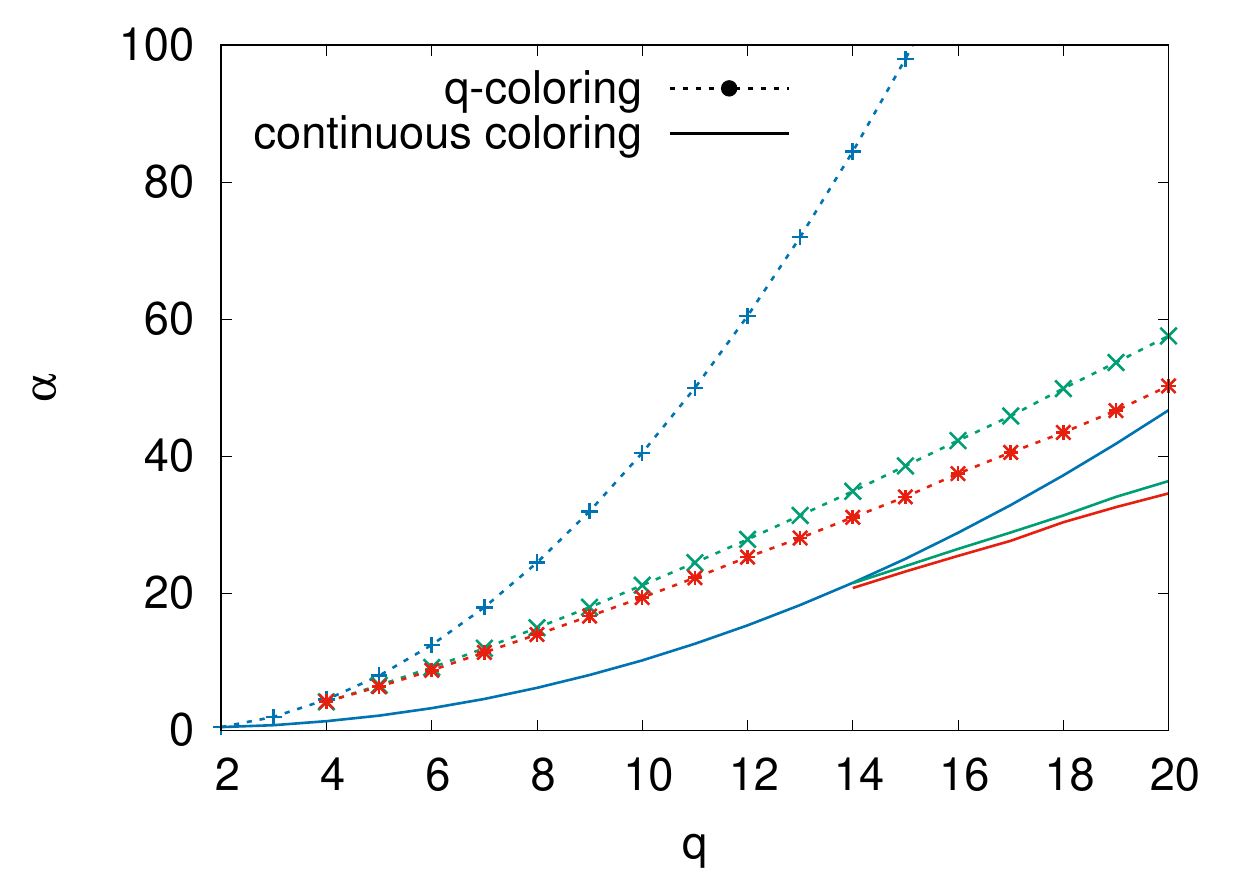}
        \caption{Comparison of the previous plots}
    \end{subfigure}
    \caption{$T=0$ phase diagram in the ($q$,\,$\alpha$) plane for both the discrete (a) and continuous coloring (b).
    Third figure (c) serves as a direct comparison between the two models.
    In the discrete coloring $q$ can only take integer values, the dotted transition lines in the figure being just meant to guide the eye. The discrete model exhibits for $q \geq 4$ a random first order transition (the three transition lines merge at $\alpha=2$ only for $q=3$)~\cite{zdeborovaPhaseTransitionsColoring2007}. 
    In the (discretized) CCP, a random first order transition is found for $q \geq 14$, where $q$ is defined by the relation $\theta=2\pi/q$. This is however just an upper bound to the actual point where the three transition lines are expected to merge. This point is likely to be located at a value of $q$ that is not an integer.
    We notice that, for a given $q$, all the transitions in the continuous model exhibit a lower value of $\alpha$ than in the $q$-coloring, see also the discussion in the text.}
    \label{fig:PhaseDiag_discr_vs_cont}
\end{figure}
In this section we discuss the phase diagram of the continuous coloring, as obtained by numerically solving the 1RSB $m=1$ population dynamics equations up to $\alpha_c$. This means that we will focus on the detection of the dynamic and condensation thresholds $\alpha_d=\alpha_d(T=0)$ and $\alpha_c=\alpha_c(T=0)$ by studying the formation of \emph{typical} clusters (with respect to the uniform measure under consideration). The same analysis will be performed in temperature in the next section, allowing the calculation of the transition lines $\alpha_d(T)$ and $\alpha_c(T)$, or equivalently $T_d(\alpha)$, $T_c(\alpha)$. All the transition lines for the continuous coloring here presented are computed using a discretization $d=10$; we refer the reader to section~\ref{section:extrapolation_continuous_limit} for an analysis on the corrections to the continuous limit, that are shown to scale as $1/d$.

The three $T=0$ transition lines $\alpha_d$, $\alpha_c$ and $\alpha_{\rm KS}$ are shown in figure~\ref{fig:PhaseDiag_discr_vs_cont} as a function of $q$ for both the $q$-coloring on the left and the discretized version of the continuous coloring ($d=10$) on the right. Looking at the values of $\alpha$ in the figure, it is evident how all the transitions appear in the continuous coloring before (\emph{i.e.} at lower $\alpha$) than in the DCP. 
This is interesting and somehow surprising, as it results that despite \emph{enlarging} the set of solutions for each value of $\alpha$ (all the solutions to the $q$-coloring also satisfy the continuous problem with $\theta=\frac{2\pi}{q}$ by definition), the clustering point is anticipated. As a practical consequence, since the ergodicity breaking point $\alpha_d$ is generally connected (even if it does not rigorously coincide) with the onset of hardness in the constraint satisfaction optimization problem,
this would imply for $q>14$ the existence of a whole $\alpha$ region where local algorithms searching for solutions to the continuous coloring get stuck and fail, while solutions to the discrete model are less and yet still easy to find. 

We interpret this fact by arguing that in the $q$-coloring the discrete prior forces the solutions to satisfy all the constraints in a more tight, \emph{i.e.} efficient, way. In this case, variables can only take $q$ values, and the angular distance between two neighbours is either 0 (violated constraint) or a multiple of $2\pi/q$. The minimum allowed distance between variables is hence $2\pi/q$, which corresponds to a \emph{contact} in the particle-system jargon. The fraction of contacts is in this case very relevant: the equilibrium probability distribution of (discrete) angular differences in the paramagnetic phase, which is valid for $\alpha<\alpha_c$, is by definition equal to 0 if two angles take the same values and it is uniform otherwise, so that the probability of having a contact is simply equal to $2/(q-1)$, where the factor 2 comes from taking into account both the possibilities to form a contact in one dimension, $x\pm2\pi/q$. This same analysis can be repeated in the case of continuous coloring. For clarity, let us consider the discretized version with $p=qd$ possible states. The fraction of exact contacts now becomes $2/(p-2d+1)\sim2/[d(q-2)]$, which is as expected heavily suppressed for $d\to\infty$. In this situation, one should rather ask for an angular interval $\Delta\equiv n\frac{2\pi}{p}$ with associated probability $1/(q-1)$ on each side of the `particle' by integrating the previous probability. In our discretized notation we can write $\frac{n}{d(q-2)+1}=\frac{1}{q-1}$, from which it follows $\Delta\approx\frac{2\pi}{q}\frac{q-2}{q-1}\approx\frac{2\pi}{q}$. In the continuous coloring, typical solutions satisfy the constraints in a very loose way: exact contacts are `smeared' over an interval comparable to the diameter of the particles. Finally, notice that reversing this perspective the discrete model can be considered, for some qualitative aspect, akin to a continuous sphere system where particles are encouraged to stick together. It is indeed very well known in the literature that, by adding a short range attraction to a hard core repulsion, one can greatly extend the liquid phase of the system~\cite{sciortinoOneLiquidTwo2002}. This phenomenon is at the very heart of the results from the biased thermodynamics approach as discussed in the following.

From figure~\ref{fig:PhaseDiag_discr_vs_cont} it is also evident that the region where the transition is continuous (small q region) results much more extended in the continuous coloring than in the discrete problem: the discrete q-coloring on Erd{\H o}s-R\'enyi random graphs is known to display a random first order transition already for $q > 3$~\cite{zdeborovaPhaseTransitionsColoring2007}, while in the continuous coloring the continuous transition ranges up to $q \approx 14$. The actual point where the transition changes its nature is in this case likely located at a value of $q$ which is not an integer, close to but smaller than $q = 14$ (we only have points for integer $q$'s). Below this point the three transition lines do coincide. In particular, since $\alpha_d = \alpha_c$, the phase dominated by an exponential number of clusters is missing.

\subsubsection{Biased measure for the CCP}
\label{phasediag-fopt}
We have performed a computation analogous in spirit to the one of ref.~\cite{maimbourgGeneratingDensePackings2018} in order to postpone as much as possible the location of the $T=0$ dynamical transition $\alpha_d$. Unfortunately, we lack in this case an analytical expression for $\alpha_d$. A possible way round is to consider a gradient descent with respect to some observable which is hopefully related with the location of $\alpha_d$: we consider the maximization of the complexity $\Sigma$, as explained in Section~\ref{sec:optimizing}. We will call the resulting optimized threshold $\alpha_d^{\rm opt}$, and the optimized interaction $f^{\rm opt}$. 
We stress, however, that $\alpha_d^{\rm opt}$ should be more safely interpreted as a lower bound to the actual optimal threshold, since there is still no rigorous proof linking the maximization of the dynamical transition with the one of $\Sigma$. On the other hand, the gradient of the complexity will be shown to be relatively simple to compute, and also accessible through Monte Carlo sampling.
If the complexity-maximization approach actually turns out to be valid in more general models with a RFOT, then we would have in our hands a very powerful and versatile tool to compute the optimized interaction also in more computer-memory consuming settings, such as in higher spatial dimensions, where BP really struggle. 

\begin{figure}[t]
    \centering
    \includegraphics[width=0.60\textwidth]{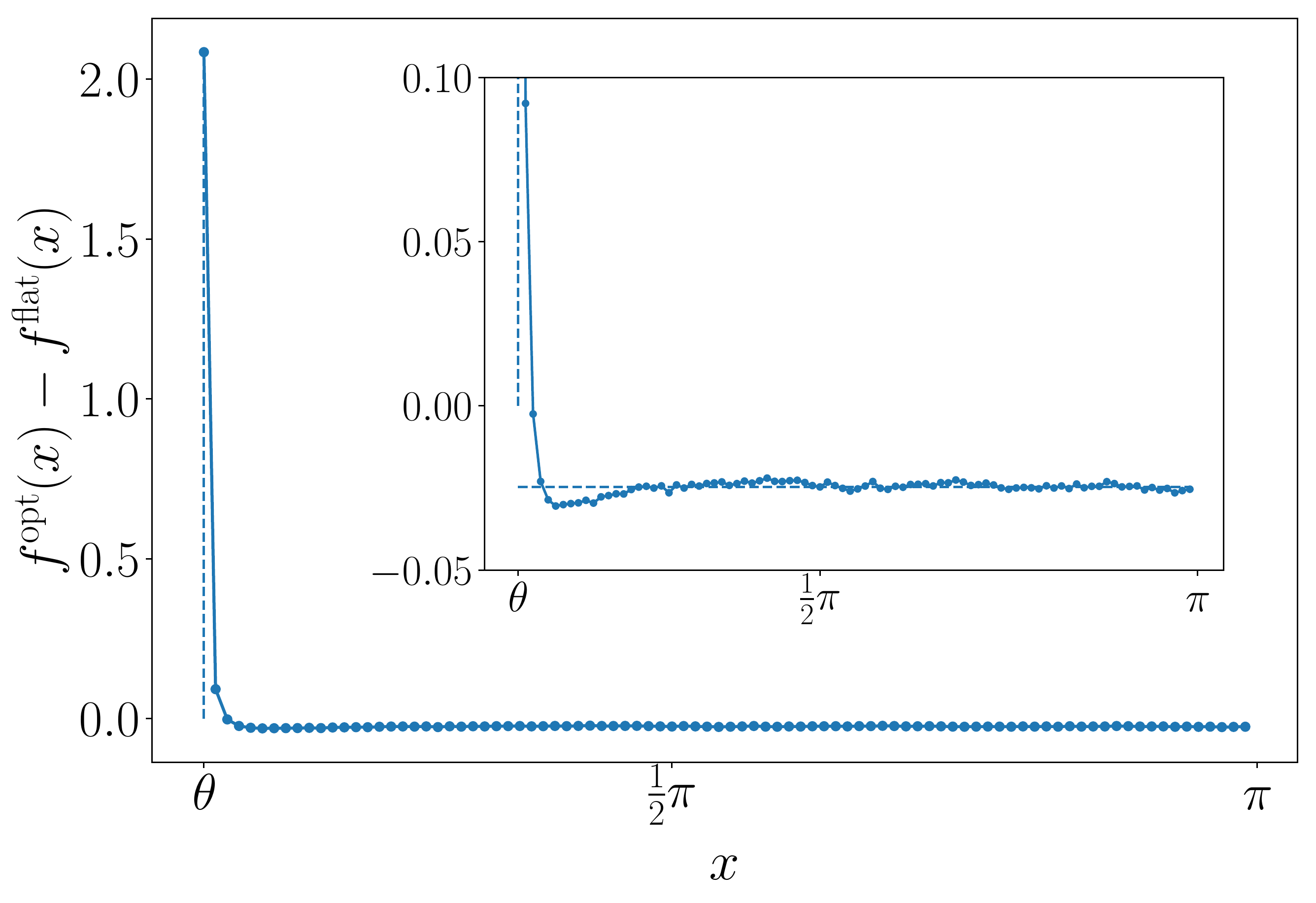}
    \caption{Soft part $\lvert x \lvert \geq \theta$ of the optimized interaction at $T=0$ for the continuous coloring ($\theta=2\pi/q$, with $q=20$) as a function of the interparticle angular distance. The function is symmetric for $x\rightarrow2\pi-x$. The difference $f^{\rm opt}-f^{\rm flat}$ ($f^{\rm flat}$ being a simple step function around $x=\theta$) does not go precisely to zero at large angles because we are subtracting two normalized probabilities, both satisfying $\sum_a f_a = p$ in the discretized case. The inset shows a zoom.}
    \label{fig:f_opt}
\end{figure}

The optimized interaction $f^{\rm opt}$ is given in figure~\ref{fig:f_opt} for the usual choice of $q=20$ and a discretization of precision $d=10$ ($p=200$ clock-states). 
The interaction $f$ is in our setting proportional to the Boltzmann-Gibbs weight, rather than to an energy, for this reason a peak in the probability for $x=\theta$ has to be interpreted as a very short range attraction favouring contacts between angles/particles. The fact that by perturbing the hard spheres Hamiltonian with a short-range attractive potential one can dramatically extend the liquid phase of the system is very well known in the literature~\cite{sciortinoOneLiquidTwo2002,dawsonHigherorderGlasstransitionSingularities2000,sellittoThermodynamicDescriptionColloidal2013,maimbourgGeneratingDensePackings2018,charbonneauPostponingDynamicalTransition2020}. This is also very relevant for soft matter colloidal systems, where such interaction potentials can be experimentally engineered~\cite{poonPhysicsModelColloid2002,eckertReentrantGlassTransition2002}. With our choice for the potential, the zero temperature dynamic threshold estimated from BP is moved from $\alpha_d^{\rm flat}=34.63(2)$ to $\alpha_d^{\rm opt}=37.71(1)$, thus exhibiting an increase of about 8\%.

The numerical optimization has been performed for $T=0$, when the hard part of the interaction $f$ (for $\lvert x \lvert < \theta$) is exactly zero. By taking the logarithm of the soft part of $f$ (for $\theta\leq x\leq2\pi-\theta$), one can obtain the pairwise contribution to the biasing Hamiltonian $H_1$ up to an additive constant, due to the arbitrariness in the normalization. When switching on temperature, the hard core excluded volume interaction is softened, the cost for violating a constraint being now proportional to $\exp(-\beta)$. The soft part of the interaction should instead be independent from $\beta$ in our approach, apart for a global $\beta$-dependent normalization. In the following, we will always adopt implicitly the definition $f^{\rm opt}(x;\beta) \propto Ae^{-\beta}\mathbb{I}[\lvert x\lvert<\theta] + f^{\rm opt}_{T=0}\mathbb{I}[\theta\leq x\leq2\pi-\theta]$. We fix the relative amplitude $A$ of the hard part with respect to the soft part of the interaction $f$ by requiring (using the notation for the discretized model)
\begin{align}
    &\frac{\sum_{(-\theta;\theta)} f^{\rm opt}_a }{\sum_{[\theta;2\pi-\theta]} f^{\rm opt}_a } =
    \frac{\sum_{(-\theta;\theta)} f^{\rm flat}_a }{\sum_{[\theta;2\pi-\theta]} f^{\rm flat}_a } =
    \frac{(2d-1)e^{-\beta}}{p-(2d-1)}.
    \label{eq:relative_amplitude_soft_interaction}
\end{align}
With this choice, the annealed energy $e_{\rm ann}(\beta)$, counting the number of violated constraints per spin as a function of temperature in the paramagnetic phase, is equal for both $f^{\rm flat}$ and $f^{\rm opt}$ to 
\begin{equation}
    e_{\rm ann}(\beta) \equiv \frac{M}{N}\frac{\sum_a f_a \mathbb{I}[\cos(\frac{2\pi a}{p})>0]}{\sum_a f_a} =\frac{\alpha (2d-1)e^{-\beta}}{(2d-1)e^{-\beta}+p-(2d-1)}.
    \label{eq:energy_annealed}
\end{equation}

\begin{figure}
    \centering
    \begin{subfigure}[b]{.49\columnwidth}
        \includegraphics[width=\textwidth]{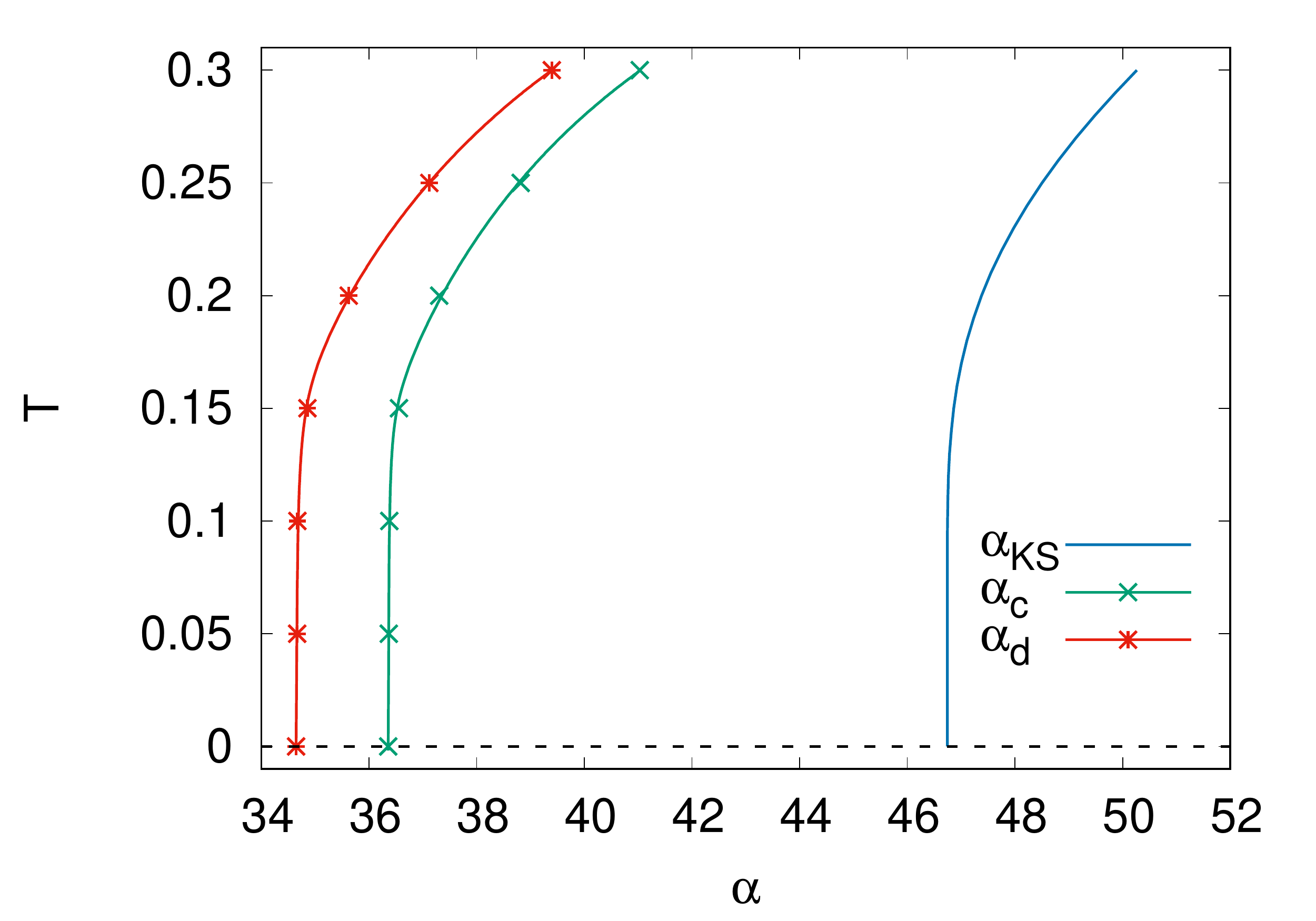}
        \caption{}
    \end{subfigure}
    \hfill
    \begin{subfigure}[b]{.49\columnwidth}
        \includegraphics[width=\textwidth]{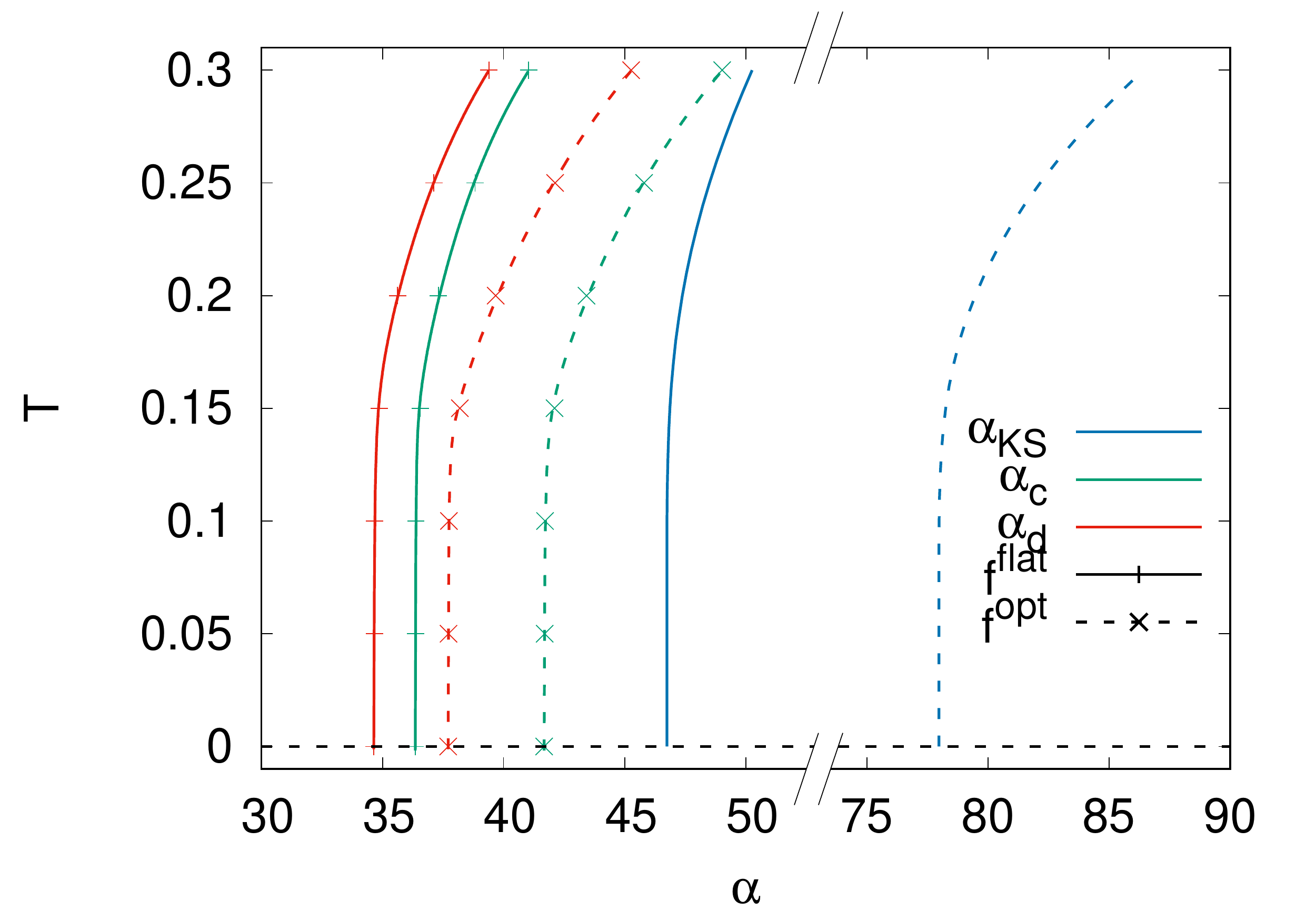}
        \caption{}
    \end{subfigure}
    \caption{Phase diagram in the $(\alpha,T)$ plane for the continuous coloring with fixed $q=20$ and discretization precision $d=10$. The dynamic and condensation thresholds are obtained by numerically solving the BP equations for different values of $T$ (points in the figure). Left: uniform measure. Right: comparison between the two choices for the interaction $f^{\rm flat}$ (solid) and $f^{\rm opt} $ (dashed). The $T=0$ dynamical transition is moved from $\alpha_d^{\rm flat}=34.63(2)$ to $\alpha_d^{\rm opt}=37.71(1)$.}
    \label{fig:flatopt_phasediag}
\end{figure}

The phase diagram in the $(\alpha,T)$ plane is shown in figure~\ref{fig:flatopt_phasediag}, solid lines corresponding to $f^{\rm flat}$, while dashed ones referring to $f^{\rm opt}$.
We stress that in our case temperature tunes the strength of the violated constraints (we recover the infinite hard sphere potential only in the limit $\beta\to\infty$), and thus plays a fundamentally different role than, for instance, in~\cite{sellittoThermodynamicDescriptionColloidal2013,maimbourgGeneratingDensePackings2018}. On the right panel we directly compare the phase diagram according to the uniform measure (solid lines) with the one obtained by using the optimized interaction $f^{\rm opt}$ (dashed lines). It is interesting to notice that also the Kesten-Stigum bound $\alpha_{\rm KS}$ is sensitive to the modification of the interaction, and it actually increases the most with respect to the other thresholds, going from the $T = 0$ value $\alpha_{\rm KS} = 46.743$ to $\alpha_{\rm KS} = 77.972$, with an increase of more than 60\%. The location of $\alpha_{\rm KS}$ is particularly relevant for the associated Bayesian inference problem, since it signals the $\alpha$-point below which inference is practically unfeasible. In order to improve the performance of inference protocols, one should probably try to optimize the interaction in the opposite direction (lowering $\alpha_{\rm KS}$).

\section{Numerical determination of the dynamic transition}
\label{sec:numerical_determination}
In this section, the question of how to obtain the most accurate numerical estimation of $\alpha_d$ is addressed. The typical way of detecting $\alpha_d$ is by verifying the existence of a non-trivial solution to the planted population dynamics BP equations: usually, one starts from $\alpha>\alpha_d$ and decreases $\alpha$ until the non-trivial solution is lost. This method, however, is prone to inaccuracies. On the one hand, because of the finite number of BP iterations one can perform, one may simply not give the algorithm enough time to forget the planted solution and conclude that the non-trivial fixed point survives for values of $\alpha$ slightly smaller than $\alpha_d$, thus leading to an underestimation of $\alpha_d$. On the other hand, fluctuations due to the finite size of the population may become relevant close to $\alpha_d$, where the non-trivial fixed point looses its stability, this leading to an early loss of the fixed point and to a consequent overestimation of $\alpha_d$. 

For these reasons, a more robust approach taking into account how things scale \emph{around} $\alpha_d$ is very desirable. Following~\cite{budzynskiBiasedLandscapesRandom2019}, we give independent, compatible estimations of $\alpha_d$ coming from both below and above the dynamic threshold, by studying respectively the (power-law) divergence of BP relaxation time and the birth of instability in the non-trivial fixed point.

The situation is more complicated when one considers real instances of the problem, where the finite size of the system determines the presence of loops in the graph and in general smooths the transitions, which are sharp only in the thermodynamic limit. For large enough sizes, however, we certainly expect the population dynamics BP predictions to be of some use. This can be directly tested through Monte Carlo simulations, where we observe a MCT-like behaviour of the dynamic correlation (the overlap) compatible with $\alpha_d\equiv\alpha_d^{\rm BP}=34.63(2)$, at least up to timescales where finite size effects are negligible. In this section we present the results of a detailed analysis at $T=0$, the same procedure being straightforwardly applicable to any temperature.

\subsection{BP population dynamics}
\begin{figure}[t]
    \begin{subfigure}[b]{0.49\columnwidth}
        \centering
        \includegraphics[width=\textwidth]{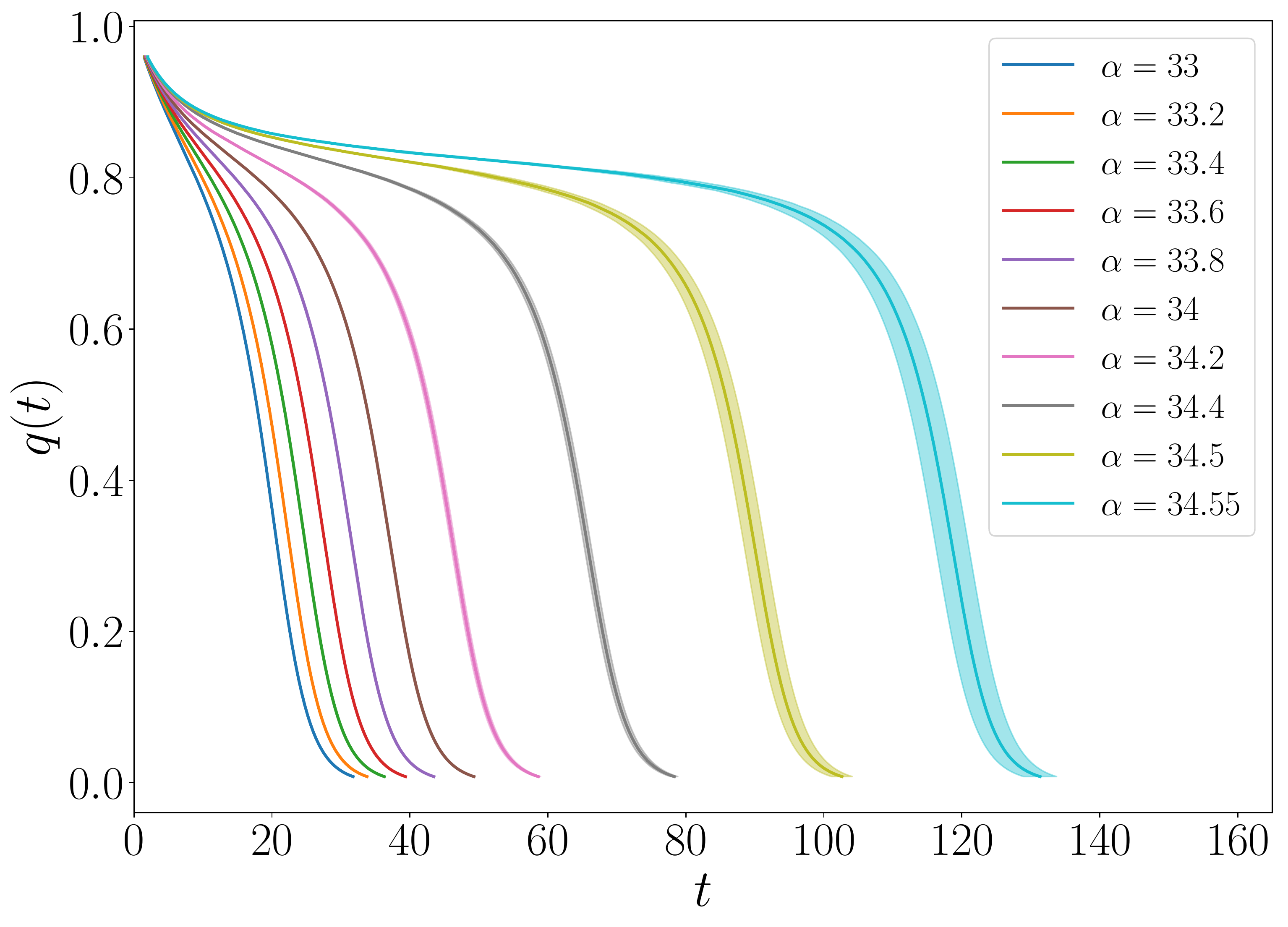}
        \caption{}
        \label{fig:q_PD_a}
    \end{subfigure}
    \hfill
    \begin{subfigure}[b]{0.49\columnwidth}
    \centering
    \includegraphics[width=\textwidth]{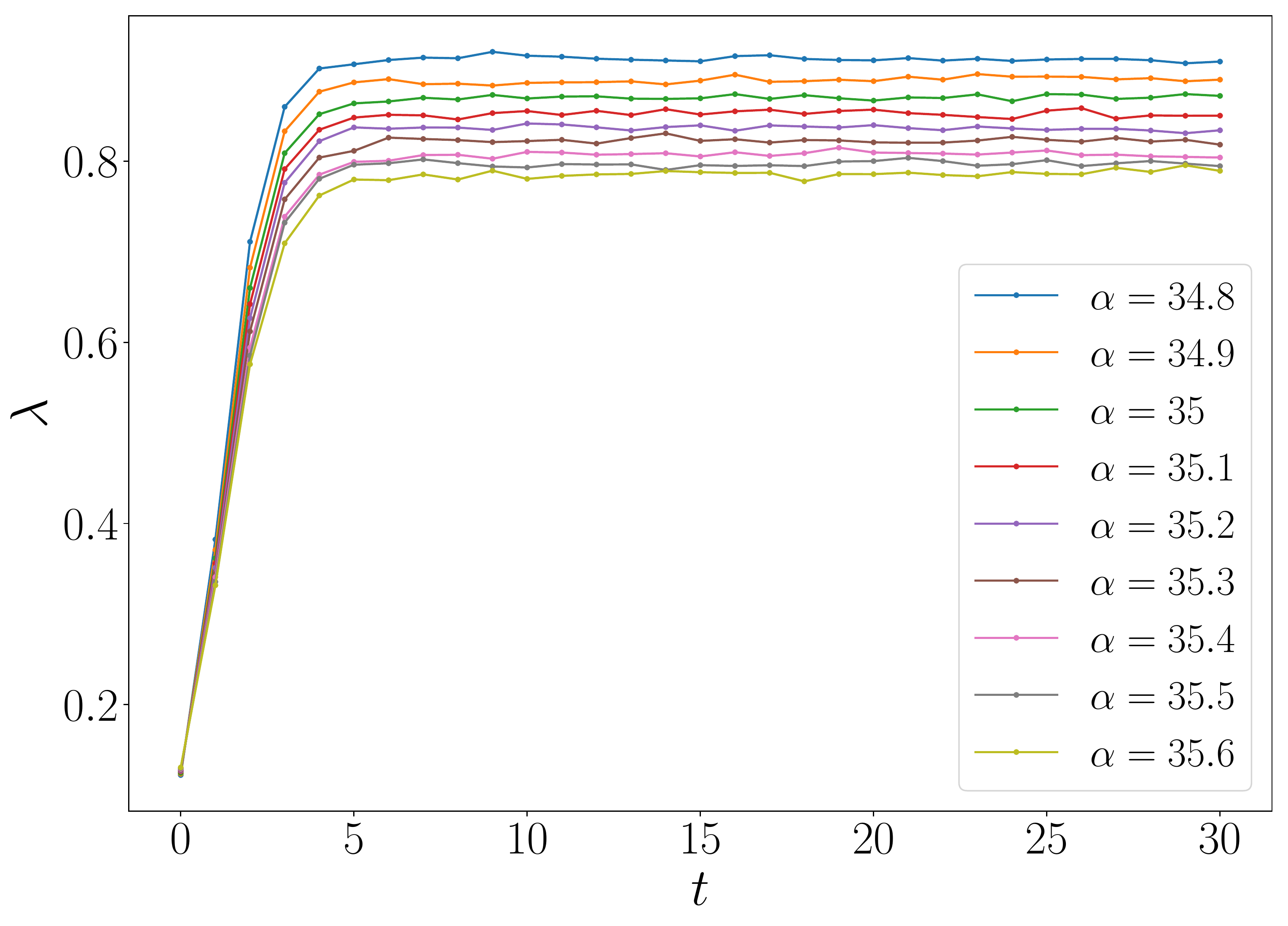}
    \caption{}
    \label{fig:q_PD_b}
    \end{subfigure}
    \caption{Left: overlap as a function of the number $t$ of BP iterations. As long as $\alpha<\alpha_d$, the overlap decays to zero, thus implying that any memory of the planted initialization is lost and one recovers the paramagnetic solution with uniform messages. Right: ratio $\lambda(t)$ between the magnitude of perturbations to the non-trivial solution at times $t+1$ and $t$ for $\alpha>\alpha_d$. After a transient, stationarity is reached, thus identifying the stability parameter $\lambda(\alpha)<1$.}
    \label{fig:q_PD}    
\end{figure}

In figure~\ref{fig:q_PD} we show the data from the numerical solution of population dynamics equations for the Erd{\H o}s-R{\'e}nyi ensemble. The size of the population here considered is $U=10^6$, each message being a probability array of $p=200$ values ($q=20$, $d=10$). We choose the value $q=20$ because, as it is suggested from figure~\ref{fig:PhaseDiag_discr_vs_cont}, the transition should be in this case appreciably first order. On the left panel the overlap decay $q(t)$ is displayed for different values of $\alpha<\alpha_d$. Given a set of $U$ marginals $\{\nu_i\}^t$, the overlap is defined as (using for convenience continuous notation)
\begin{align}
    q(t)&=\frac{1}{U}\sum_{i=1}^U \int dx_1 dx_2 \,\nu_i(x_1)\nu_i(x_2)\cos(x_1-x_2) \nonumber \\
    &= \frac{1}{U}\sum_{i=1}^U \left[\langle\cos(x)\rangle_i^2+\langle\sin(x)\rangle_i^2\right],
\end{align}
where $\langle \cdot \rangle_i$ represents the average over the marginal distribution $\nu_i$. We considered 10 different inizializations of the random generator, which we refer to as samples. Each sample has a very smooth behaviour, almost identical to the others except for a slightly difference in the time it leaves the forming plateau. For this reason, the most natural way to present the data is by averaging over $t$ at fixed $q$, once a spline interpolation of the single sample is performed.

On the right panel, we consider the case $\alpha>\alpha_d$, where the BP equations admit a non-trivial solution with $\lim_{t\to\infty}q(t)\neq0$. Once convergence in the overlap is reached, we compute the ratio of the magnitude between two subsequent perturbations around the fixed point (by using the linearized version of the BP equations derived in Appendix~\ref{section:stability_analysis}), as a function of time
\begin{equation}
    \lambda(t)=\frac{\Vert \epsilon(t+1) \Vert}{\Vert \epsilon(t) \Vert},
\end{equation}
where we adopted the $L_2$ norm of the perturbation parameters averaged over the population, $\Vert \epsilon \Vert = \frac{1}{pU}\sum_{i=1}^U\sum_{a=0}^{p-1} \epsilon_{i,a}^2$. After a sufficient but finite number of steps, $\lambda(t)$ is actually selecting the contribution from the biggest in modulus eigenvalue of the linearized matrix which enforces the BP iterative procedure (see~\cite{budzynskiBiasedLandscapesRandom2019} for a detailed discussion), being the slowest mode to decay. We define the stability parameter $\lambda(\alpha)$ as essentially the modulus of this eigenvalue, with $\lambda(\alpha)<1$ for the solution to be stable. In practice, we average over $t$ the $\lambda(t)$ data of figure~\ref{fig:q_PD_b} once the plateau is reached. In this case we consider just one sample for each value of $\alpha>\alpha_d$.

Let us first consider the region $\alpha<\alpha_d$. From figure~\ref{fig:q_PD_a}, it is clear that approaching $\alpha_d$ the overlap relaxation becomes increasingly slower. Following~\cite{budzynskiBiasedLandscapesRandom2019}, this behaviour can be directly connected to the discontinuous bifurcation occurring to the fixed-point solution of the BP equations, which is of the form $\mathbb{P}=\mathcal{F}(\mathbb{P},\alpha)$. To fix the ideas, it is convenient to consider an analogous relation in the scalar case, $q=f(q,\alpha)$, where $q$ is real (the similarity with the overlap, which is indeed our scalar representative for $\mathbb{P}$, will be readily evident). Assuming that $q\geq 0$ and that $q=0$ is always a solution, the dynamic threshold $\alpha_d$ is defined as the subitaneous birth of a second solution $q_d=f(q_d,\alpha_d)$, with $q_d>0$. Our problem reduces to the study of the dynamical process $q^{(t+1)}=f(q^{(t)},\alpha)$, with initial condition $q^{(t=0)}>q_d$ (in analogy with the planted initialization). In this context, it can be proved that the number of steps $t$ around the plateau $q_d$ diverges as $\tau\sim K(\alpha_d-\alpha)^{-\frac{1}{2}}$ to leading order for $\alpha\to\alpha_d^-$, where $K$ is fixed by some derivative of $f(q,\alpha)$ at the special point $(q_d,\alpha_d)$. 

\begin{figure}[t]
    \begin{subfigure}[b]{0.49\columnwidth}
        \centering
        \includegraphics[width=\textwidth]{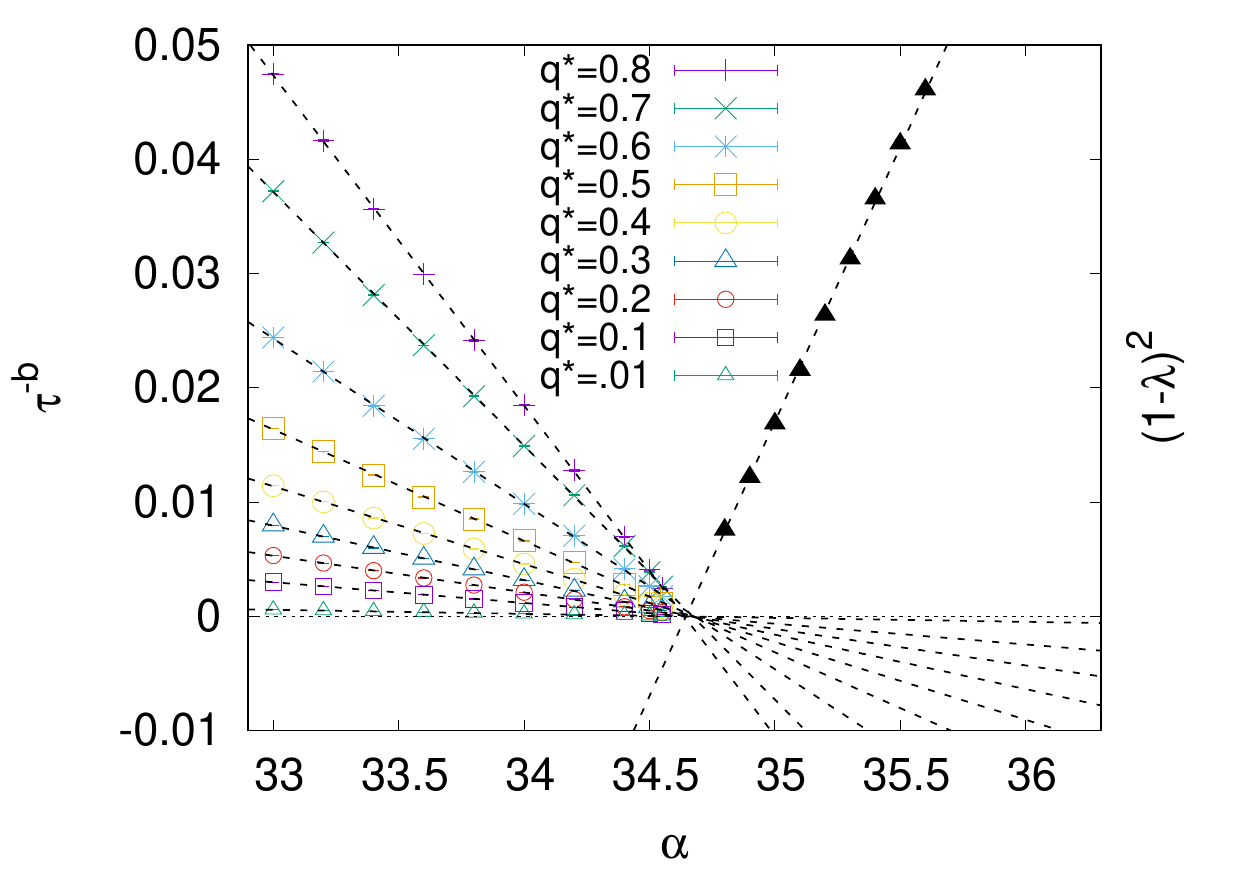}
        \caption{}
        \label{fig:tau_PD_a}
    \end{subfigure}
    \hfill
    \begin{subfigure}[b]{0.49\columnwidth}
    \centering
    \includegraphics[width=\textwidth]{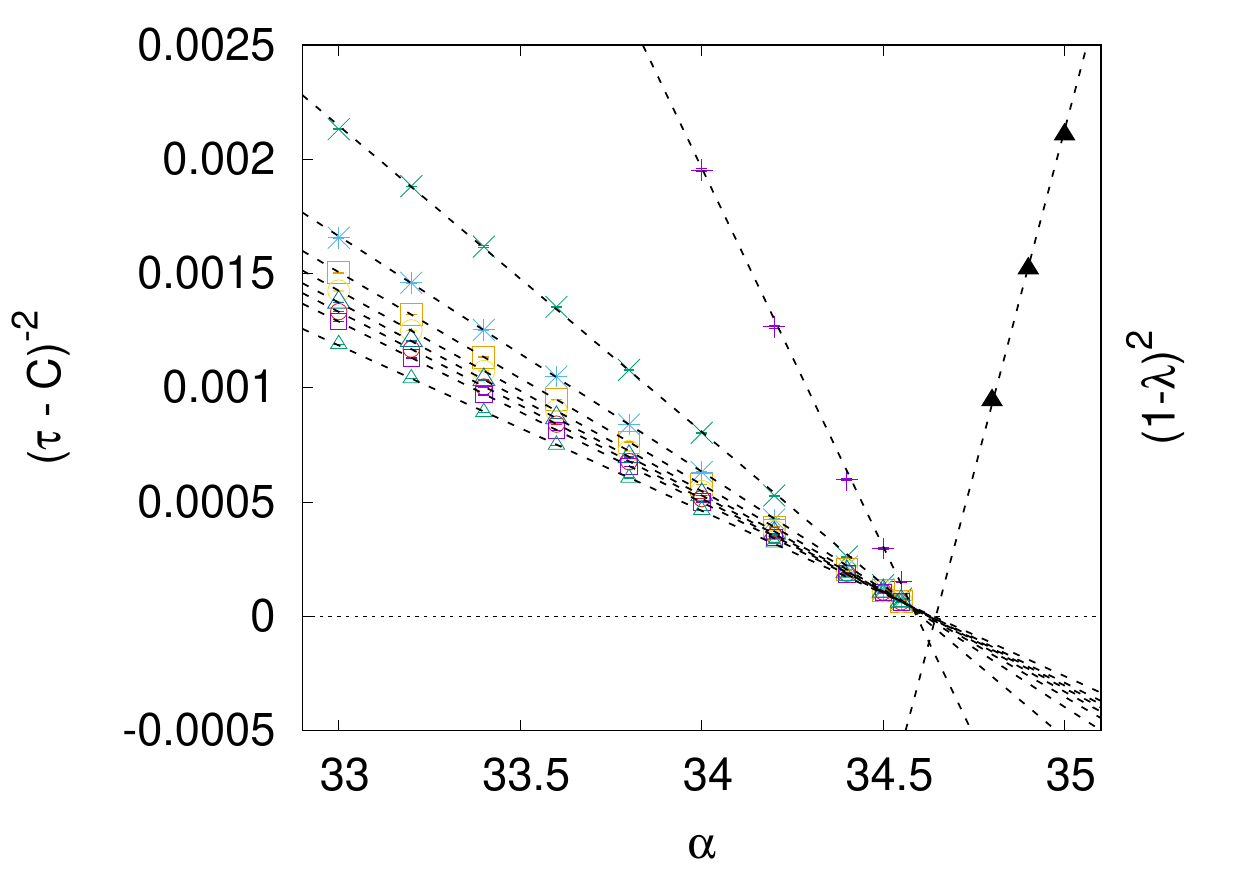}
    \caption{}
    \label{fig:tau_PD_b}
    \end{subfigure}
    \hfill
    \centering
    \begin{subfigure}[b]{0.49\columnwidth}
    \centering
    \includegraphics[width=\textwidth]{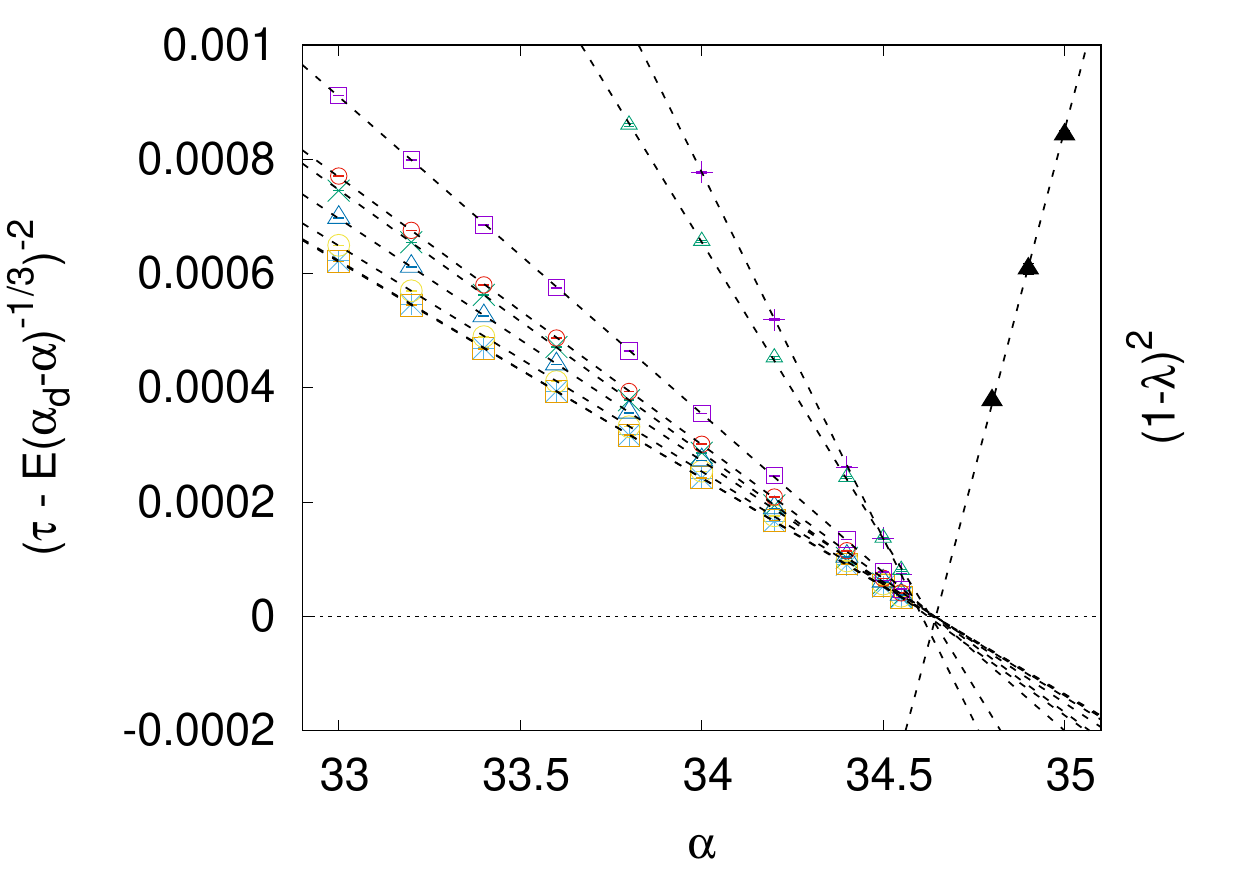}
    \caption{}
    \label{fig:tau_PD_c}
    \end{subfigure}
    \caption{Colors: divergence of the overlap relaxation time $\tau_{q^*}(\alpha)$, defined as the time needed for $q(t)$ to decrease down to $q^*$. Figures (a), (b) and (c) show the data rescaled according to functions~\eqref{first_fit_func},~\eqref{second_fit_func} and~\eqref{third_fit_func} respectively. Black: behaviour of $(1-\lambda(\alpha))^2$, where $\lambda(\alpha)$ is the stability parameter of the non-trivial solution for $\alpha>\alpha_d$. The correct scale for $\lambda$ is given in the first figure (in both figures (b) and (c), black points follow a different y-scale than colored ones and are just meant to guide the eye towards the transition point). The dynamic transition $\alpha_d$ is located at the intersection of the color and black lines with the x-axis.}
    \label{fig:tau_PD}    
\end{figure} 

In order to test this prediction on our real data, we operatively define a family of relaxation times $\tau_{q^*}(\alpha)$ representing, for any given threshold $q^*<q_d$, the time needed for the overlap $q(t)$ to decrease down to $q^*$. By looking at the data in figure~\ref{fig:q_PD_a}, one notices that different sets of $\tau_{q^*}(\alpha)$ for different $q^*$ should essentially differ by only a finite amount of steps, which is reasonably constant in $\alpha$ for any given $q^*$ since the curves have approximately the same behaviour (same slope) on a linear x-scale. However, if we also consider the relatively small values of $t$ needed to complete the decay and hence of  $\tau_{q^*}(\alpha)$, it follows that relevant subleading corrections to the asymptotic scaling may be due just to the choice of $q^*$. This is evidenced by the behaviour of the first function that we use to fit the data, neglecting any scaling correction,
\begin{equation}
    \tau^{(1)}(\alpha) = \frac{A}{(\alpha_d-\alpha)^{1/b}}.
    \label{first_fit_func}
\end{equation}
This function depends on three free parameters: $\alpha_d$, $b$ and $A$. Even though at first glance it may appear to reasonably fit the data, see figure~\ref{fig:tau_PD_a} (colors), all the values of the parameters, presented in table~\ref{tab:fitresult_PD} (left), exhibit some dependence on $q^*$, which is very noticeable in the case of the exponent $b$ (that was argued to be $b=2$). To account for subleading corrections, we also consider the two following power laws, still depending on three parameters,
\begin{align}
    \tau^{(2)}(\alpha) &= \frac{B}{(\alpha_d-\alpha)^{1/2}} + C, 
    \label{second_fit_func}
    \\
    \tau^{(3)}(\alpha) &= \frac{D}{(\alpha_d-\alpha)^{1/2}} + \frac{E}{(\alpha_d-\alpha)^{1/3}}.
    \label{third_fit_func}
\end{align}
\begin{table}[t]
		\centering
		\begin{tabular}{|c|c|c|c|c|}
			\hline
			\multicolumn{5}{|c|}{$\tau^{(1)}$} \\
			\hline
			$q^*$ & $\alpha_d$ & $b$ & $A$ & $\chi^2/\mathrm{dof}$ \\
			\hline\hline
			0.8 & 34.639(2) & 1.384(5) & 12.94(2) & 4.39 \\
			0.7 & 34.672(2) & 1.280(4) & 19.57(4) & 10.83 \\
			0.6 & 34.680(2) & 1.345(4) & 23.31(4) & 21.21 \\
			0.5 & 34.677(2) & 1.427(4) & 25.65(4) & 23.81 \\
			0.4 & 34.671(2) & 1.507(4) & 27.43(4) & 23.41 \\
			0.3 & 34.664(2) & 1.588(5) & 28.99(4) & 20.64 \\
			0.2 & 34.657(2) & 1.679(5) & 30.63(4) & 17.18 \\
			0.1 & 34.647(2) & 1.807(5) & 32.85(3) & 12.21 \\
			0.01 & 34.627(1) & 2.164(5) & 39.48(3) & 3.83 \\
			\hline
		\end{tabular}
		\begin{tabular}{|c|c|c|c|c|}
		    \hline
		    \multicolumn{5}{|c|}{$\tau^{(2)}$} \\
			\hline
			$q^*$ & $\alpha_d$ & $B$ & $C$ & $\chi^2/\mathrm{dof}$ \\
			\hline\hline
			0.8 & 34.592(1) & 17.4(1) & -4.77(5) & 22.81\\
			0.7 & 34.604(1) & 27.3(1) & -8.58(6) & 30.19\\
			0.6 & 34.615(1) & 31.1(1)  & -8.76(7) & 6.62\\
			0.5 & 34.621(1) & 32.8(1) & -7.99(7) & 1.46\\
			0.4 & 34.624(1) & 33.7(1) & -7.05(7) & 0.42\\
			0.3 & 34.627(1) & 34.4(1) & -6.01(7) & 0.80\\
			0.2 & 34.629(1) & 35.0(1) & -4.75(7) & 1.68\\
			0.1 & 34.632(1) & 35.6(1) & -2.94(8) & 3.42\\
			0.01 & 34.638(1) & 37.2(1) & 2.51(8) & 9.80\\
			\hline
		\end{tabular}
		\begin{tabular}{|c|c|c|c|c|}
			\hline
			\multicolumn{5}{|c|}{$\tau^{(3)}$} \\
			\hline
			$q^*$ & $\alpha_d$ & $D$ & $E$ & $\chi^2/\mathrm{dof}$ \\
			\hline\hline
			0.8 & 34.607(1) & 27.9(2) & -15.2(2) & 2.89\\
			0.7 & 34.623(1) & 46.6(2) & -27.7(2) & 3.20\\
			0.6 & 34.636(1) & 51.3(3) & -28.6(2) & 1.69\\
			0.5 & 34.641(1) & 51.4(3) & -26.3(3) & 4.37\\
			0.4 & 34.643(1) & 50.3(3) & -23.3(3) & 6.64\\
			0.3 & 34.643(1) & 48.6(3) & -19.9(3) & 7.79\\
			0.2 & 34.642(1) & 46.2(3) & -15.8(3) & 8.26\\
			0.1 & 34.639(1) & 42.4(3) & -9.7(3) & 7.99\\
			0.01 & 34.631(1) & 31.0(3) & 8.5(3) & 5.34\\
			\hline
		\end{tabular}
		\caption{Results from fit~\eqref{first_fit_func} (on the left),~\eqref{second_fit_func} (center) and~\eqref{third_fit_func} (right) for the divergence of the overlap relaxation time $\tau_{q^*}(\alpha)$ as obtained from population dynamics.}
		\label{tab:fitresult_PD}
\end{table}
\begin{table}[]
    \centering
		\begin{tabular}{|c|c|}
			\hline
			$\alpha_d$ & $\chi^2/\mathrm{dof}$ \\
			\hline\hline
			34.645(2) & 1.39 \\
			\hline
		\end{tabular}
		\caption{Results from the fit on the stability parameter $\lambda(\alpha)$.}
		\label{tab:fitresult_PD_stability}
\end{table}
The rescaled data is presented respectively in figures~\ref{fig:tau_PD_b} and~\ref{fig:tau_PD_c}. For what concerns the results of fit $\tau^{(2)}$, see again table~\ref{tab:fitresult_PD}, we obtain good $\chi^2$ and a much less $q^*$-dependent slope (\emph{i.e.} $B$) especially for intermediate values of $0.1\leq q^*\leq 0.5$. We conclude that in this range the scalar bifurcation scaling form with exponent $b=2$ adequately describes our $\tau_{q^*}(\alpha)$ data. On the other hand, the function $\tau^{(2)}$ is shown to fail for values of $q^*$ close to the plateau, $q^*=0.8,0.7$ (compare the values of $\chi^2$ in table~\ref{tab:fitresult_PD}). In this case fit $\tau^{(3)}$ does a better job, thus suggesting that the scaling $b=2$ is just apparently altered by some non-trivial subleading corrections. In general, we observe that function $\tau^{(3)}$, even though outperformed by $\tau^{(2)}$ when $q^*$ takes the central values, always performs better than fit $\tau^{(1)}$, apart for the case $q^*=0.01$ for which already $\tau^{(1)}$ produces a value of $b$ close to $b=2$. This corroborates the argument supporting $b=2$; a cautious estimate for $\alpha_d$ from our data would be at this point $\alpha_d=34.625(20)$.

A second independent estimation of $\alpha_d$ can be obtained by studying the loss of stability for the non-trivial solution when approaching $\alpha_d$ from above. A quantitative prediction from scalar bifurcation theory~\cite{budzynskiBiasedLandscapesRandom2019} states that $(1-\lambda(\alpha))^2\sim (\alpha-\alpha_d)$ vanishes linearly for $\alpha\to\alpha_d^+$. This is shown in figure~\ref{fig:tau_PD_a} in black points (beware that points reproduced in figures~\ref{fig:tau_PD_b} and~\ref{fig:tau_PD_c} are out of scale), while the estimated $\alpha_d$ from the fit is given in table~\ref{tab:fitresult_PD_stability}. Comparing it with the result from the analysis of the overlap decay for $\alpha<\alpha_d$, we conclude $\alpha_d=34.64(2)$. The quoted uncertainty
comes from the reasonable assumption that the difference in the two independent estimations is mostly due to systematic rather than statistical errors.

\subsubsection{Numerical determination of \texorpdfstring{$\alpha_c$}{alphac}}
\begin{figure}[t]
    \centering
    \includegraphics[width=0.55\textwidth]{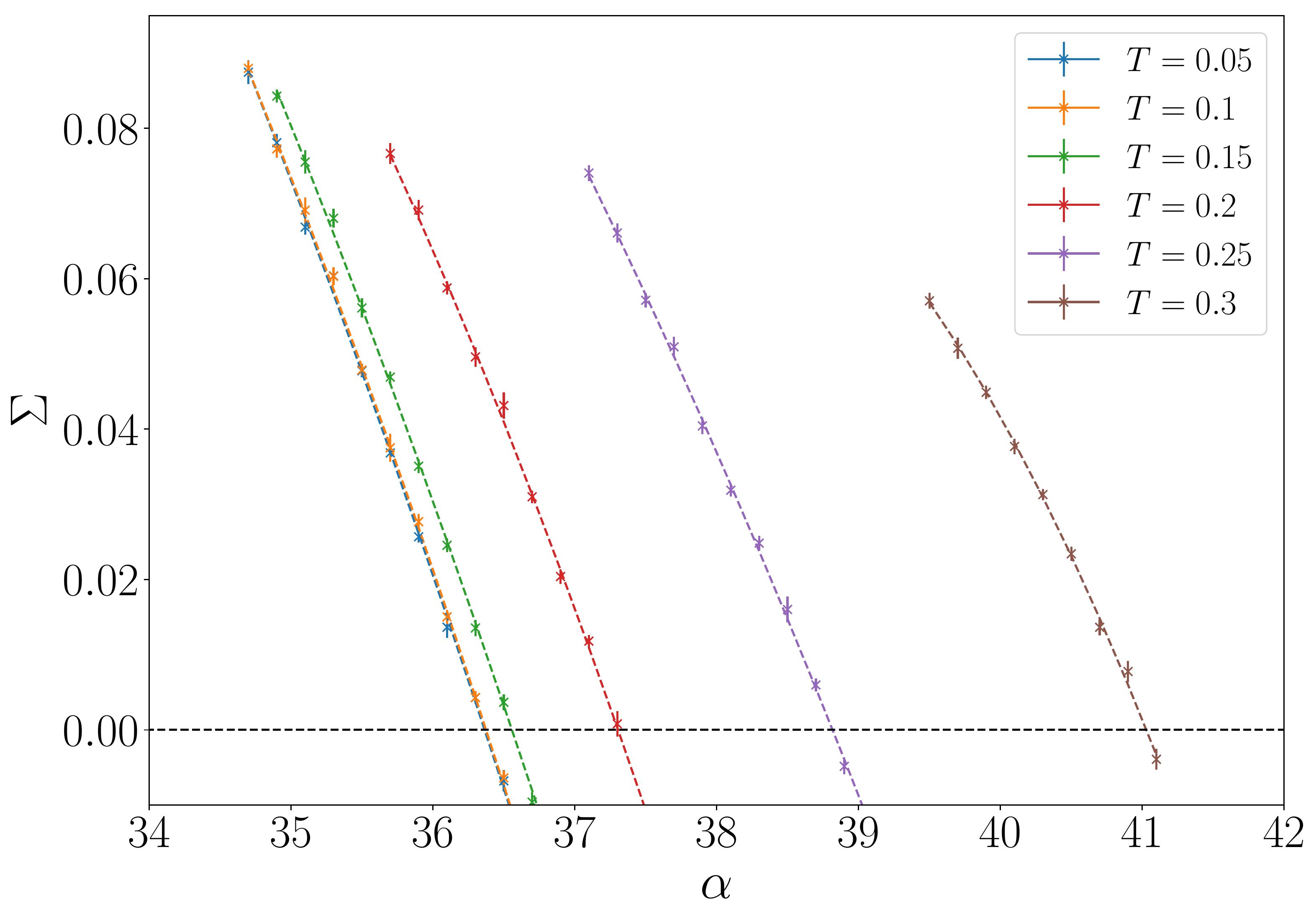}
    \caption{Complexity curves $\Sigma(\alpha,T)$. Dashed lines are a second order polynomial fit. The intersection of each curve with the x-axis returns the value of the condensation transition $\alpha_c(T)$.}
    \label{fig:alphac_complexities}
\end{figure}

The condensation transition line $\alpha_c(T)$ is defined by the condition $\Sigma(\alpha,T)=0$, where $\Sigma$ is the complexity of the states dominating the partition function at temperature $T$. Recalling eq.~\eqref{eq:complexity_def}, $\Sigma$ can be written as the difference of Bethe free entropies
\begin{equation}
    \Sigma = \frac{1}{N} \left[F_{\rm Rec\, II}({\rm para}) - F_{\rm Rec\, II}({\rm planted})\right],
    \label{eq:sigma_def_BP}
\end{equation}
where $F_{\rm Rec\, II}$ is obtained for any BP fixed point from eq.~\eqref{FreeEnergy_Equation_Reconstruction_Sym}, the temperature dependence being encoded into the affinity function $f_a$. In particular, $F_{\rm Rec\, II}({\rm para})=-\alpha$. A number of curves $\Sigma(\alpha,T)$ for different values of $T$ are plotted in figure~\ref{fig:alphac_complexities} as a function of $\alpha$.
The $T=0$ limit is under control thanks to the fact that the curves for the two lowest values of $T$ are practically indistinguishable. Notice that the step $\Delta\alpha = 0.2$ in the data is quite large, and the first point in each curve does not exactly correspond to $\alpha_d(T)$. The estimation of $\Sigma$ is on the contrary much more precise and allows for a reliable determination of $\alpha_c(T)$. We also observe that the maximum of the complexity appears to decrease when increasing temperature. We cannot exclude the possibility that it goes to zero for some finite temperature, \emph{i.e.} that the transition becomes continuous for high enough $\alpha$.

\subsection{Equilibrium autocorrelation time from Monte Carlo}
\label{sec:ZeroEnergyMC}
\begin{figure}[t]
    \begin{subfigure}[b]{0.49\columnwidth}
    \centering
    \includegraphics[width=\textwidth]{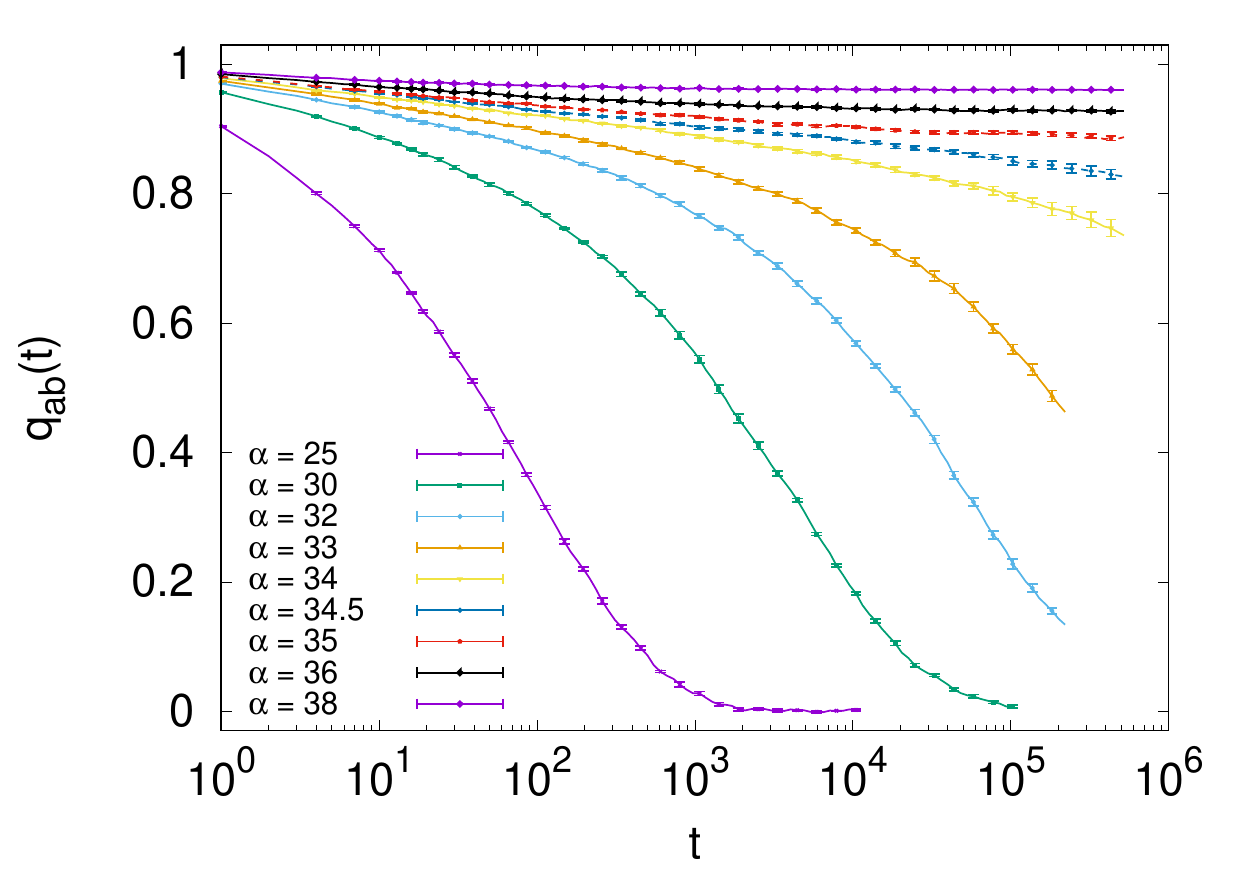}
    \caption{}
    \label{fig:q-d20-decay-a}
    \end{subfigure}
    \hfill
    \begin{subfigure}[b]{0.49\columnwidth}
    \centering
    \includegraphics[width=\textwidth]{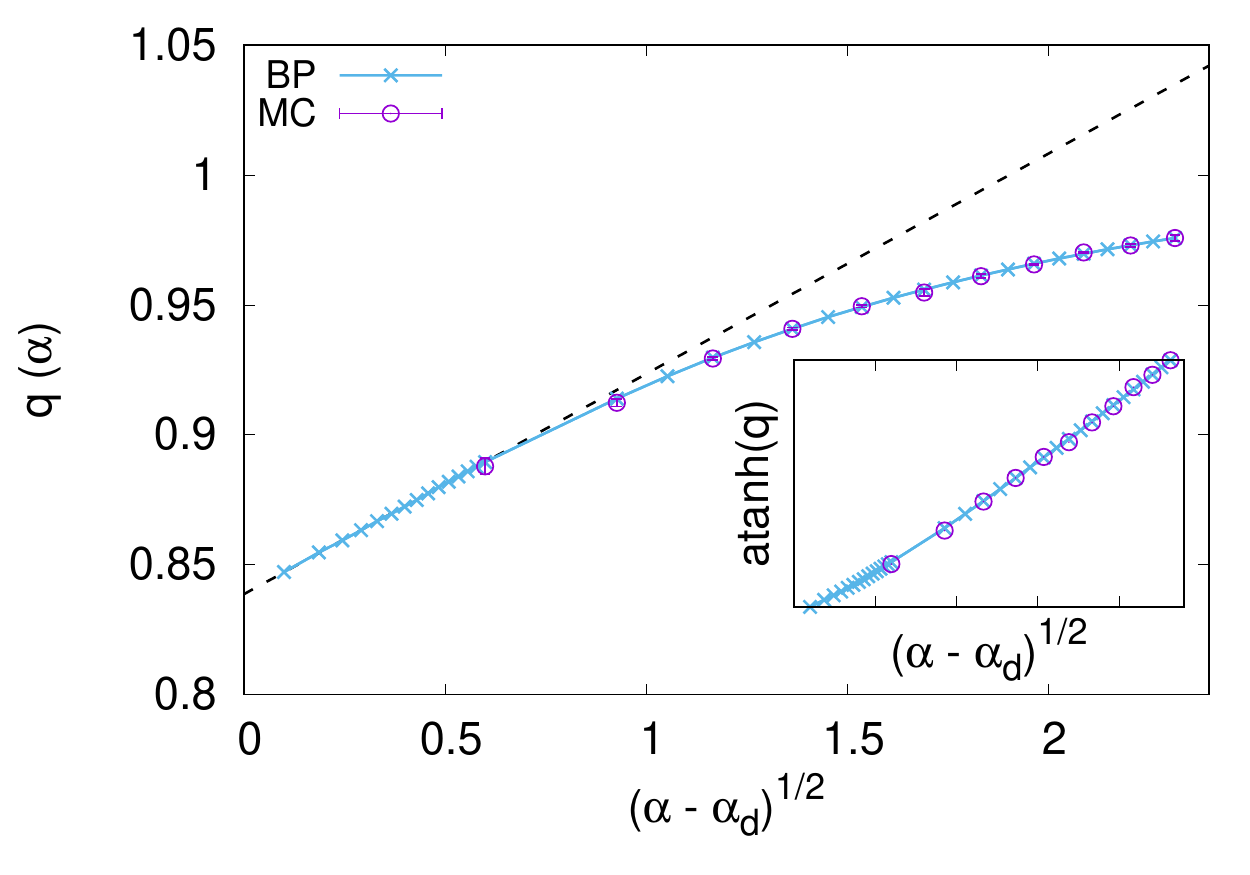}
    \caption{}
    \label{fig:q-d20-decay-b}
    \end{subfigure}
	\caption{Left: overlap as a function of time between two independently evolving replicas starting from the same equilibrated (planted) initial condition. Different curves correspond to different values of $\alpha$. Dashed lines for $\alpha=34.5$ and $\alpha=35$ indicate the region where the overlap supposedly starts to develop a plateau with $q=q_d$. Right: long time value of the overlap for $\alpha>\alpha_d$ from MC and population dynamics BP simulations. The birth of the non-trivial solution follows a square root singularity, allowing us to extrapolate from the BP data $q_d=0.838(3)$. The data is scaled according to $\alpha_d=34.63$ from BP population dynamics. Inset: the trivial bending of data points for high $\alpha$ can be straightened by plotting $\tanh^{-1}\bigl(q(\alpha)\bigr)$}
	\label{fig:q-d20-decay}
\end{figure}

Monte Carlo (MC) simulations have the advantage of being much less eager of computer memory as compared to solving the BP equations, where one needs to work with entire distributions (for each of the messages) even in our RS-like case. On the other hand, relaxation times can be very long (we are approaching a MCT-like glass transition, indeed). This can be noticed by looking at  figure~\ref{fig:q-d20-decay-a}, where we plot the overlap decay for several orders of magnitude in the number of MC sweeps. Before analyzing the results, we give a precise definition of the overlap and discuss the details of the MC simulations at zero temperature.

The overlap between two configurations of angles $\underline{x}_a$ and $\underline{x}_b$ needs to be maximized over rotations due to the global rotational symmetry of the model, which becomes continuous in the limit $d\to\infty$ of infinite discretization precision,
\begin{align}
    q(t)=\max_{A\in[0,2\pi)} \langle \cos(x_a-x_b-A)\rangle = \sqrt{\langle \cos(x_a-x_b)\rangle^2 + \langle \sin(x_a-x_b)\rangle^2}, 
    \label{eq:overlap_MC}
\end{align}
where $\langle\cdot\rangle = \frac{1}{N}\sum_{i=1}^N$ stands for the average over the system. We consider both the overlap $q_{ab}(t)$ between two independent replicas of the same system, in which case $x_a=x_i^{(a)}(t)$ and $x_b=x_i^{(b)}(t)$, and the overlap $q(t)$ of each replica with the starting configuration, in which case $x_a=x_i(t)$ and $x_b=x_i(t=0)$. The latter situation can be further simplified by initially planting the system in a configuration with all the variables identically set equal to zero, which is possible thanks to the introduction of random shifts associated with each edge of the graph (that are required in the first place in order to suppress crystallization). Of course, we expect the equilibrium long time limit of both the overlaps $q(t)$ and $q_{ab}(t)$ to be identical. However, it is also clear that $q_{ab}(t)$ will in general decorrelate faster.

For what concerns the algorithm, we adopt in this case a simple zero temperature heat bath rule: at each step, a randomly selected spin is updated with uniform probability among the Potts states which satisfy all the constraints with the neighbours. Since we start from a planted configurations of exactly zero energy, at least one such a state is guaranteed to exist for every spin, and the procedure is well posed. The advantage with respect to standard Metropolis is that moves are always accepted, this coming at the price of computing the local energy for each of the $p$ states a variable can assume. Notice, however, that (for $f^{\rm flat}$) this can be performed in a smart way by running just one time over the neighbours of the variable to update and increasing by 1 the energy of the $2d-1$ states inside the excluded region. 

Obtaining a precise estimate of $\alpha_d$ from MC data is more difficult than from population dynamics. Figure~\ref{fig:q-d20-decay-a} shows as a function of time (MC sweeps) the average of the overlap over 10 different systems of size $N=10^4$. We checked finite size corrections to be smaller than the statistical errors by simulating a few samples of size $N=10^5$. For $\alpha>\alpha_d$, planting allows us to initialize the system \emph{inside} the planted cluster, which is indistinguishable from a typical cluster for the random model, up to $\alpha_c$. Unfortunately, the $\alpha>\alpha_d$ data is of little use in the estimation of $\alpha_d$, since we do not have in this case an analogous of the BP stability parameter. A possible way round would be to consider the long time limit $q(\alpha)$ of the overlap at the plateau, which is expected~\cite{budzynskiBiasedLandscapesRandom2019} to approach a non zero value $q_d>0$ with a square root singularity at $\alpha_d$, as shown in figure~\ref{fig:q-d20-decay-b}. Obtaining $\alpha_d$ from the $q(\alpha)$ data, on the other hand, would imply a fit where also the parameter $q_d$ is unknown. For this reason we prefer first to estimate $\alpha_d$ from the overlap decay for $\alpha<\alpha_d$, and then use it to verify the square root singularity and eventually obtain $q_d$. Notice that the behaviour $(q(\alpha)-q_d)\sim(\alpha-\alpha_d)^{1/2}$ can be valid only in the vicinity of $\alpha_d$, since we know by definition that $q(\alpha)<1$. The strong deviation from a straight line behaviour in figure~\ref{fig:q-d20-decay-b} is in this sense totally expected. What is inconvenient, though, is the fact that MC points are located in the region where these corrections are already relevant. This is due to the fact that obtaining a reliable, size independent estimation of $q(\alpha)$ by MC simulations becomes very difficult close to $\alpha_d$, where one has to increase both the time of the simulation and $N$. For this reason, we can only rely on the BP data, which fit reasonably well the square root singularity, to obtain $q_d=0.838(3)$. 

\begin{figure}[t]
    \begin{subfigure}[b]{0.49\columnwidth}
    \centering
    \includegraphics[width=\textwidth]{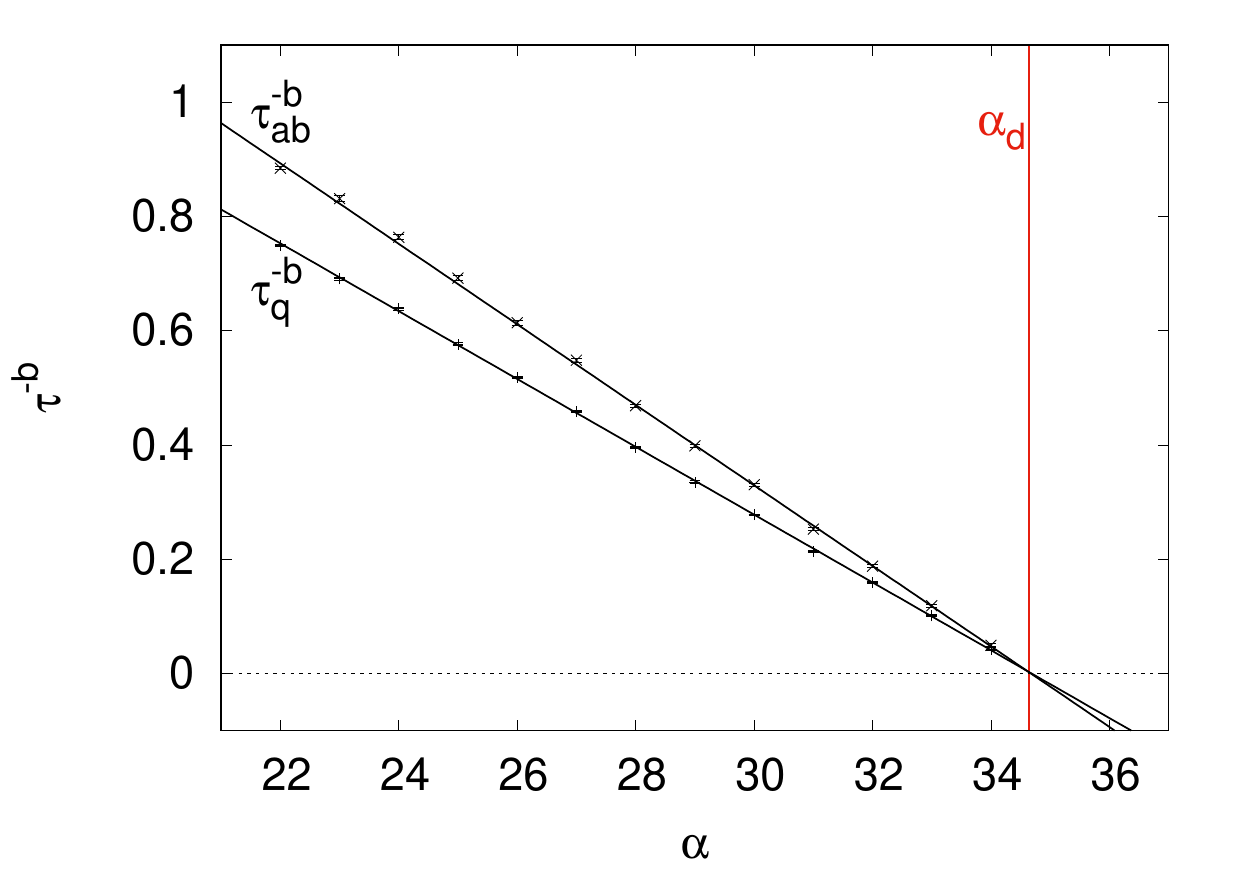}
    \caption{}
    \label{fig:q-d20-fit-a}
    \end{subfigure}
    \hfill
    \begin{subfigure}[b]{0.49\columnwidth}
    \centering
    \includegraphics[width=\textwidth]{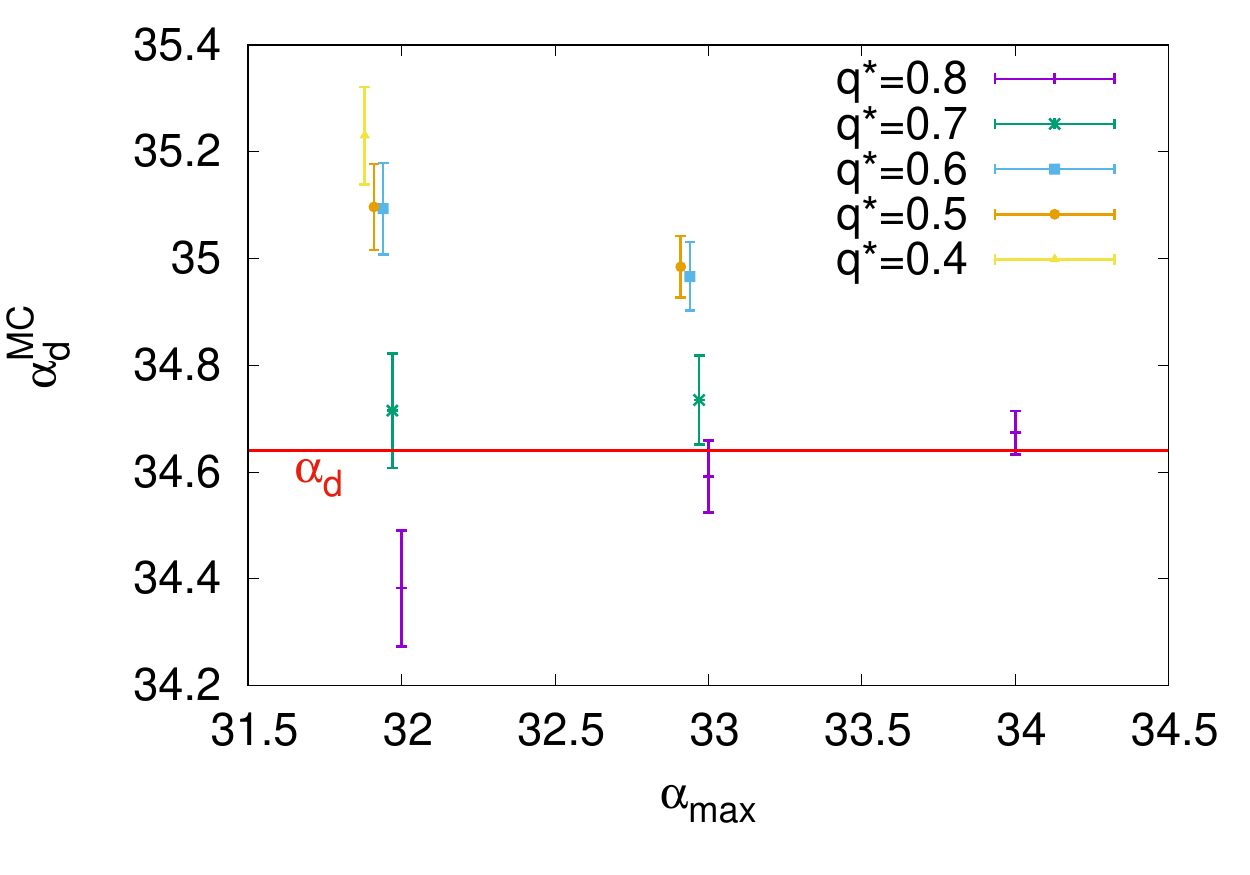}
    \caption{}
    \label{fig:q-d20-fit-b}
    \end{subfigure}
	\caption{Left: power-law divergence of $\tau_{q^*}(\alpha)$, for $q^*=0.8$. Both the overlap with the initial configuration and the one between two independent replicas are considered (the correlation times are named respectively $\tau_q$ and $\tau_{ab}$). In this case we obtain $b\approx0.26$, corresponding to a MCT-like exponent $\gamma_0\approx3.85$. Notice however that the value of $b$ strongly depends on $q^*$, see table~\ref{tab:fitresult_MC}. Right: the value of $\alpha_d$ resulting from the fit depends both on $q^*$ and on $\alpha_{\rm max}$, corresponding to the biggest value of $\alpha$ included in the fit. The red line indicates in both figures the estimate of $\alpha_d$ as obtained from BP in the previous section.}
	\label{fig:q-d20-fit}
\end{figure}

The last approach we can resort to is the study of the overlap correlation time for $\alpha<\alpha_d$. To this end, we once again define $\tau_{q^*}(\alpha)$ as the time needed for $q(t)$ or $q_{ab}(t)$ to decrease down to $q^*$, with $q^*<q_d\approx0.83$. The $T=0$ approach to $\alpha_d$ is essentially governed by MCT-like laws~\cite{montanariDynamicsGlassTransition2006}, predicting a power-law divergence $\tau(\alpha)\sim (\alpha_d-\alpha)^{-\gamma_0}$ for $\alpha\to\alpha_d^{-}$, where $\gamma_0$ is a specific $T=0$ exponent. Here we prefer to use instead the parameter $b=1/\gamma_0$, and fit the data according to eq.~\eqref{first_fit_func}. The outcome is given in figure~\ref{fig:q-d20-fit}. On the left panel we exhibit the rescaled data for $q^*=0.8$, showing a value of $\alpha_d$ compatible with the BP prediction. On the other hand, from figure~\ref{fig:q-d20-decay-a} it is already evident that in order to decrease $q^*$ one may be forced to exclude some of the biggest values of $\alpha$. For this reason, in figure~\ref{fig:q-d20-fit-b} we study how the choice of the maximum value $\alpha_{\rm max}$ included in the fit affects the estimation of $\alpha_d$, for different values of $q^*$. From the picture, one can notice that estimates on $\alpha_d$ are the lesser precise the lower $\alpha_{\rm max}$; in particular, the dispersion of data points over $q^*$ at fixed $\alpha_{\rm max}$ and the size of error bars visibly decrease from $\alpha_{\rm max}=32$ to $\alpha_{\rm max}=33$ (for $\alpha_{\rm max}=34$ we have only one point, so we cannot make any strong assertion). It is interesting, by the way, to highlight how the high dependence of $\alpha_d$ from $q^*$, for $\alpha_{\rm max}$ fixed, is accompanied by a relevant variation of also the exponent $b$ with $q^*$, see table~\ref{tab:fitresult_MC}, similarly to what happened when analyzing BP data in the previous section. In this case, however, we lack at the present moment an analytical prediction for $b$, and an accurate treatment of subleading corrections to the simple power-law behaviour is beyond our reach. For this reason, we can only conclude from figure~\ref{fig:q-d20-fit-b} that the behaviour of real MC data is reasonably well described by the population dynamics BP prediction, considering the level of the systematic biases affecting our determination of $\alpha_d$ from MC simulations.

\begin{table}[b]
		\centering
		\begin{tabular}{|c|c|c|r|}
			\hline
			$q^*$ & $\alpha_{\rm max}$ & $b$~~($\tau_q$) & $b$~~($\tau_{ab}$) \\
			\hline\hline
			0.8 & 34 & 0.260(2) & 0.263(2) \\
			    & 33 & 0.262(3) & 0.266(3) \\
			    & 32 & 0.266(4) & 0.274(5) \\[3pt]
            0.7 & 33 & 0.223(2) & 0.225(3) \\
                & 32 & 0.221(3) & 0.226(4) \\[3pt]
            0.6 & 33 & 0.206(2) & 0.204(2) \\
                & 32 & 0.205(2) & 0.200(3) \\[3pt]
            0.5 & 33 & 0.197(2) & 0.197(2) \\
                & 32 & 0.197(2) & 0.194(2) \\[3pt]
            0.4 & 32 & 0.188(2) & 0.186(2) \\ 
			\hline
		\end{tabular} 
		\caption{Value of the MCT-like exponent $b=1/\gamma_0$ from Monte Carlo simulations.}
		\label{tab:fitresult_MC}
\end{table}

\begin{figure}[t]
\centering
    \begin{subfigure}[t]{0.49\columnwidth}
        \centering
        \includegraphics[width=\textwidth]{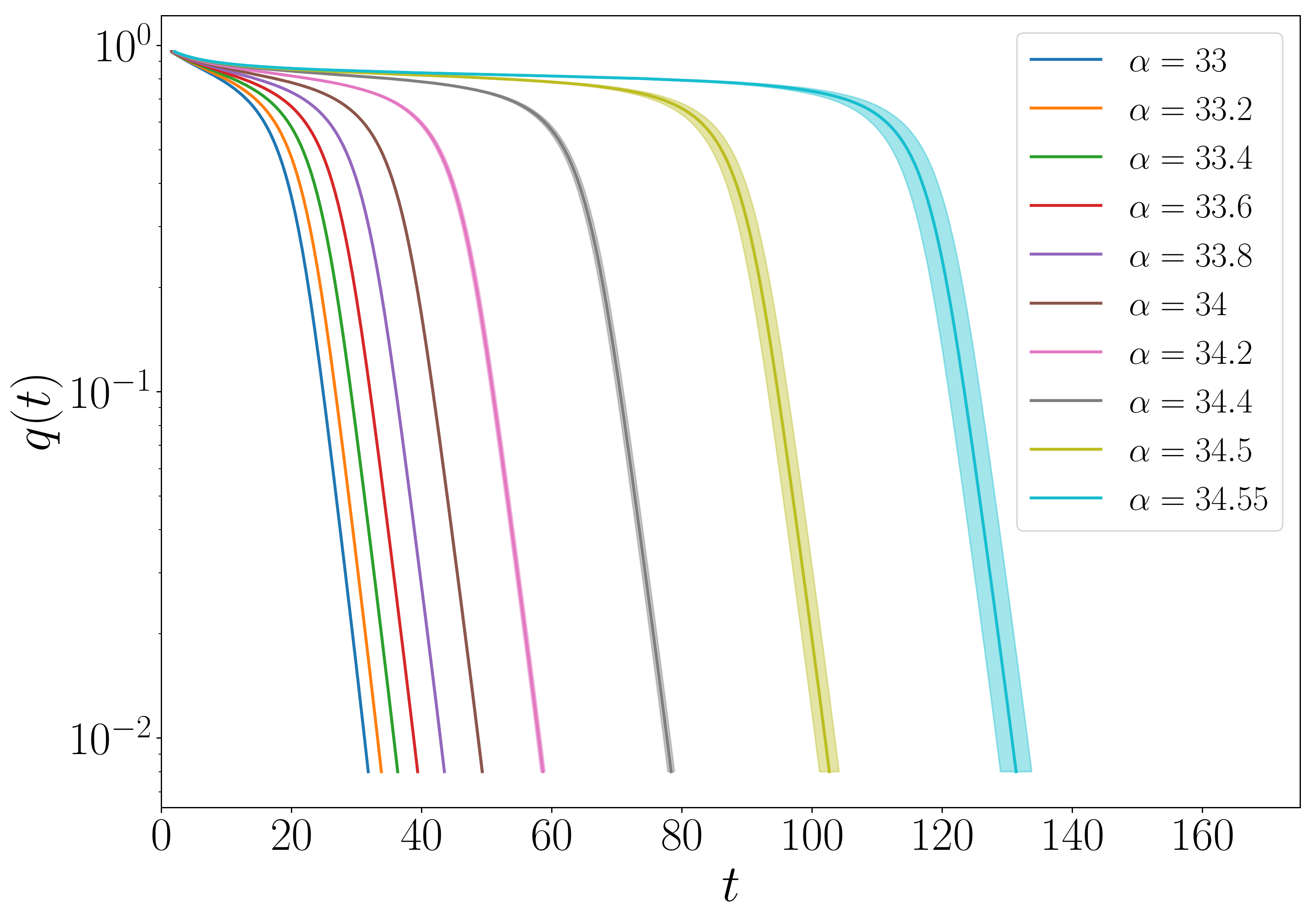}
        \caption{}
        \label{fig:q_PD_logy}
    \end{subfigure}\\ \par\bigskip
    \begin{subfigure}[b]{0.49\columnwidth}
        \centering
        \includegraphics[width=\textwidth]{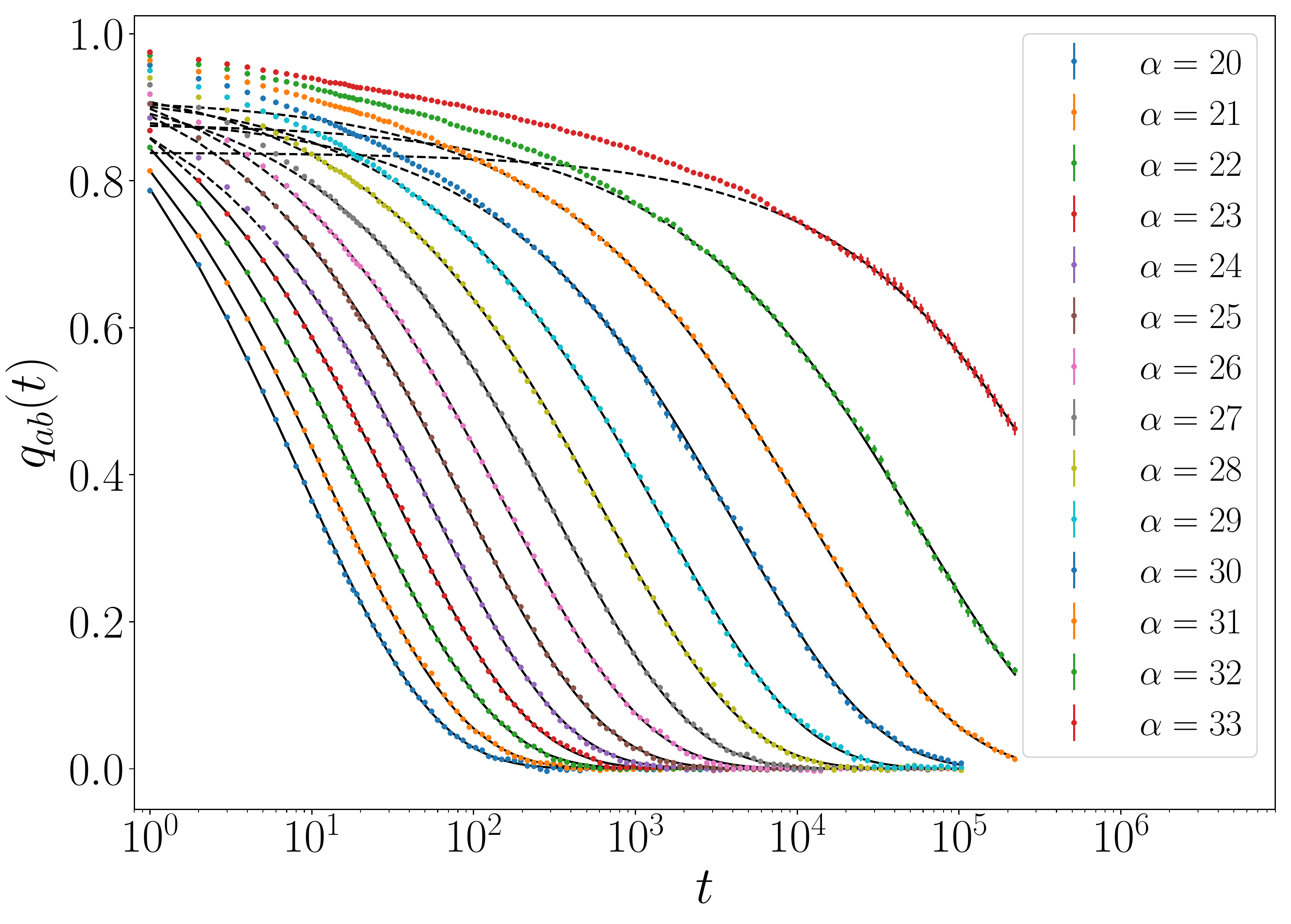}
        \caption{}
        \label{fig:q-d20-stretched}
    \end{subfigure}
    \hfill
    \begin{subfigure}[b]{0.49\columnwidth}
        \centering
        \includegraphics[width=\textwidth]{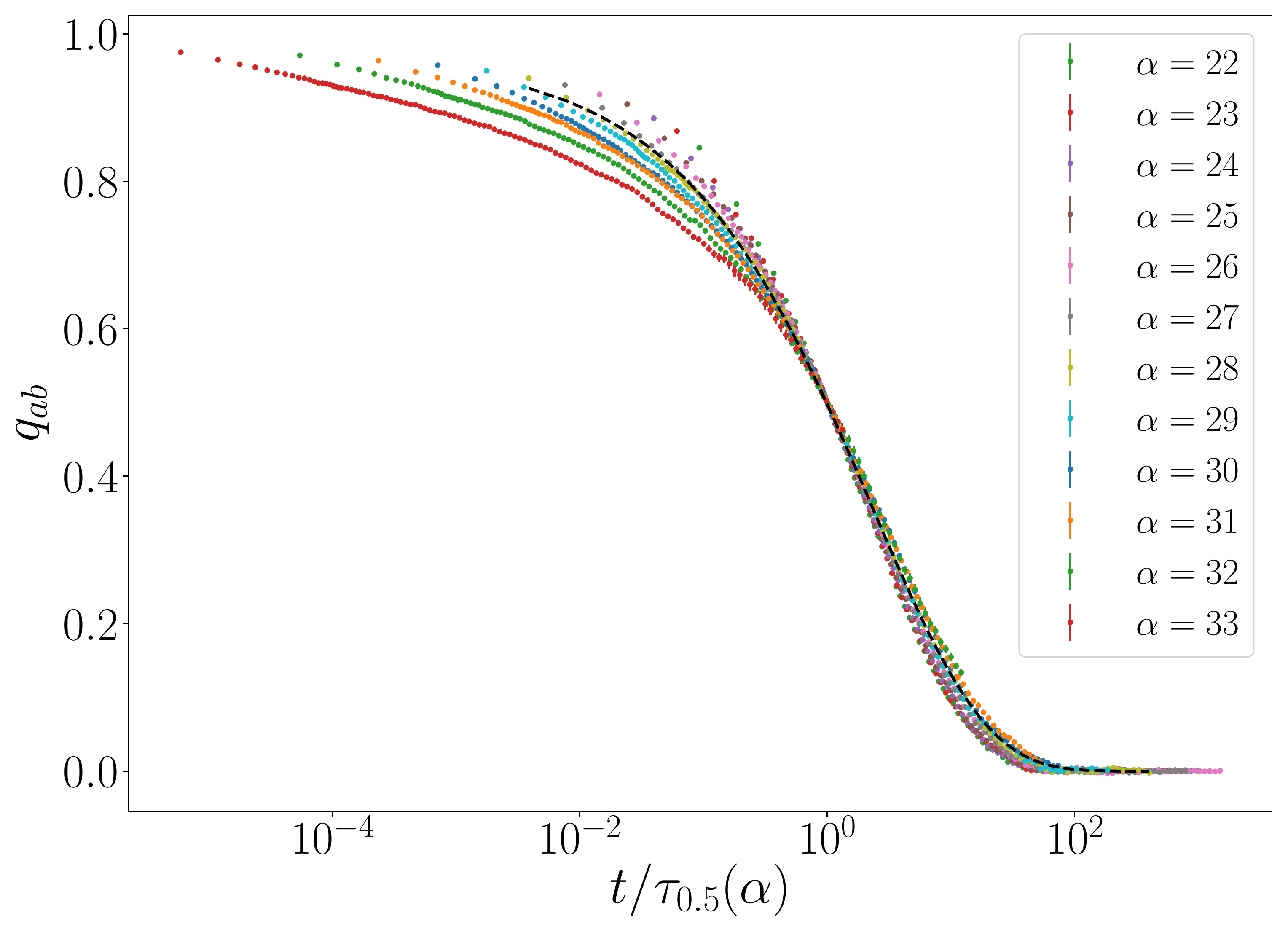}
        \caption{}
        \label{fig:q-d20-stretched-collapse}
    \end{subfigure}
    \caption{Top: overlap decay from population dynamics BP equations for $\alpha<\alpha_d$, using the same data of figure~\ref{fig:q_PD_a}. The log-scale on the y-axis highlights a purely exponential tail for the decay at small overlap. Bottom left: overlap decay from MC data. Black lines represent fits according to a stretched exponential of the form $A e^{-(t/\hat{\tau})^{\beta}}$. The values of $\beta$ vary randomly between 0.48 and 0.53. Bottom right: the same data is plotted against $t/\tau_{q^*=0.5}(\alpha)$. Black dashed line is a stretched exponential of exponent $\beta=0.48$.}
    \label{fig:q-stretched-compare}
\end{figure}

Finally, it is interesting to consider in figure~\ref{fig:q-stretched-compare} the long time tail of the overlap decay as approaching $\alpha_d$ from below. In the top panel, the BP data is shown to possess a simple exponential behaviour. The MC dynamics of real systems, on the contrary, exhibits a stretched exponential form $A e^{-(t/\hat{\tau})^{\beta}}$, typical of MCT systems close to the MCT transition. The value of the exponent $\beta$ from the different fits (black lines in fig.~\ref{fig:q-d20-stretched}) is reasonably constant on the whole considered $\alpha$-interval, varying between 0.48 and 0.53 without any recognizable trend. The parameter $\hat{\tau}(\alpha)$ from the fit is in principle proportional to the structural relaxation time of the system $\tau(\alpha)$. However, we believe the definition of $\tau(\alpha)$ in terms of $\tau_{q^*}(\alpha)$ to be somewhat more reliable, since it does not depend on the quality of a phenomenological fit and on the considered fitting interval. On the right panel we show the correlation for different values of $\alpha$ as a function of the rescaled time $t/\tau_{q^*=0.5}(\alpha)$, resulting in a decent collapse of all the curves. The collapse is not perfect as the tail becomes slightly more pronounced when increasing $\alpha$.

\subsection{Effects of discretization}
\label{section:extrapolation_continuous_limit}
All the analysis performed so far was only limited to a discretized version of the model, with precision $d=10$ to be exact. This coincides, having chosen for definiteness to focus on the fixed $q=20$ case, for which the transition is appreciably first order, with approximating the continuous interval $[0,2\pi)$ with $p=qd=200$ discrete clock-states. It is thus essential to assess how much of the quantitative predictions given so far, in particular for what concerns the computation of the transition lines, is still valid in the limit $p\to\infty$. 
\begin{figure}[t]
    \centering
    \centering
    \begin{subfigure}[b]{.49\columnwidth}
        \includegraphics[width=\textwidth]{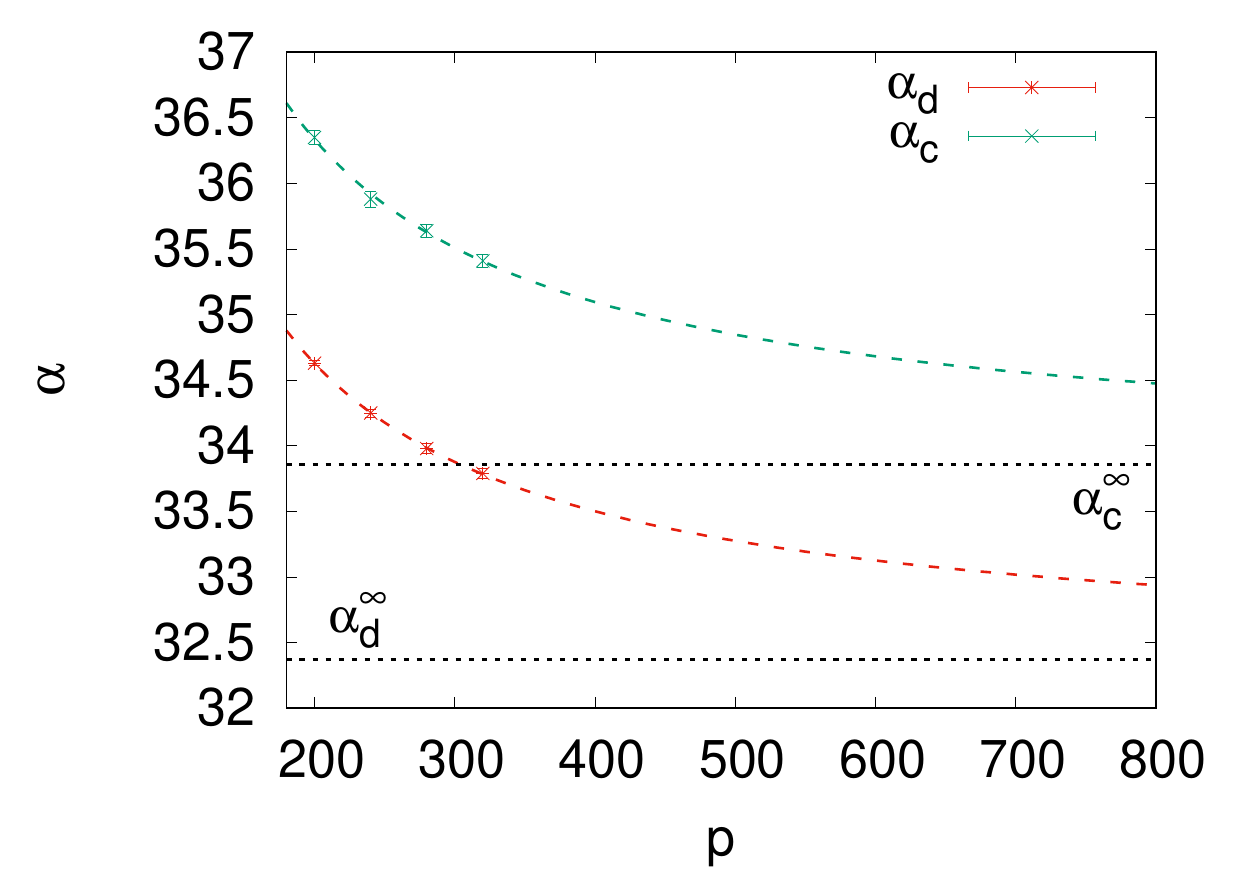}
        \caption{}
    \end{subfigure}
    \hfill
    \begin{subfigure}[b]{.49\columnwidth}
        \includegraphics[width=\textwidth]{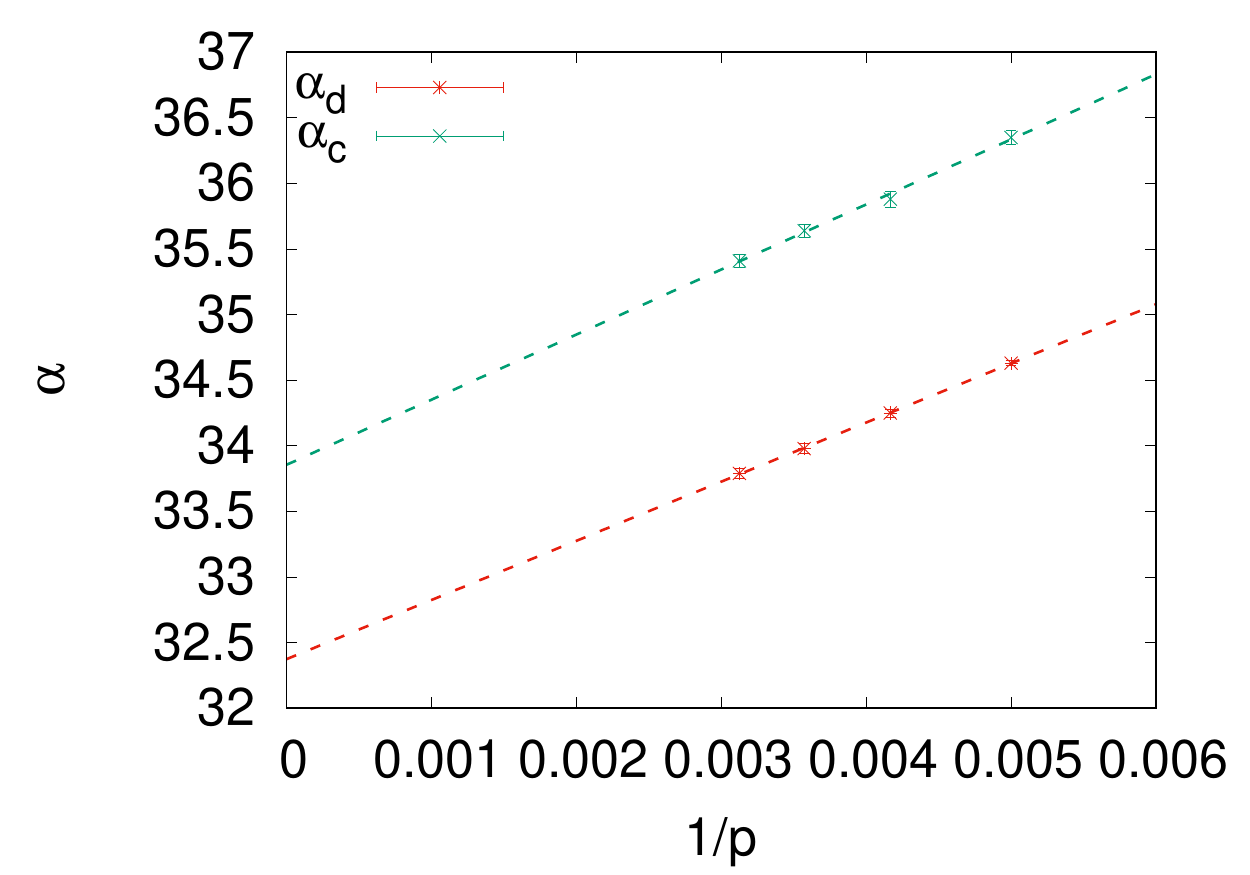}
        \caption{}
    \end{subfigure}
    \caption{Corrections to the $T=0$ transition thresholds scale as $1/p$.}
    \label{fig:discretization_study}
\end{figure}
The answer is provided in figure~\ref{fig:discretization_study}, where we plot the $T=0$ thresholds $\alpha_d(p)$ and $\alpha_c(p)$ for $p=200,240,280,320$, as obtained from the population dynamics BP numerical procedure. The corrections, for the values of $p$ considered, appear to be very relevant. Even worse, they show a very slow $1/p$ scaling. For these reasons, we believe a precise extrapolation of the continuous limit of the thresholds, starting only from the solution of the discretized BP equations, to be rather delicate. In this spirit we report the results from the fits of figure~\ref{fig:discretization_study}, to be taken with caution: $\alpha_d^{\infty}\approx32.37$ and $\alpha_c^{\infty}\approx33.86$. On a more bright side, however, some way to better control the extrapolation error could be hopefully provided by the study of the $p$ dependence of $\alpha_{\rm KS}$, which is known analytically~\eqref{KS_DiscreteModel}, and that also shows an approximate $1/p$ behaviour.

The $1/p$ scaling of discretization corrections for the model at hand was already reported in~\cite{mezardSolutionSolvableModel2011}. Unfortunately, the situation is very different from other planar spin-models on random graphs, exhibiting a very fast exponential convergence~\cite{lupoApproximatingXYModel2017}. This is presumably due to the peculiarity of the excluded volume interaction, that sharply depends on the particle diameter $\theta$. Among the effects of discretization, we have indeed that the forbidden states are contained by all means into an effective diameter which is smaller than $\theta$ by a quantity of order $1/p$ (see for example the red points in figure~\ref{fig:discretization}). When we increase discretization, we are hence actually increasing also the effective diameter (\emph{i.e.} decreasing the effective $q$), which may play a direct role in the $1/p$ decrease of the transition thresholds, since the system is more constrained for bigger diameters, see for instance figure~\ref{fig:PhaseDiag_discr_vs_cont}.

\section{Optimizing the dynamical transition}
\label{sec:optimizing}

\subsection{Complexity maximization}
\label{section:complexity_maximization}
\begin{figure}[t]
    \begin{subfigure}[b]{0.49\columnwidth}
    \centering
    \includegraphics[width=\textwidth]{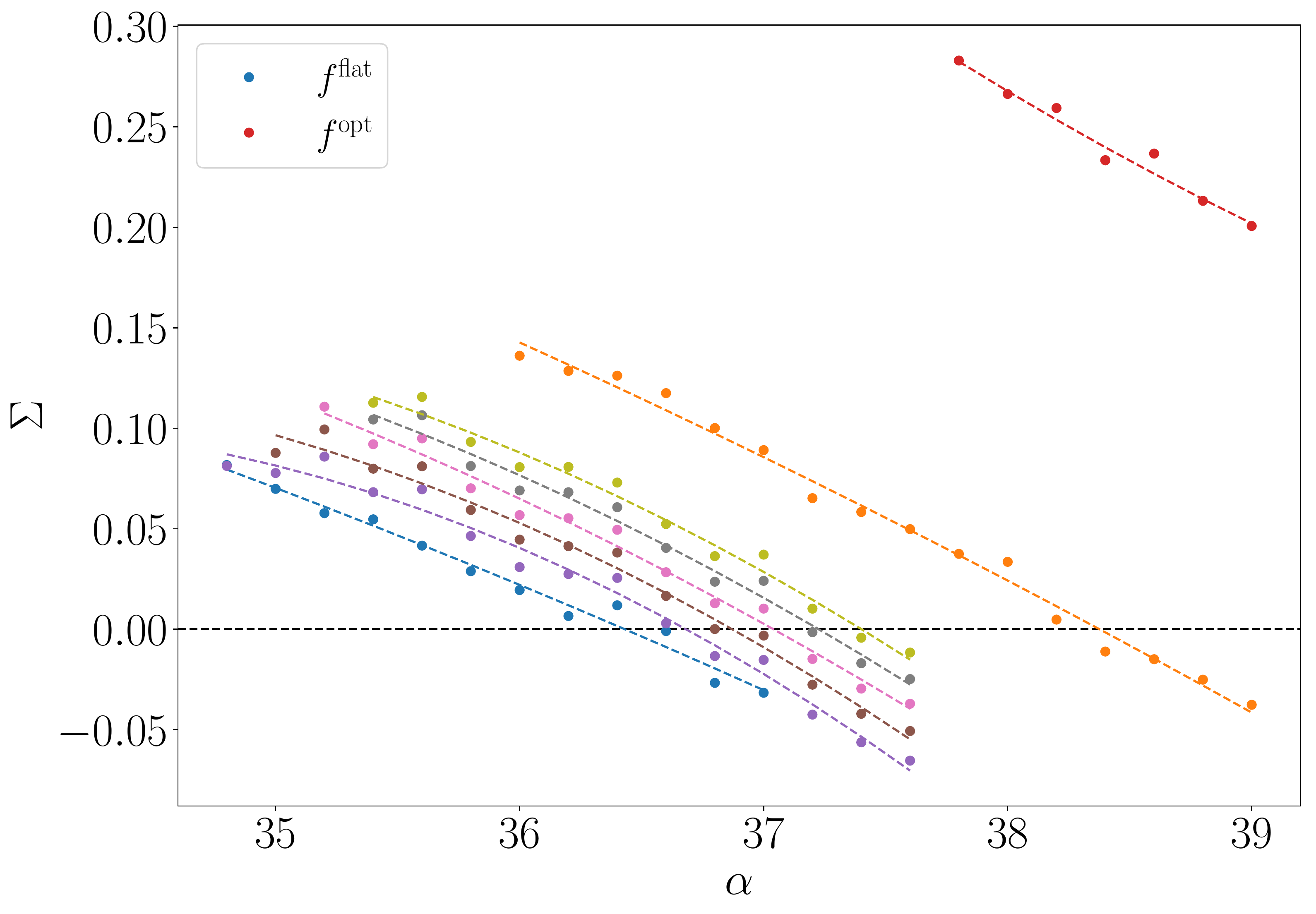}
    \caption{}
    \label{fig:sigmas_optimize_1}
    \end{subfigure}
    \hfill
    \begin{subfigure}[b]{0.49\columnwidth}
    \centering
    \includegraphics[width=\textwidth]{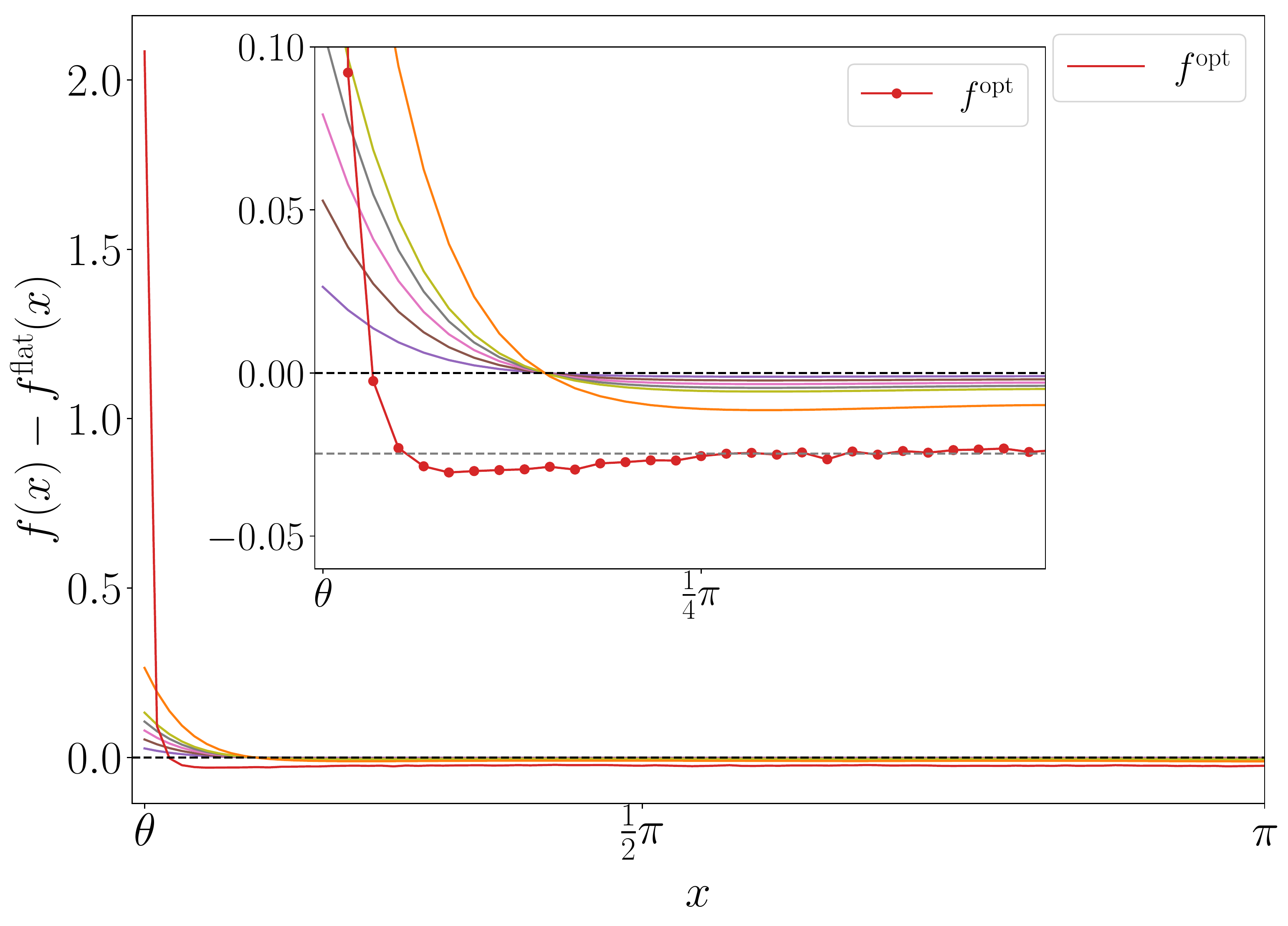}
    \caption{}
    \label{fig:sigmas_optimize_potentials}
    \end{subfigure}
	\caption{Left: complexity curves as a function of $\alpha$ for different interactions (that are depicted on the right panel with the corresponding color) along the optimization procedure. Dotted lines are meant only to guide the eye. Blue curve refers to the starting point $f^{\rm flat}$, while red curve  to the result of the optimization $f^{\rm opt}$. Curves in between all refer to the first step in the procedure (we move along the direction of the gradient of the complexity starting from $f^{\rm flat}$), for increasing sizes of the `step' in the direction of the gradient, \emph{i.e.} the speed rate of the gradient descent.
	}
	\label{fig:sigmas_optimize}
\end{figure}

The `typical' complexity $\Sigma$ (\emph{i.e.} the one associated with dominant states) is an observable that can be estimated via BP as a difference of free entropies, see eq.~\eqref{eq:sigma_def_BP}. The general behaviour of $\Sigma$ as a function of $\alpha$ in the case of a discontinuous transition, as shown in figure~\ref{fig:alphac_complexities}, is the following: $\Sigma=0$ below $\alpha_d$, then $\Sigma$ has a jump at $\alpha_d$ to a finite (maximum) value, and hence starts to monotonically decrease with $\alpha$ up to $\alpha_c$, where $\Sigma=0$. One may then wonder if there is a way to postpone the dynamical threshold $\alpha_d$ by directly acting on the shape of the complexity. A strategy that we found to be successful is to iteratively maximize the value of the complexity close to its maximum at $\alpha_d$: the idea is that by shifting the complexity curve to higher values of $\alpha$, also the value of $\Sigma$ at fixed $\alpha > \alpha_d$ will increase, see figure~\ref{fig:sigmas_optimize_1}. However, since the maximum value of the complexity can take any strictly positive value, there is no reason to exclude the existence of particular functions $f$ for which the complexity curve is still postponed, but it is also lowered. In this sense, we lack a physical intuition (not to mention a rigorous proof) supporting the fact that the interaction $f$ (not to be confused with a free energy) for which $\alpha_d(f)$ has its maximum should coincide with the function $f$ for which also $\Sigma(\alpha_d(f);f)$ has its maximum.

Having said that, this method is the best chance to our knowledge to automatically optimize the potential when an analytical expression for $\alpha_d$ is not known, and it yields promising results. We compared it with other strategies, such as  the maximization of the stability parameter $\lambda$ of the BP non-trivial solution, which is $\lambda=1$ at the dynamical transition and then decreases for $\alpha>\alpha_d$ (the solution being more stable). Since it takes a fixed value at $\alpha_d$ for any $f$, then the maximization procedure is in this case justified. Another strategy is the minimization of the so called Kullack-Leibler divergence, measuring the distance of the non-trivial distribution of messages $\mathbb{P}^{\rm eq}(\nu)$ from the paramagnetic one, which in the continuous notation reads
\begin{align}
\mathbb{E}_{\mathbb{P}^{\rm eq}(\nu)} \left[ D_{\rm KL}(\nu^{\rm para}\Vert\nu) \right] = \mathbb{E}_{\mathbb{P}^{\rm eq}(\nu)} \left[  -\frac{1}{2\pi}\int_0^{2\pi} dx \log\left(\frac{\nu(x)}{(2\pi)^{-1}}\right) \right]. \label{KL_Cost_Function}
\end{align}
For $\alpha<\alpha_d$ this quantity is zero, since the only solution is $\mathbb{P}^{\rm eq}=\mathbb{P}^{\rm para}$, it takes a non-trivial value at $\alpha_d$ and then increases for $\alpha>\alpha_d$. Following the same line of thought as in the previous cases, one should minimize it. Furthermore, the minimization of this quantity has a nice physical implication: since~\eqref{KL_Cost_Function} becomes bigger the more the fixed point messages depart from the uniform solution, this implies that solutions containing frozen fields (which are highly irregular distributions) are strongly penalized by the procedure. However, the numerical estimation of the gradient of $\lambda$ and of the Kullback-Leibler divergence is much more involving than computing $\partial \Sigma/\partial f$, which reduces to computing the derivative of a free energy. Moreover, the $\lambda$ estimator becomes very noisy as the optimization proceeds, thus requiring the use of an increasing number of samples, while the Kullback-Leibler divergence appears also to decrease the condensation threshold $\alpha_c$, making convergence difficult near the tri-critical point. We did not further investigate this phenomenon. 

In the end, the best result is obtained through the complexity maximization. Despite the caveats already expressed, it has some practical advantages. In particular, this method does not rely exclusively on BP to estimate messages. Being a derivative of a difference of free energies, the gradient $\partial \Sigma/\partial f$ can be computed also as some two-point correlation function, \emph{e.g.} via Monte Carlo sampling, as outlined in the following. Moreover, the method is completely general, and it can in principle be applied also to mean-field hard-spheres systems as studied in~\cite{maimbourgGeneratingDensePackings2018}, where an exact method can be independently developed, hopefully contributing to shed some light on the physics behind the complexity maximization.

The maximization is implemented through a gradient descent: we move in the space of $f$ (subjected to the normalization constraint $\sum_a f_a=p$) along the gradient $\partial\Sigma/\partial f$ at fixed $\alpha=\alpha_d+\delta\alpha$, compute the new $\alpha_d$ and repeat the procedure until stationarity is reached. Thanks to discretization, the gradient descent in the space of functions $f$ (which depend only on one argument, the angular distance) can be numerically treated as an optimization in the finite set of parameters $\{f_a\}$.
The procedure is the following
\begin{align}
f_a^{(t+1)} = f^{(t)}_a + \mu \left(\frac{\partial \Sigma \left(\alpha_d( f^{(t)} ) + \delta\alpha;f^{(t)}\right)}{\partial f_a} + \Lambda\right) \mathbb{I} \big(  \cos(2\pi a/p) \leq \cos(2\pi/q)    \big), \label{Update_func_f}
\end{align}
where $\Lambda=-[p-(2d-1)]^{-1}\sum_{a=d}^{p-(d-1)}\partial\Sigma^{(t)}/\partial f_a$ is a Lagrange multiplier ensuring that $\{f_a^{(t+1)}\}$ is correctly normalized at each step,  and $\mu$ is the speed rate of the gradient descent. We found empirically that computing the gradient close to the dynamical transition results in less noise. In all the computations we have used $\delta\alpha = 0.5$.
As the optimization proceeds, the support of the gradient concentrates on the forbidden values that encode the constraints of the problem. The optimization procedure ends, when the only way to further increase the complexity would be to relax the constraints, something that is forbidden thanks to the factor $\mathbb{I} \left(  \cos(2\pi (a-1)/p) \leq \cos(2\pi/q)\right)$ in \eqref{Update_func_f}.

\subsection{Computation of the gradient of the complexity}
The complexity we want to maximize is defined as
\begin{equation}
    \Sigma = \frac{1}{N} \left[F_{\rm Rec\,II}(\mathbb{P}^{\rm Para},f) - F_{\rm Rec\,II}(\mathbb{P}^{\rm eq},f)\right],
\end{equation}
where $\mathbb{P}^{\rm Para}(\nu)=\prod_{a=1}^p \delta(\nu_a-n)$, ~$n=\frac{1}{p}$, is the paramagnetic BP fixed point and $\mathbb{P}^{\rm eq}$ is the non-trivial one arising for $\alpha>\alpha_d(f)$. In our discretized setting, $f=\{f_a\}$ is a symmetric function under $a\to p-a$ and periodic of period $p$ (we use in this sense the convention $f_{-a}=f_{p-a}$). The population dynamics definition of the free entropy $F_{\rm Rec\, II}$ is given by eq.~(\ref{FreeEnergy_Equation_Reconstruction_Sym}), which we recall here for convenience as
\begin{align}
\frac{1}{N}F_{\rm Rec\,II}(\mathbb{P},f^{\rm Fixed},f^{\rm State}) =& \sum_d P_{\rm Node}(d) \int\prod_{k = 1}^d  \left[d\nu_k \sum_b f^{\rm State}_{b} n \,\mathbb{P}\left(T(-b)\nu_k\right) \right] \log \left[Z(\{\nu_k\}_d,f^{\rm Fixed}) \right] \nonumber 
\\&- \alpha\int d\nu_1 d\nu_2  \,\mathbb{P}(\nu_1)  \sum_{b} f_{b}^{\rm State} n \,\mathbb{P}(T(-b)\nu_2)  \log\left[ \sum_{r,s} \nu_{1,r} f_{s-r}^{\rm Fixed} \nu_{2,s} \right].
\label{eq:F1RSB_complexity}
\end{align}
The distinction between $f^{\rm State}$ and $f^{\rm Fixed}$ is here only introduced in order to discuss separately the different terms coming from the derivative with respect to $f_a$. Of course in the end one has to evaluate every derivative at $f^{\rm State}=f^{\rm Fixed}=f$, with $\sum_a n f_{a}=1$, $n=\frac{1}{p}$. Finally, we also recall that 
\begin{equation}
    Z(\{\nu_k\}_d,f) = \sum_{a=0}^{p-1} n \prod_{k=1}^d \sum_{b=0}^{p-1}f_{b-a}\,\nu_{k,b}.
\end{equation}

The derivative  $\partial\Sigma/\partial f_a$ contains terms coming from the explicit dependence of $\Sigma$ from both $f^{\rm State}$ and $f^{\rm Fixed}$, but also from the change in the non-trivial fixed point distribution $\mathbb{P}^{\rm eq}$. The paramagnetic fixed point distribution, on the contrary, is given by $\mathbb{P}^{\rm Para}\equiv \prod_a\delta(\nu_a-n)$ for any choice of $f$ satisfying $\sum_a f_a n = 1$, as can be immediately noticed by a direct inspection of the BP fixed point equation~(\ref{Reconstruction_BP_II}). We can thus write
\begin{align}
\frac{N\partial \Sigma}{\partial f_a} =& \frac{\partial F_{\rm Rec\,II}(\mathbb{P}^{\rm Para},f^{\rm Fixed},f^{\rm State})}{\partial f_a^{\rm Fixed}} - \frac{\partial F_{\rm Rec\,II}(\mathbb{P}^{\rm eq},f^{\rm Fixed},f^{\rm State})}{\partial f_a^{\rm Fixed}} \nonumber
\\
&+ \frac{\partial F_{\rm Rec\,II}(\mathbb{P}^{\rm Para},f^{\rm Fixed},f^{\rm State})}{\partial f_a^{\rm State}} - \frac{\partial F_{\rm Rec\,II}(\mathbb{P}^{\rm eq},f^{\rm Fixed},f^{\rm State})}{\partial f_a^{\rm State}} \nonumber
\\
&-\frac{\partial F_{\rm Rec\,II}}{\partial \mathbb{P}}(\mathbb{P}^{\rm eq},f^{\rm Fixed},f^{\rm State}) \left[\frac{\partial \mathbb{P}^{\rm eq}}{\partial f^{\rm Fixed}_a} + \frac{\partial \mathbb{P}^{\rm eq}}{\partial f^{\rm State}_a} \right].
\label{eq:gradient_complexity_all_terms}
\end{align}
Most of the terms either simplify or are equal to zero. The first two terms cancel, as we are going to comment in a moment. The third term is trivially equal to zero, since both the logarithms in eq.~(\ref{eq:F1RSB_complexity}) vanish when one considers $\nu_a=n$ and the fact that $n f^{\rm Fixed}_a$ is correctly normalized. The fifth term is zero since the distribution $\mathbb{P}^{\rm eq}$ extremizes the Bethe free entropy. In the end, the only remaining term is
\begin{align}
\frac{\partial \Sigma}{\partial f_a} =&  - \frac{\partial N^{-1} F_{\rm Rec\,II}(\mathbb{P}^{\rm eq},f^{\rm Fixed},f^{\rm State})}{\partial f_a^{\rm State}} \Bigg\lvert_{f^{\rm State}=f^{\rm Fixed}=f}.
\label{eq:gradient_complexity_remaining_term}
\end{align}

Before proceeding with the computation, let us show that the first two terms indeed cancel. The derivative $\partial N^{-1}F_{\rm Rec\,II}/\partial f_a^{\rm Fixed}$, for generic $\mathbb{P}$, can be written after some manipulations\footnote{One can use the decomposition~\eqref{eq:logZ_d+1} in order to recast the derivative of the first term in the rhs of~\eqref{eq:F1RSB_complexity} in the form reported in the text (the derivative of the second term is trivially proportional and they can be added).}~as
\begin{equation}
    \frac{\partial N^{-1}F_{\rm Rec\,II}}{\partial f_a^{\rm Fixed}}\Bigg\lvert_f = \alpha \int d\nu_1d\nu_2\,\mathbb{P}(\nu_1)\sum_bf_b^{\rm State}n\,\mathbb{P}(T(-b)\nu_2)\frac{\sum_r\nu_{1,r}\,\nu_{2,r+a}}{\sum_{r,s}\nu_{1,r}f_{s-r}\,\nu_{2,s}}.
\end{equation}
If we multiply and divide by the same amount $f_a$, we can make a term $\nu_{1,r}f_a\,\nu_{2,r+a}$ appear, which is proportional by the definition of messages to the marginal probability $p(x_1=r,x_2=r+a)$, as a function of $r$, of two neighbouring variables at a fixed distance $a$. The whole integral thus represents the total equilibrium probability for two neighbours to be at distance $a$, which is essentially the pair correlation function of the system. In the entire range $\alpha<\alpha_c$, though, the pair correlation function of the paramagnetic and of the planted fixed point is the same and is simply equal (whenever short loops are absent) to the normalized interaction $nf_a$, that is essentially the Boltzmann-Gibbs exponential of the pairwise potential. Simplifying the extra $1/f_a$ factor, we obtain that the first two terms in eq.~\eqref{eq:gradient_complexity_all_terms} cancel as they are both equal to a constant $\alpha/p$. As a final remark, this derivation is clearly valid for any $T\neq 0$, when also $f_a\neq0$ ~$\forall a$. The case $f_a=0$ can nevertheless be recovered from the previous one by taking the limit $T\to0$.

Coming back to our computation, plugging into eq.~\eqref{eq:gradient_complexity_remaining_term} the definition of $F_{\rm Rec\,II}$ given by eq.~(\ref{eq:F1RSB_complexity}), we obtain
\begin{align}
&\frac{\partial \Sigma}{\partial f_a} = \alpha n \int d\nu_1d\nu_2  \,\mathbb{P}^{\rm eq}(\nu_1)  \mathbb{P}^{\rm eq}(T(-a)\nu_2)  \log\left[ \sum_{r,s} \nu_{1,r} f_{s-r} \nu_{2,s} \right] +\nonumber \\
&
-2\alpha n \sum_d P_{\rm Node}(d) \int d\nu_{d+1} \,\mathbb{P}^{\rm eq}(T(-a)\nu_{d+1})   \prod_{k = 1}^{d} \left[ d\nu_k  \sum_b f_{b}n \,\mathbb{P}^{\rm eq}(T(-b)\nu_k)\right] \log \left[ Z(\{\nu_k\}_d \cup \nu_{d+1},f) \right],
 \label{eq:Gradient_Complexity_1}
\end{align}
where we have used the fact that $d \cdot P_{\rm Node}(d) = 2\alpha P_{\rm Node}(d-1)$ and replaced a dummy index $d$ in the second term with $d+1$. The quantity $\log(Z)$ appearing in the previous equation can be iteratively expressed in terms of the update function $\Phi^{\rm update}$ given by eq.~(\ref{Update_Function_F}) as 
\begin{align}
\log \left[Z(\{\nu_k\}_d \cup  \nu_{d+1} , f ) \right] &= \log \left[ \sum_a n \prod_{1 \leq k \leq d+1} \left(\sum_b f_{b-a} \,\nu_{k,b} \right) \right] = \nonumber\\
&=\log \left[\sum_a \Phi^{\rm update}_a(\{\nu_k\}_d, f)Z(\{\nu_k\}_d, f )\sum_b f_{b-a} \,\nu_{d+1,b}\right] = \nonumber \\
&=\log \left[Z(\{\nu_k\}_d, f ) \right] + \log\left[\sum_{ab} \Phi^{\rm update}_a(\{\nu_k\}_d, f) f_{b-a} \,\nu_{d+1,b} \right].
\label{eq:logZ_d+1}
\end{align}
Plugging~\eqref{eq:logZ_d+1} into~\eqref{eq:Gradient_Complexity_1}, one recognizes\footnote{Formally, one can multiply~\eqref{eq:Gradient_Complexity_1} by a factor $1=\int d\nu \,\delta\!\left(\nu - \Phi_a^{\rm update}(\{\nu_k\}_d,f)\right)$ in order to make appear the BP expression~\eqref{Reconstruction_BP_II} for $\mathbb{P}^{\rm eq}(\nu)$.}~that $\Phi^{\rm update}$ can be interpreted as a random message $\nu$ distributed according to $\mathbb{P}^{\rm eq}(\nu)$. The second term from~\eqref{eq:logZ_d+1} has then the same form of the first term in~\eqref{eq:Gradient_Complexity_1} and they can be directly added. Therefore one obtains
\begin{align}
\frac{\partial \Sigma}{\partial f_a} =& - \alpha n \int d\nu_1 d\nu_2  \,\mathbb{P}^{\rm eq}(\nu_1)  \mathbb{P}^{\rm eq}(T(-a)\nu_2)  \log\left[ \sum_{rs} \nu_{1,r} f_{s-r} \nu_{2,s} \right] + \nonumber \\
 &-2\alpha n \sum_d P_{\rm Node}(d) \int \prod_{k = 1}^{d} \left[ d\nu_k \sum_b f_{b} n \,\mathbb{P}^{\rm eq}(T(-b)\nu_k) \right] \log \left[Z(\{\nu_k\}_d, f) \right].
\end{align}
We notice that the second term does not depend on the index $a$. This term is trivially constant and can hence be disregarded, since at each step we already enforce the normalization condition $\sum_a f_a n = 1$. Finally one gets
\begin{align}
\frac{\partial \Sigma}{\partial f_a} =&  -\frac{\alpha}{p} \int d\nu_1 d\nu_2 \,\mathbb{P}^{\rm eq}(\nu_1)  \mathbb{P}^{\rm eq}(T(-a)\nu_2)  \log\left[ \sum_{rs} \nu_{1,r} f_{s-r} \nu_{2,s} \right]  \label{Gradient_Complexity}
\end{align}

\subsection{Monte Carlo estimator of the gradient of the complexity}
\label{sec:grad_complex_MC}

Remarkably, the previous formula~\eqref{Gradient_Complexity} can also be estimated using Monte Carlo sampling. This is very convenient, since it might be the only numerically feasible approach in the case of higher spatial dimensions and/or high discretization. To this purpose, one needs to generate a planted graph $A$ along with a planted configuration, denoted $\{x_i^0\}_{1 \leq i \leq N}$, and then run a Monte Carlo algorithm initialized in the planted configuration to create samples. Let us assume we have generated $T$ samples $\{x_i^t\}_{1 \leq i \leq N,\, 1\leq t \leq T} \in \{0, \dots, p-1 \}^{N \times T}$ of the system. The two cavity messages $\nu_1$ and $\nu_2$ in eq.~\eqref{Gradient_Complexity} are random variables drawn independently from $\mathbb{P}^{\rm eq}(\nu)$. We can use at this point the fact that the distribution of cavity messages conincides, for the Erd{\H o}s-R\'enyi ensemble, also with the distribution of single-variable local marginal probabilities for the complete graph. This allows one to recast an average over $\nu_i$ of the kind of $\sum_a \nu_{i,a} \,g(a)$ inside the integral in eq.~\eqref{Gradient_Complexity}, where $g$ is a generic function, as the time average over the generated samples $\{x_i^t\}_{1\leq t\leq T}$, for each fixed site $i$ (eventually averaging over the choice of site $i$): $\frac{1}{T}\sum_{t=1}^T g(x_i^t)$. Since $\nu_1$ and $\nu_2$ should be extracted independently, we simulate two different graphs (the distribution $\mathbb{P}^{\rm eq}$ itself encodes the average over the graphs ensemble).
\begin{figure}[t]
    \centering
    \includegraphics[width=0.55\textwidth]{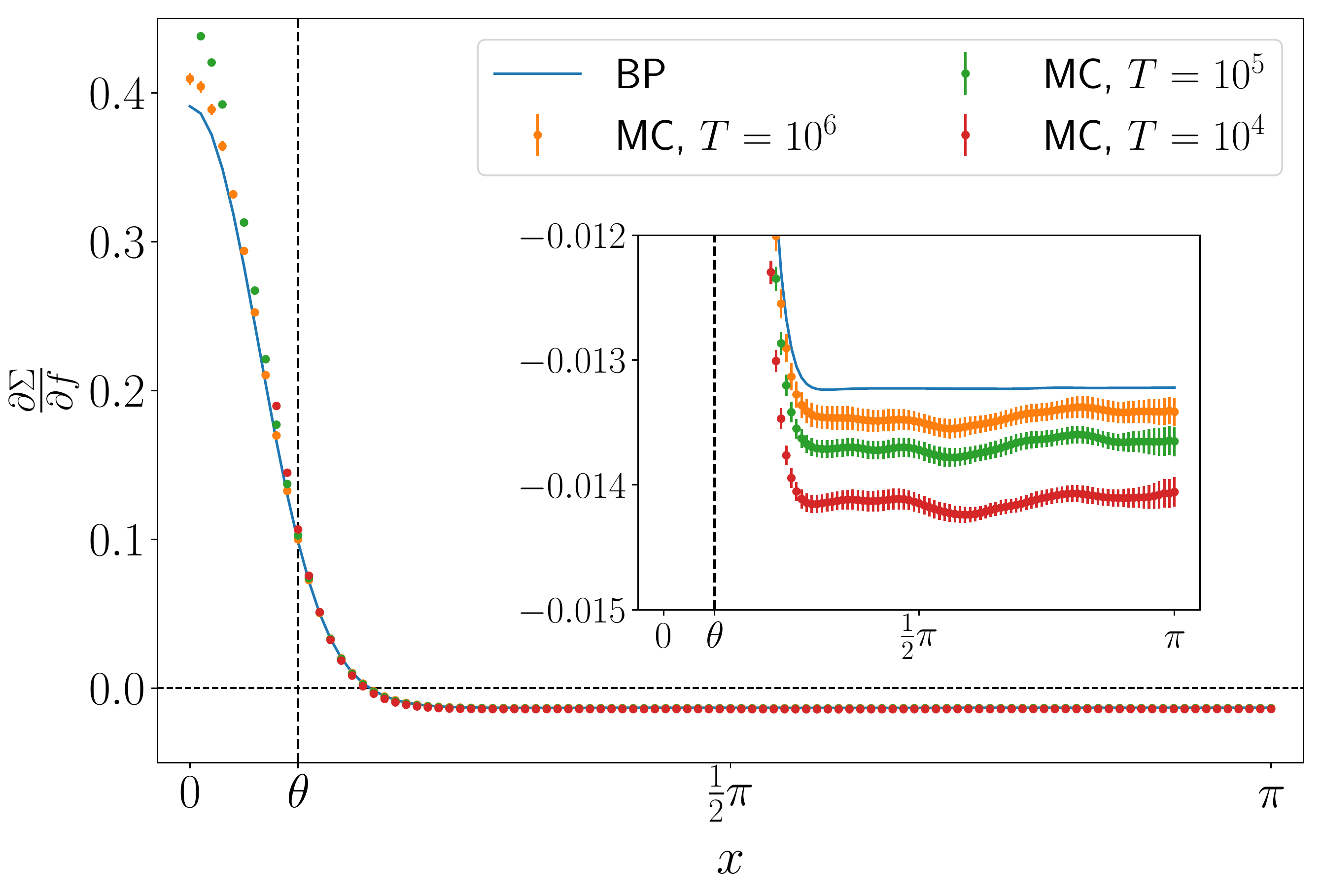}
    \caption{Comparison of the gradient of the complexity around $f=f^{\rm flat}$ and for $\alpha=\alpha_d+0.5$, estimated from BP population dynamics using eq.~\eqref{Gradient_Complexity} and MC simulations using~\eqref{Gradient_Complexity_MCMC}. In the latter case, we considered systems of size $N=10^4$, while error bars accounts for 10 repetitions of the experiment on different graphs. MC data for $x\geq\theta$ is less sensible to the total simulation time $T$ (this is convenient as long as one is interested in optimizing the soft part of the interaction only, as in our case). In particular, not all the data points in the region $x<\theta$ are well defined for small $T$. Inset: fine detail of the soft region $x>\theta$. }
    \label{fig:MC_grad_complexity}
\end{figure}
Finally, notice that the presence of the shift $\mathbb{P}^{\rm eq}(T(-a)\nu_2)$ translates into considering for $\nu_2$ a new set of generated samples given by $\{x_i^t + a\}_{1\leq i \leq N,\, 1\leq t\leq T}$.
The gradient of the complexity is then estimated using the following formula
\begin{align}
\frac{\partial \Sigma}{\partial f_a} =&  -\frac{\alpha}{p}\frac{1}{U}\sum_{(i,j)} \log\left[ \frac{1}{T^2}\sum_{ 1 \leq t_1 , t_2 \leq T} f_{(x_j^{t_2}-x_j^0 + a) - (x_i^{t_1}-x_i^0) }  \right],  \label{Gradient_Complexity_MCMC} 
\end{align}
where $U$ is the number of pairs $( i,j )$ in the sum. We take $U=N$, by running one time over the index $i=1,\dots,N$, and considering $j=i$ to belong to a second different system. Since the MC simulations are subject to coherent rotations of the spin variables due to the global rotational symmetry of the model, which becomes continuous in the $p\to\infty$ limit, one should also maximize at each step the overlap with the planted configuration over a global rotation of the system.

In figure~\ref{fig:MC_grad_complexity} we compare the gradient estimated from MC simulations following eq.~\eqref{Gradient_Complexity_MCMC} with the one obtained from eq.~\eqref{Gradient_Complexity}. The gradient is computed by fixing $\alpha=\alpha_d+0.5$ and in correspondence of $f=f^{\rm flat}$. For $x<\theta$ not all the data points are defined if the simulation time $T$ is not long enough. This can be understood by directly inspecting eq.~\eqref{Gradient_Complexity_MCMC}. In general, above $\alpha_d$ variables will display a distribution biased towards their initial planted state. However, once we subtract the initial position, the distributions are all shifted on top of each other. This can create problems as for $a$ small the sum inside square brackets in eq.~\eqref{Gradient_Complexity_MCMC} gets contributions different from zero only from the tail of the distributions, \emph{i.e.} when $x_i$ or $x_j$ are sufficiently distant from $x_i^0$, $x_j^0$. For this reason, one should give the Monte Carlo enough time to correctly sample the tails of the distributions. We stress however that the determination of the gradient of the complexity in the region $x<\theta$ is not required in order to perform the complexity maximization procedure, since one is interested in this case in following the gradient only relatively to the soft range $x\geq \theta$.

\section{Conclusions and outlook}

The continuous coloring problem we have studied in detail in the present work has several features that make it a unique model: it is a CSP with continuous variables defined on a sparse random graph and presenting a RFOT. To the best of our knowledge this is the only model possessing all these features.
The model can be seen as a random CSP, as a mean-field model for the jamming transition and as a solvable glassy model with a RFOT. So it may be useful in several fields of research.

Being defined on a random graph, the model is solvable via the cavity method, although the solution turns out to be quite involved. The first part of this work has been dedicated to the identification of the thermodynamical phase transitions. Previous works already attempted at computing critical thresholds, but under approximations valid in specific limits. Here we provide analytically exact or numerically very accurate values for the critical lines as a function of the ``number of colors''. We clearly identify when the thermodynamical phase transition has a continuous nature and when it is preceded by a dynamical phase transition.

The comparison of the phase diagram of the continuous coloring problem with that of the much better known discrete coloring problem reveals some surprises. The space of solutions of the CCP is much larger and contains the space of solutions to the DCP: this could suggest that finding a solution to the CCP is much easier.
The critical lines we have computed say the contrary!
The CCP undergoes a phase transition at a ratio $\alpha$ of constraints per variables much smaller than for the DCP.
The explanation of this counter-intuitive observation can be found in the way constraints are satisfied in the two problems. While in the DCP the constraints are satisfied in a tight way, in a typical solution to the CCP small gaps remains between variables and this makes harder to satisfy all constraints.
Obviously the solution with all constraints satisfied tightly exists also in CCP, but it has a much smaller entropy and so it does not dominate the thermodynamic measure over the solutions, and it is not found by actual algorithms that try to solve the CCP.

Having understood that the phase transitions take place essentially because the uniform measure over the space of solution undergoes a breaking of ergodicity due to entropic reasons, in the second part of this work we have tried to re-weight solutions in order to postpone the phase transitions. To this aim we have modified the interaction potential and found the optimal one, that is the one with phase transitions taking place at the largest $\alpha$ values.
The optimization of the interaction potential has been carried on by maximizing the complexity in a fully automatized way (other procedures have been tried and discarded as more noisy and less effective).

We believe this procedure can be of general applicability in any model undergoing a RFOT: indeed any model of that kind in the relevant range of parameters has a non-zero complexity that can be used to optimize the interaction potential.
It is worth noticing that the derivative of the complexity with respect to the interaction potential can be written as a correlation function and thus estimated also via Monte Carlo simulations.
This is one more important ingredient to allow generalization of this optimization process to other models with a RFOT.

Interestingly enough the optimized potential has a strongly attractive part at short distances that has the main effect of closing the smallest gaps and allows for better packing (i.e.\ larger $\alpha$ values at the transitions).
In other words, the typical configurations obtained using the optimized potential have much more contacts and fewer small gaps with respect to the original flat measure over the solution space.
Preliminary simulations suggest that configurations of this kind (with an excess of contacts with respect to the flat measure) are those often found by smart algorithms trying to find solutions at the largest possible $\alpha$ values.
The connection between the algorithmic threshold for smart searching algorithms and the dynamical transition for the optimized interaction potential is currently under study.

The sparseness of the model allows to run efficient Monte Carlo simulations, in order to measure directly the breaking of ergodicity that takes place at the dynamical transition $\alpha_d$. Our results are compatible with a divergence at $\alpha_d$ of the timescale controlling the correlation decay.
However, the decay of the correlation function during the Monte Carlo simulation is quite slow: the tail of the correlation is well fitted by a stretched exponential with exponent close to 0.5 and the decay timescale becomes very large approaching $\alpha_d$ (the power law divergence has an exponent close to 4).
This makes the estimate of $\alpha_d$ from Monte Carlo data very noisy, especially if compared with data obtained by running the Belief propagation algorithm.

We have presented an accurate comparison between the actual behavior of the relaxation dynamics simulated via Monte Carlo algorithms and the analytical predictions obtained via the cavity method and the BP algorithm. This comparison has not been achieved before in other models (to the best of our knowledge). The reason is simple: most mean-field solvable models are defined on fully connected graphs and thus their Monte Carlo simulations are very demanding, and could not achieve sizes and timescales useful for a reliable comparison.

We conclude this work commenting on possible future applications of the present model. Having continuous variables, the model can be used to study the behavior of continuous relaxation dynamics like the stochastic gradient descent; particularly interesting would be to study how the dynamical phase transitions affects this kind of dynamics. Moreover, once a global or local minimum is reached, it would be very interesting to compute the spectral properties of the Hessian. We expect marginal states to play an important role as attracting fixed point for relaxation dynamics and the study of the Hessian spectrum can support this hypothesis. Finally, the jamming transition in this model is still to be studied in detail and connecting it to the behavior of relaxation algorithms would be very enlightening.

\section*{Acknowledgements}
The authors thank for finantial support the European Research Council under the European Unions Horizon 2020 research and innovation programme (grant No. 694925, G. Parisi) and the Italian Ministry of Foreign Affairs and International Cooperation through the Adinmat project. AGC also thanks Rafael D\'iaz Hern\'andez Rojas for useful discussions during the preparation of this work.

\appendix
\section{Planted CCP and DCP as Bayesian inference problems}
\label{appendix:inference_def}
\subsection{Community detection definition (Stochastic block model)}
In the planting procedure (without random shifts), one adds an edge $(ij)$ to the graph according to some probability\footnote{In the following, we approximate $N-1$ in the denominator with $N$, which is reasonable as $N\to\infty$.}~$2\alpha f(x_i^0,x_j^0;\beta)/N$, which depends on the values taken by $i$ and $j$ in the planted configuration $\{x_i^0\}$. This generative model for random graphs is also known as the stochastic block model~\cite{decelleAsymptoticAnalysisStochastic2011}. We may interpret the different values each variable can take as signaling their membership to a specific `community', and the stochastic block model (planting) as a rule to establish connections between these communities. A natural question is to understand under what circumstances one is able to recover some knowledge on $\{x_i^0\}$ (original community structure) from the observation of the generated graph $A$. In the simplest case (Bayes optimal), one also knows the parameters of the model, namely the prior $P_X$ from which $\{x_i^0\}$ was extracted, and the function $f(x,y;\beta)$. 

Following~\cite{decelleAsymptoticAnalysisStochastic2011}, this problem can be addressed through Bayesian inference. The conditional probability for the adjacency matrix of the graph ($A_{ij}=1$ if an edge is drawn between nodes $i$ and $j$, $A_{ij}=0$ otherwise), reads
\begin{align}
P(A|\{x_i^0\})  &= \prod\limits_{1 \leq i < j \leq N} \left( 2\alpha f(x_i^0,x_j^0;\beta)/N \right)^{A_{ij}} (1 - 2\alpha f(x_i^0,x_j^0;\beta)/N)^{1-A_{ij}}.
\label{eq:graph-ensembles}
\end{align}
The function $2\alpha f(x,y;\beta)$ is usually called the affinity function, and in our case is a symmetric function. 
The Bayesian belief for the original group assignment, conditional on the particular realization of the graph, is thus 
\begin{align}
    P(\{x_i\}\lvert A) &\propto P(A\lvert\{x_i\})P_X(\{x_i\}) = \nonumber \\
    &= \prod_{i=1}^NP_X(x_i)\prod_{1\leq i<j\leq N} \left( \frac{2\alpha f(x_i,x_j;\beta)}{N} \right)^{A_{ij}} \left(1 - \frac{2 \alpha f(x_i,x_j;\beta)}{N}\right)^{1-A_{ij}} \,.
\end{align}
We can finally define an extensive Hamiltonian by considering
\begin{align}
    &\beta \hat{\mathcal{H}}(\{x_i\}\lvert A) = -\ln P(\{x_i\}\lvert A) - M\ln N = \nonumber
    \\
    &=-\sum_{i=1}^N\ln P_X(x_i)-\sum_{1\leq i<j\leq N}\left[A_{ij}\ln\ 2\alpha f(x_i,x_j;\beta) + (1-A_{ij})\ln\left(1-\frac{2\alpha f(x_i,x_j;\beta)}{N}\right) \right] \,.
    \label{eq:H_inference}
\end{align}

We notice, in addition to the expected edge-term proportional to $A_{ij}$, the presence of a weak $1/N$ interaction between non-edges making the graph fully connected, and an external field given by the prior $P_X$. However, as was shown in~\cite{decelleAsymptoticAnalysisStochastic2011}, the non-edge interaction can be treated in the large $N$ limit in a mean-field way, thus recovering the sparsity of the graph. The BP equations associated to the planted Hamiltonian model~(\ref{eq:H_inference}) are discussed in Appendix~\ref{section:BP_CCP}.

\subsection{Mixed model}
\label{sec:mixed_model}
Using the community detection formalism, we can recast both the discrete and continuous coloring as inference problems, also defining a mixed model that allows us to interpolate between the two.

\subsubsection{Discrete and continuous coloring as inference problems}

Both the discrete and continuous coloring can be treated in a unified way by defining the following affinity function
\begin{equation}
    f(x-y;\beta) =  \frac{\exp\left( - \beta\mathbb{I}( \cos(x-y) > \cos\theta) \right)}{   \int dx dy P_X(x) P_X(y) \exp\left( - \beta\mathbb{I}( \cos(x-y) > \cos\theta) \right)},
\end{equation}
the only difference being in the prior distributions: $P_X(x) = \frac{1}{2\pi}$ for the continuous coloring and $P_X(x) = \frac{1}{q} \sum_{k = 0}^{q-1} \delta\left(x - \frac{2\pi k}{q} \right)$ for $q$-coloring, with $x_i \in [0;2\pi)$. In the limit $\beta\to\infty$, this reads
\begin{align}
    f_{q\rm -col}^{^{T=0}}(x-y) = \begin{cases}
0, &{\rm if}\,\cos( x- y) > \cos(2\pi/q) \\
\frac{q}{q - 1}, &{\rm otherwise},
\end{cases}
\end{align}
and
\begin{align}
    f_{\rm CCP}^{^{T=0}}(x-y) = \begin{cases}
0, &{\rm if}\,\cos( x- y) > \cos(2\pi/q) \\
\frac{q}{q - 2}, &{\rm otherwise}.
\end{cases}
\end{align}

\subsubsection{Interpolating (mixed) model}
\begin{figure}[t]
    \centering
    \includegraphics[width=0.60\textwidth]{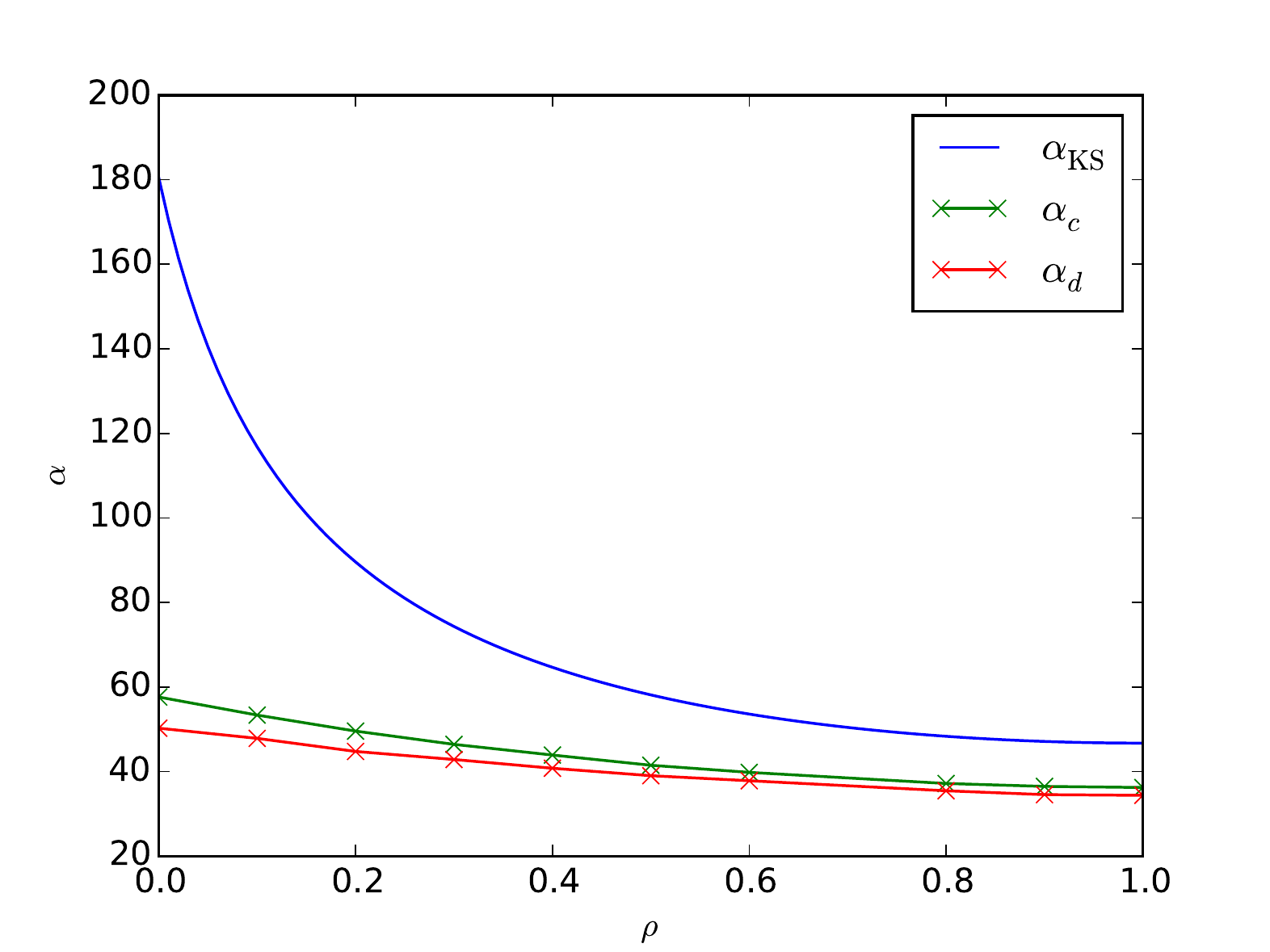}
    \caption{Phase diagram of the mixed model, eqs.~(\ref{discr-mix-continuous}-\ref{discr-mix-continuous-prior}) and~(\ref{discr-mix-4}-\ref{discr-mix-4-prior}), interpolating between discrete $(\rho = 0)$ and continuous $(\rho = 1)$ coloring. Parameters are set to $q=20$ and $d=10$.}
    \label{fig:diagram_mixedmodel_rho}
\end{figure}

We also consider a mixed model that interpolates between the discrete and the continuous coloring for each value of $\theta=\frac{2\pi}{q}$, $q\in\mathbb{N}$. To this end, we modify the prior by introducing a continuous parameter $\rho \in[0,1]$, so that for $\rho=0$ one recovers the $q$-coloring, while for $\rho=1$ one recovers the continuous version.
We will be interested in the $\beta \to \infty$ limit of the interaction function, which reads
\begin{align}
&P_X(x) = \frac{\rho}{2\pi} + \frac{1-\rho}{q} \sum_{k = 0}^{q-1} \delta\left(x - \frac{2\pi k}{q} \right), \quad\quad \rho\in[0,1]
\label{discr-mix-continuous-prior}\\
&f_{\rm Mixed}^{^{T=0}}(x,y) = \begin{cases}
0, &{\rm if}\,\cos( x- y) > \cos(2\pi/q) \\
\frac{q}{q-1}, &{\rm if}\, x,y \in \{ 2\pi k/q , \,k \in \mathbb{Z} \} \,\,  {\rm and}\,\, x \neq y
\\\frac{q}{q - 2}, &{\rm otherwise}.
\end{cases}
\label{discr-mix-continuous}
\end{align}

This peculiar shape for $f(x,y)$ is due to the fact
that, in order to ensure the normalization condition $\int {\rm d}x P_X(x) f(x,y)=1$ $\forall y\in[0;2\pi)$, a simpler translationally invariant form is not enough. This can be understood by considering that variables are effectively divided into two classes, depending on whether they can or cannot be written as $y=2\pi k/q$, for $k\in\mathbb{Z}$. If variable $y$ belongs to the former class, then it is compatible with $(q-1)$ of the $q$ possible `discrete' values that neighbours assume with probability $(1-\rho)$. Conversely, if variable $y$ belongs to the continuous background, by geometrical reasons it is then compatible with only $(q-2)$ of the $q$ possible discrete angles. If the linking probability was a function of only the difference $x-y$, this would induce a difference in the average degree of variable nodes depending on the class, and hence the normalization condition would not hold. One can still work with a uniform average degree by increasing the linking probability of variables belonging to the continuous class with respect to the discrete one. Notice that when $\rho=0$ or $\rho=1$, there is just one kind of population (the probability for continuous variables with uniform distribution to take a specific set of discrete values being negligible), and the linking probability correctly matches the $\beta\to\infty$ limits $f_{q\rm -col}^{^{T=0}}(x-y)$ and $f_{\rm CCP}^{^{T=0}}(x-y)$ given above.

\subsubsection{Discretized version of the inference problems}
\label{sec:inference_discretized_version}
In the case of discretization, variables $x$'s can take only $p$ values. We will denote the prior $P_X(x)$ as $\{n_a\}\in\mathbb{R}^p$, with $0\leq n_a\leq 1$ and $\sum_{a=0}^{p-1} n_a = 1$. The function $f(x,y)$ becomes now in general a matrix $\{f_{ab}\}\in\mathbb{R}^{p\times p}$, with $\sum_{b}n_b f_{ab} = 1$ ~$\forall a\in\{0,\dots,p-1\}$.

\begin{itemize}
    \item {Discretized version of the continuous model}
    \begin{align}
    &n_a = \frac{1}{p},
    \\
    &f^{\rm CCP}_{ab}(\beta) =  \frac{\exp\left( - \beta\mathbb{I}( \cos( 2\pi (a-b)/p) > \cos(2\pi/q)) \right)}{\left[(2d-1)e^{-\beta} + p - (2d-1) \right] /p}
    \end{align}

    \item {Discretized $q$-coloring}
    \begin{align}
    &n_a = \frac{1}{q}\sum_{k = 0}^{q-1} {\delta_{a}^{kd}}\\
    &f^{q{\rm -col}}_{ab}(\beta) =  \frac{\exp\left( - \beta\mathbb{I}( \cos( 2\pi (a-b)/p) = 1) \right)}{\left[\exp(-\beta) + (q-1) \right] /q }
    \end{align}

    \item {Discretized mixed model ($\beta\to\infty$)}
    \begin{align}
    &n_a = \frac{\rho}{p} +\frac{(1-\rho)}{q} \sum_{k = 0}^{q-1}  \delta_{a}^{kd}, \quad \rho \in [0,1]
    \label{discr-mix-4-prior}
    \\
    &f^{\rm Mixed}_{ab} = \begin{cases}
    0, &{\rm if}\,\,\cos(2 \pi (a-b)/p) > \cos(2\pi/q) 
    \\
    A=\frac{q \left[ (1-\rho)(q-2)+ \frac{\rho}{d}(q-1) \right] } {(q-1)(q-2 + \frac{\rho}{d})(1-\rho+\frac{\rho}{d})}   , &{\rm if} \,\,\sin(\pi a/d)=\sin(\pi b/d) = 0\,\, {\rm and}\,\, a \neq b
    \\
    B=\frac{q}{q-2+\frac{\rho}{d}}  &{\rm otherwise}.
    \label{discr-mix-4}
    \end{cases}
    \end{align}
    % where $\lambda$ is a constant that ensures that the average connectivity is equal to $2\alpha$,
    % \begin{align}
    % \frac{1}{\lambda} = \sum_{ab} K_{ab} n_{a} n_{b}.
    % \end{align}
    In the limit $d \to \infty$ the continuous formulation is correctly recovered, while for $\rho=0$ and $\rho=1$ we obtain the $\beta\to\infty$ limit of $f^{q{\rm -col}}_{ab}(\beta$) and $f^{{\rm CCP}}_{ab}(\beta)$, respectively. Expression~\eqref{discr-mix-4} is obtained in the same way as~\eqref{discr-mix-continuous} by solving for $A,B$ the system $\sum_a n_a f_{ab}^{\rm Mixed}(A,B)=1$ ~$\forall b$.
\end{itemize}

\section{Belief propagation for CCP}
\label{section:BP_CCP}

In this section we derive the belief propagation equations for the continuous coloring problem in the Bayes optimal inference setting, following the discussion of~\cite{decelleAsymptoticAnalysisStochastic2011}.
We denote $\nu^{i \rightarrow j}$ the message sent from $i$ to $j$ and $\nu^{i}$ the marginal probability of variable $i$. Both messages and local marginals are in principle distributions over the continuous interval $[0,2\pi)$, but thanks to discretization we treat them as $p$-components arrays $\nu_a^{i \rightarrow j}$ and $\nu_a^{i}$, with $a\in\{0,\dotso, p-1\}$. The belief propagation equations for $\nu^{i\to j}$ associated to the planted Hamiltonian model~\eqref{eq:H_inference} reads then
\begin{equation}
\nu_a^{i \rightarrow j} = \frac{1}{Z_{i \rightarrow j}} \,n_a \prod_{k \neq j} \left[ \sum_{b = 1} ^{p}  (2 \alpha f_{a b})^{A_{ik}} \left(1 - 2\alpha f_{a b}/N\right)^{1-A_{ik}}{\nu}_{b}^{k \rightarrow i} \right],
\end{equation}
while local marginals are computed in terms of the messages as
\begin{equation}
\nu_a^{i} = \frac{1}{Z_{i}} \,n_a \prod_{k=1}^N \left[ \sum_{b = 1}^{p}  (2\alpha f_{a b})^{A_{ik}} \left(1 - 2\alpha f_{a b}/N\right)^{1-A_{ik}}\nu_b^{k \rightarrow i} \right],
\end{equation}
where $Z_{i \rightarrow j}$ and $Z_i$ ensures that $\nu^{i \rightarrow j}$ and $\nu^{i}$ are normalized probability distributions.
In the following, we choose to get rid of a term $2\alpha$ multiplying $f_{ab}$ in the edge-term since it factorizes and just shifts the free energy by a fixed amount.

Since the factor graph of this system is fully connected, one should in principle have to keep track of $N(N-1)$ messages. However, one can notice that the messages $\nu^{i \rightarrow j}$ depend weakly on the indices $j$ as long as $A_{ij} = 0$,
\begin{align}
\nu_a^{i \rightarrow j} - \nu_a^{i} = O(1/N) \;\;\;\; \forall (ij) \notin E.
\label{eq:non-edges_weak_dependence}
\end{align}
In the end~\cite{decelleAsymptoticAnalysisStochastic2011}, this allows to simplify the cavity equations to only keep track of the messages between connected nodes by introducing an average external field $h_a$ which accounts for the global message coming on each site from non-edges,
\begin{align}
{\nu_a^{i \rightarrow j}}^{(t+1)} &= \frac{1}{Z_{i \rightarrow j}} \,n_a \,e^{- h_a^{(t)}} \prod_{k \in \partial i\setminus j} \left[ \sum_{b = 1}^p  f_{ab} \,{\nu_b^{k \rightarrow i}}^{(t)} \right]
\label{BP_1}
\\
{\nu_a^{i}}^{(t+1)} &= \frac{1}{Z_{i}} \,n_a \,e^{- h_a^{(t)}} \prod_{k\in\partial i} \left[ \sum_{b = 1}^p  f_{ab} \,{\nu_b^{k \rightarrow i}}^{(t+1)} \right]
\label{BP_2}
\\
h_a^{(t+1)} &= \frac{2\alpha}{N}\sum_{k=1}^N \sum_{b=1}^p f_{a b} \,{\nu_{b}^{k}}^{(t+1)},
\label{BP_3}
\end{align}
where we have added time indices in order to define update equations. For future convenience, we can also introduce an update function $\Phi^{\rm update}(\{\nu_k\}_{1 \leq k \leq d}, f,h) \in\mathbb{R}^p$ and a normalization factor $Z(\{\nu_k\}_{1 \leq k \leq d},f,h) \in \mathbb{R}$, that both take as inputs $d$ generic \emph{messages} and return
\begin{align}
Z(\{\nu_k\}_{1 \leq k \leq d},f,h) =& \sum_{a=1}^p n_a\,e^{- h_{a}} \prod_{1 \leq k \leq d} \left[\sum_{b=1}^p f_{ab} \,\nu_{k,b} \right],
\label{Update_Function_Z}
\\
\Phi^{\rm update}_a(\{\nu_k\}_{1 \leq k \leq d},f,h) =&  \frac{ n_a \,e^{- h_{a}}}{Z(\{\nu_k\}_{1 \leq k \leq d}),f,h)} \prod_{1 \leq k \leq d} \left[\sum_{b=1}^p f_{ab} \,\nu_{k,b} \right]. \label{Update_Function_F}
\end{align}

Once a fixed point to the cavity equations is found, its corresponding Bethe free entropy can be computed as
\begin{equation}
\frac{1}{N}F_{\rm Bethe}(\{ \nu^{i \rightarrow j} \},\{ \nu^i \}) = \frac{1}{N}\sum_{i=1}^N \log (Z_i) - \frac{1}{N}\sum_{1 \leq i  < j \leq N,\,A_{ij} = 1} \log (Z_{ij}) - \frac{1}{N}\sum_{1 \leq i  < j \leq N,\,A_{ij} = 0} \log(\tilde{Z}_{ij}),
\end{equation}
where
\begin{align}
Z_i &= \sum_{a=1}^p n_a e^{- h_a} \prod_{k \in \partial i} \left[ \sum_{b = 1}^p  f_{a b} \,\nu_{b}^{k \rightarrow i} \right],
\\
Z_{ij} &= \sum_{a,b}  \nu_a^{i \rightarrow j} f_{ab} \,\nu_b^{j \rightarrow i},
\\
\tilde{Z}_{ij} &= \sum_{a,b}  \nu_a^i\left(1 - 2\alpha f_{ab}/N\right) \nu_b^j = 1-\frac{2\alpha}{N}\sum_{a,b}\nu_a^i \,f_{ab}\,\nu_b^j.
\end{align}
The extra term $\tilde{Z}_{ij}$ comes from the contribution of non-edges, and has the same form of $Z_{ij}$ (messages are replaced with local marginals thanks to relation~(\ref{eq:non-edges_weak_dependence})).
The sum $\sum \log(\tilde{Z}_{ij})$ can be further simplified. By expanding the logarithm to order $1/N$ and extending the summation over all the $j\in V$, one gets in the limit $N\to\infty$
\begin{align}
\frac{1}{N}\sum_{1 \leq i  < j \leq N,A_{ij} = 0}  \log(\tilde{Z}_{ij}) &= - \frac{2 \alpha}{2}\sum_{a,b} \left[\frac{1}{N}\sum_{1 \leq i \leq N}\nu_a^i \right] f_{ab}   \left[\frac{1}{N}\sum_{1 \leq i \leq N}\nu_b^i \right] + o(1) = -\alpha + o(1).
\end{align}
The last equality comes from the fact that $N^{-1}\sum_i\nu_a^i$ converges to the prior $n_a$ (the numerical solutions to the BP equations from the planted initialization being either $\nu^i_a=n_a$ $\forall i$, or a non-trivial one characterized by a large overlap with the planted configuration, $\nu^i_a\approx\delta_{a,x_i^0}$, where $x_i^0$ is extracted according to $n_a$) and from the normalization condition $\sum_a n_a f_{ab}=1$.
%In this whole article will be always equal to \eqref{FreeEnergyAntiferro} will be always equal to $\alpha$ and $h$ will always be up to finite size correction equal to
Moreover, from this also follows that the fixed-point auxiliary field $h_a$ defined in~(\ref{BP_3}) will always be, up to finite size correction, equal to a constant $h_a = 2\alpha$, $\forall a\in\{0,\cdots, p-1\}$. The contribution to the Bethe free entropy from $\Tilde{Z}_{ij}$ and $h_a$ (inside $Z_i$) is hence only an additive constant globally amounting to $-\alpha$, and we could disregard it, as already done before. We choose, however, to keep it in the definition of $F_{\rm Bethe}$, therefore having
\begin{align}
\frac{1}{N}F_{\rm Bethe}(\{\nu^{i \rightarrow j}\},\{\nu^i\}) &= \frac{1}{N}\sum_{1 \leq i \leq N} \log(Z_i) - \frac{1}{N}\sum_{1 \leq i  < j \leq N,A_{ij} = 1} \log(Z_{ij}) - \alpha.
\label{BetheFreeEnergy_CCP}
\end{align}
The presence of the site-independent field $h_a$ can be useful to cure some instability of the BP equations on a given graph. Since $h_a$ becomes equal to a constant in the thermodynamic limit, we will in the following just remove it from every equation and assume $h_a=0$ everywhere.

\subsection{RS population dynamics}
In order to analyze the properties of a system in the thermodynamic limit, one is usually interested in computing the quenched average over an ensemble of random graphs. In the Replica Symmetric cavity equations one assumes that messages incoming to a variable node are independent, and one can therefore describe the properties of a typical fixed point in the large $N$ limit using a unique distribution of messages $\mathbb{P}(\nu)$. The update equation~\eqref{BP_1} can then be generalized to
\begin{align}
\mathbb{P}^{(t+1)}(\nu) =& \sum_d P_{\rm Edge}(d) \int\prod_{k = 1}^d d\nu_k \mathbb{P}^{(t)}(\nu_k) \,\delta\left[\nu - \Phi^{\rm update}(\{\nu_k\}_d,f)\right] \label{Cavity_Equation_RS},
\end{align}
where $\Phi^{\rm update}$ is given by eq.~\eqref{Update_Function_F}, and for Erd\H{o}s-R\'enyi random graphs one has
\begin{align}
P_{\rm Node}(d) =& \exp(-2\alpha)\frac{ (2\alpha)^d }{d!}
\\
P_{\rm Edge}(d) =& \frac{(d+1) P_{\rm Node}(d+1)}{\sum_{k = 0}^{+\infty} k P_{\rm Node}(k) }  = P_{\rm Node}(d).
\end{align}
From the equivalence of $P_{\rm Node}(d)$ and $P_{\rm Edge}(d)$, it also follows that for Erd\H{o}s-R\'enyi random graphs messages $\nu^{i \rightarrow j}$ and variable marginals $\nu^i$ are subjected to the same statistics $\mathbb{P}(\nu)$ given by eq.~\eqref{Cavity_Equation_RS}. In the following we will thus neglect the superscripts and work with a unique family $\{\nu\}$ of messages.
The Bethe free energy~\eqref{BetheFreeEnergy_CCP} can be rewritten as
\begin{align}
\frac{1}{N}F_{\rm RS}(\mathbb{P}) =& \sum_d P_{\rm Node}(d) \int\prod_{k = 1}^d d\nu_k \mathbb{P}(\nu_k) \log \left[Z(\{\nu_k\}_d,f)\right]  \nonumber
\\
&-\alpha \int d\nu_1 d\nu_2  \,\mathbb{P}(\nu_1)  \mathbb{P}(\nu_2)  \log\left[ \sum_{ab} \nu_{1,a} f_{ab}  \,\nu_{2,b} \right] - \alpha,
\label{F_Cavity_Equation_RS}
\end{align}
where $Z$ is given by~\eqref{Update_Function_Z}.

In practice, one represents $\mathbb{P}(\nu)$ by a finite population of $U$ messages,  with $U$ large enough (typically, $U=O(10^6)$ messages), and estimates \eqref{Cavity_Equation_RS} by sampling messages from this set. The following algorithm (Algorithm~\ref{alg:Cavity_RS}) updates the starting population $\mathbb{P} = \{ \nu_k \}_{1 \leq k \leq U}$ according to equation~\eqref{Cavity_Equation_RS}.

\begin{algorithm}[hbtp!]
\caption{Replica symmetric population dynamics}
\label{alg:Cavity_RS}
\begin{algorithmic}[1]
\State $\mathbb{P}^{\rm input} = \{ \nu_k \}_{1 \leq k \leq U}$; 
\State $\mathbb{P}^{\rm output} \Leftarrow \{ \}$;
\For{$i = 1\cdots U$} 
    \State sample $d$ according to $P_{\rm Edge}(d)$;
    \State $S \Leftarrow \{\}$;
    \For{$j = 1 \cdots d$}
        \State sample $\nu_j$ uniformly from $\mathbb{P}^{\rm input}$;
        \State $S \Leftarrow S \cup \{ \nu_j \} $;
    \EndFor
    \State $\nu \Leftarrow \Phi^{\rm update}(S,f)$ ;
    \State $\mathbb{P}^{\rm output} \Leftarrow \mathbb{P}^{\rm output} \cup \{ \nu \} $;
\EndFor
\State \Return $\mathbb{P}_{\rm output}$;
\end{algorithmic}
\end{algorithm}

\subsection{Reconstruction equations for \texorpdfstring{$m=1$}{Lg}}

The assumption that incoming messages are independent can end up being false if the density of the constraints $\alpha$ is too large. In a factor graph with locally tree-like structure, this is usually due to the fact that variables start to develop correlations on large distances, togheter with the presence of long loops, which may then become relevant. When this is the case, one has to resort to the 1RSB cavity equations to study the system. The 1RSB population dynamics equations generally describe the distribution of distributions of messages, and are therefore computationally much heavier to implement. Luckily, limiting oneself to the case of Parisi replica symmetry breaking parameter $m=1$, and when the BP equations exhibit a para-magnetic fixed point, one can perform a simplification of the 1RSB equations that gives them a RS structure. Since the 1RSB computation at $m=1$ is correct up to $\alpha_c$, this allow us to derive the dynamical and Kauzmann transitions.

As a result of this simplification, rather than having only one  distribution of messages $\mathbb{P}(\nu)$, one keeps track of $p$ distributions $\mathbb{P}_a(\nu), \,a \in \{0,\dots, p-1\}$, where $p$ is the size of the (discrete) alphabet that variables can assume~\cite{mezardInformationPhysicsComputation2012}. The update equation~\eqref{Cavity_Equation_RS} takes the following form $\forall a$
\begin{align}
\mathbb{P}^{(t+1)}_a(\nu) =& \sum_d P_{\rm Edge}(d) \int\prod_{k = 1}^d \left[d\nu_k \sum_b n_b f_{ab} \,\mathbb{P}^{(t)}_b(\nu_k) \right]\delta\left(\nu - \Phi^{\rm update}(\{\nu_k\}_d,f)\right) .
\label{Reconstruction_BP}
\end{align}
The Bethe free entropy becomes now
\begin{align}
\frac{1}{N}F_{\rm Rec}(\{\mathbb{P}_a\})  = & \sum_d P_{\rm Node}(d) \int\prod_{k = 1}^d \left[ d\nu_k \sum_b n_bf_{ab} \,\mathbb{P}_b(\nu_k) \right] \log \left[Z(\{\nu_k\}_d,f) \right]
\nonumber \\
&- \alpha \sum_{a,b} n_a n_b f_{ab}  \int d\nu_1 d\nu_2  \,\mathbb{P}_a(\nu_1)  \mathbb{P}_b(\nu_2)  \log\left[ \sum_{r,s} \nu_{1,r} f_{rs} \,\nu_{2,s} \right] - \alpha \label{FreeEnergy_Equation_Reconstruction}
\end{align}
To iteratively find a non-trivial fixed point to~\eqref{Reconstruction_BP}, one starts with $p$ populations $\mathbb{P}^{\rm Init}_a=\{\nu_k\}_{1\leq k\leq U}$ initialized in the planted solution
\begin{align}
\nu_{k,b} = \delta_{a,b} \quad\quad \forall \nu_k \in \mathbb{P}^{\rm Init}_a, \; a\in\{0,\dots,p-1\} 
%\mathbb{P}^{\rm Init}_a(m) = \prod_{b = 1}^p\delta(m_b  - \delta_b^a).
\label{Init_Reconstruction_Equation}
\end{align}
The algorithm implementing eq.~\eqref{Reconstruction_BP} is given in Algorithm~\ref{alg:Cavity_General_Reconstruction_EQ}. 

\begin{algorithm}[H]
\caption{General reconstruction equation population dynamics}
\label{alg:Cavity_General_Reconstruction_EQ}
\begin{algorithmic}[1]
\State $\mathbb{P}^{\rm input}_a = \{ \nu_k \}_{1 \leq k \leq U}$ ~$\forall a \in \{0,\dots, p-1\}$;
\State $\mathbb{P}^{\rm output}_a \Leftarrow \{ \}$ ~$\forall a \in \{0,\dots, p-1\}$;
\For{$a = 0 \cdots (p-1)$} 
    \For{$i = 1 \cdots U$} 
        \State sample $d$ according to $P_{\rm Edge}(d)$;
        \State $S \Leftarrow \{\}$;
        \For{$j = 1 \cdots d$} 
            \State sample $b \in \{0, \dots, p-1\}$  with probability $f_{ab}n_b$;
            \State sample $\nu_j$ uniformly from $\mathbb{P}^{\rm input}_b$;
            \State $S \Leftarrow S \cup \{ \nu_j \} $;
        \EndFor
        \State $\nu \Leftarrow \Phi^{\rm update}(S,f)$;
        \State $\mathbb{P}^{\rm output}_a \Leftarrow \mathbb{P}^{\rm output}_a \cup \{ \nu \}$;
    \EndFor  
\EndFor
\State \Return $\mathbb{P}^{\rm output}_a$;
\end{algorithmic}
\end{algorithm}

As long as one works with a flat prior, as $P_X(x) = \frac{1}{2	\pi}$ or  $n_a = \frac{1}{p}\equiv n$ ~$\forall a\in\{0,\dots,p-1\}$, and the matrix $f_{ab}$ depends only on the modulus of the difference between variables
$f_{ab} = f_{0,(b-a+p)\mathrm{mod}\,p}$, the system has a global rotational symmetry and one can avoid keeping track of $p$ populations $P_a$, while can just focus on messages centered around zero and then shift them. The presence of a rotational symmetry implies
\begin{align}
\mathbb{P}_b(\nu) = \mathbb{P}_a(T(a-b)\nu),
\end{align}
where $T(\Delta x)$ is the linear transformation that shifts a message $\nu(x)$ by an amount $\Delta x$. This allows us to simplify the equations, by keeping track of one distribution of messages $\mathbb{P}$ only, as in the RS case. For convenience, we fix the index $a$ in equation~\eqref{Reconstruction_BP} to $a=0$, corresponding to messages centered around $x=0$. We also name $f_a\equiv f_{0a}$ and $f_{b-a}\equiv f_{0,(b-a+p)\mathrm{mod}\,p}$ (the matrix $f_{ab}$ being a single-variable function $f_a$ for $a\in\{0,\dots p-1\}$, with parity $f_{a}=f_{p-a}$). The update equation for the distribution of messages $\mathbb{P}$, along with the planted initial condition, then becomes
\begin{equation}
\mathbb{P}^{(t+1)}(\nu) = \sum_d P_{\rm Edge}(d) \int\prod_{k = 1}^d  \left[d\nu_k \sum_b n f_b  \,\mathbb{P}^{(t)}\left(T(-b)\nu_k\right) \right] \delta\left(\nu - \Phi^{\rm update}(\{\nu_k\}_d,f) \right),
\label{Reconstruction_BP_II}
\end{equation}
\begin{equation}
\nu_{k,b} = \delta_{0,b} \quad\quad \forall \nu_k \in \mathbb{P}^{\rm Init}.
%\mathbb{P}^{\rm Init}(\nu) =& \prod_{b = 1}^p\delta(\nu_b  - \delta_b^0)
\label{Init_Reconstruction_Equation_II}
\end{equation}

The implementation of the precedent equation is displayed in Algorithm~\ref{alg:Cavity_RotSym_Reconstruction_EQ}. The procedure formally consists in a RS population dynamics, where a planting is enforced through the introduction of random shifts for the messages extracted according to the interaction probability $nf_b$. Finally, the 1RSB $m=1$ Bethe free entropy takes the form
\begin{align}
\frac{1}{N}F_{\rm Rec\, II}(\mathbb{P}) =& \sum_d P_{\rm Node}(d) \int\prod_{k = 1}^d  \left[d\nu_k \sum_b n f_b \,\mathbb{P}^{(t)}\left(T(-b)\nu_k\right) \right] \log \left[Z(\{\nu_k\},f) \right] \nonumber 
\\&- \alpha\int d\nu_1 d\nu_2  \,\mathbb{P}^{(t)}(\nu_1)  \sum_{b} n f_b \,\mathbb{P}^{(t)}(T(-b)\nu_2)  \log\left[ \sum_{r,s} \nu_{1,r} f_{s-r} \,\nu_{2,s} \right] - \alpha .
\label{FreeEnergy_Equation_Reconstruction_Sym}
\end{align}

\begin{algorithm}[H]
\caption{Simplified reconstruction equation population dynamics ($\rm II$)}
\label{alg:Cavity_RotSym_Reconstruction_EQ}
\begin{algorithmic}[1]
\State $\mathbb{P}^{\rm input} = \{ \nu_k \}_{1 \leq k \leq U}$;
\State $\mathbb{P}^{\rm output} \Leftarrow \{ \}$;
\For{$i = 1 \cdots U$} 
    \State sample $d$ according to $P_{\rm Edge}(d)$;
    \State $S \Leftarrow \{\}$;
    \For{$j = 1 \cdots d$} 
        \State sample $b \in \{0,\dots,(p-1)\}$  with probability $n f_b$;
        \State sample $\nu_j$ uniformly from $\mathbb{P}^{\rm input}$;
        \State $S \Leftarrow S \cup \{ T(-b) \nu_j \} $;
    \EndFor
    \State $\nu \Leftarrow \Phi^{\rm update}(S,f)$;
    \State $\mathbb{P}^{\rm output} \Leftarrow \mathbb{P}^{\rm output} \cup \{ \nu \}$;
\EndFor
\State \Return $\mathbb{P}^{\rm output}$;
\end{algorithmic}
\end{algorithm}

\subsection{Stability analysis}
\label{section:stability_analysis}
The stability of the BP fixed points can be analyzed by using the population dynamics equations. One way to do this is by creating two copies of the same population and then by slightly perturbing one of the two, \emph{i.e.} $P_1 = \{ \nu_k \}_{1 \leq k \leq U}$ and $P_2 = \{ \nu_k + \epsilon_k\}_{1 \leq k \leq U}$, where the $\epsilon_k$'s are small perturbations to the fixed point messages.
One then keep updating in parallel both of them, using the same choice of messages, and tracks whether and how the difference between the two populations grows or decreases. A cleaner way to perform this task is by only keeping track of the difference between the original and perturbed messages to first order. This leads to some modified version of the reconstruction equations, where one simultaneously evolve both the messages $\nu_k$ and their linear perturbations $\epsilon_k$,
\begin{align}
\mathbb{P}^{(t+1)}(\nu,\epsilon) =& \sum_d P_{\rm Edge}(d) \int\prod_{k = 1}^d \left[ d\nu_k d\epsilon_k \sum_b n f_{b} \,\mathbb{P}^{(t)}\left(T(-b)\nu_k,T(-b)\epsilon_k\right) \right] \nonumber  \\
&\delta\left(\nu - \Phi^{\rm update}(\{\nu_k\},f) \right) \delta\left(\epsilon- \sum_{j = 1}^d \frac{\partial \Phi^{\rm update}(\{\nu_k\},f)}{\partial \nu_j} \epsilon_j\right).
\label{eq:reconstruction_messages_perts}
\end{align}
In the argument of the second $\delta$ function we have used a shorthand notation for the following matrix product
\begin{align}
    &\left(\frac{\partial \Phi^{\rm update}(\{\nu_k\},f)}{\partial \nu_j} \epsilon_j \right)_a = \sum_b \frac{\partial \Phi_a^{\rm update}(\{\nu_k\},f)}{\partial m_{j,b}} \epsilon_{j,b} = \nonumber \\
    &= \frac{n}{Z(\{\nu_k\},f)} \left(\prod_{1\leq k \leq d, k\neq j} \hat{\nu}_{k,a} \right) \hat{\epsilon}_{j,a} - \frac{\Phi_a^{\rm update}(\{\nu_k\},f)}{Z(\{\nu_k\},f)} \sum_c n \left(\prod_{1\leq k \leq d, k\neq j} \hat{\nu}_{k,c} \right) \hat{\epsilon}_{j,c},
\end{align}
where we have introduced for convenience the auxiliary (unnormalized) messages
\begin{align}
    \hat{\nu}_{k,a}(\nu_k, f) &= \sum_b \nu_{k,b}f_{b-a}, \\
    \hat{\epsilon}_{k,a}(\epsilon_k, f) &= \sum_b \epsilon_{k,b}f_{b-a}.
\end{align}
\begin{algorithm}[H]
\caption{Simplified reconstruction equation with tracking of first order difference}
\label{alg:Cavity_RotSym_Stab_Reconstruction_EQ}
\begin{algorithmic}[1]
\State $\mathbb{P}^{\rm input} = \{ (\nu_k,\epsilon_k) \}_{1 \leq k \leq U}$;
\State $\mathbb{P}^{\rm output} \Leftarrow \{ \}$;
\For{$i = 1 \cdots U$} 
    \State sample $d$ according to $P_{\rm Edge}(d)$;
    \State $S \Leftarrow \{\}$;
    \For{$j = 1 \cdots d$} 
        \State sample $b \in \{0,\dots, p-1\}$  with probability $n f_{b}$;
        \State sample $(\nu_j,\epsilon_j)$ uniformly from $\mathbb{P}^{\rm input}$;
        \State $S \Leftarrow S \cup \{ T(-b) \nu_j,T(-b) \epsilon_j \} $;
    \EndFor
    \State $\nu \Leftarrow \Phi^{\rm update}(S,f)$;
    \State $\epsilon \Leftarrow \sum_{k=1}^d \frac{\partial \Phi^{\rm update}(S,f)}{\partial \nu_k} \epsilon_k $;
    \State $P^{\rm output} \Leftarrow P^{\rm output} \cup \{ \nu,\epsilon \} $;
\EndFor
\State \Return $\mathbb{P}^{\rm output}$;
\end{algorithmic}
\end{algorithm}

\newpage

\bibliographystyle{unsrt}
\bibliography{biblio}

\end{document}